\documentclass[a4paper,12pt,twoside]{article}
\usepackage{psfig}
\usepackage{epsfig}
\usepackage{amssymb,wasysym,bbm,feynmf,rotating,subfigure,here,citesort}
\usepackage{graphicx}
\newcommand{\alphaS}{\alpha_s}
\newcommand{\alphaEM}{\alpha}
\newcommand{\SVM}{SVM}
\newcommand{\pert}{P}
\newcommand{\nprt}{N\!P}
\newcommand{\pbar}{{\bar{p}}}
\newcommand{\qbar}{{\bar{q}}}

\newcommand{\Reggeon}{I\!\!R}
\newcommand{\Pomeron}{I\!\!P}
\newcommand{\be}{\begin{equation}}
\newcommand{\ee}{\end{equation}}
\newcommand{\bea}{\begin{eqnarray}}
\newcommand{\eea}{\end{eqnarray}}
\newcommand{\benn}{\begin{displaymath}}
\newcommand{\eenn}{\end{displaymath}}
\newcommand{\beann}{\begin{eqnarray*}}
\newcommand{\eeann}{\end{eqnarray*}}
\newcommand{\barray}{\begin{array}}
\newcommand{\earray}{\end{array}}
\newcommand{\inv}{\frac{1}}
\newcommand{\gtsim}{\gtrsim}
\newcommand{\ltsim}{\lesssim}
\newcommand{\fm}{\mbox{fm}}
\newcommand{\mb}{\mbox{mb}}
\newcommand{\nb}{\mbox{nb}}
\newcommand{\MeV}{\mbox{MeV}}
\newcommand{\GeV}{\mbox{GeV}}
\newcommand{\TeV}{\mbox{TeV}}
\newcommand{\G}{{\cal G}}       
\newcommand{\GG}{\hat{\cal{G}}} 
\newcommand{\Identity}{{1\!\rm l}}

\newcommand{\impactT}{{\cal T}}
\newcommand{\Pc}{{\cal P}}      
\newcommand{\Ps}{{\cal P}_S}    
\newcommand{\Tr}{\mbox{Tr}}    
\newcommand{\im}{\mbox{Im}}    
\newcommand{\Order}{{\cal O}}   
\newcommand{\befig}{\begin{figure}}
\newcommand{\efig}{\end{figure}}
\newcommand{\betab}{\begin{table}}
\newcommand{\etab}{\end{table}}
\begin{document}
%
%
%
\pagestyle{empty}
%
%

\title{ \vspace*{-1cm} {\normalsize\rightline{HD-THEP-02-04}}
  {\normalsize\rightline{hep-ph/0202012}}  \vspace*{0.cm} {\Large
\bf  
\boldmath$S$-Matrix Unitarity, Impact Parameter Profiles, \\ 
Gluon Saturation and High-Energy Scattering
}}
\author{}
\date{} \maketitle

\vspace*{-2.5cm}

\begin{center}

\renewcommand{\thefootnote}{\alph{footnote}}

{\large 
A.~I.~Shoshi$^{1,}$\footnote{shoshi@tphys.uni-heidelberg.de},
F.~D.~Steffen$^{1,}$\footnote{Frank.D.Steffen@thphys.uni-heidelberg.de}, and
H.~J.~Pirner$^{1,2,}$\footnote{pir@tphys.uni-heidelberg.de}}


{\it $^1$Institut f\"ur Theoretische Physik, Universit\"at Heidelberg,\\
Philosophenweg 16 {\sl \&}\,19, D-69120 Heidelberg, Germany}


{\it $^2$Max-Planck-Institut f\"ur Kernphysik, Postfach 103980, \\
D-69029 Heidelberg, Germany}


\end{center}


\begin{abstract}
  
  A model combining perturbative and non-perturbative QCD is developed
  to compute high-energy reactions of hadrons and photons and to
  investigate saturation effects that manifest the $S$-matrix
  unitarity. Following a functional integral approach, the $S$-matrix
  factorizes into light-cone wave functions and the universal
  amplitude for the scattering of two color-dipoles which are
  represented by Wegner-Wilson loops. In the framework of the
  non-perturbative stochastic vacuum model of QCD supplemented by
  perturbative gluon exchange, the loop-loop correlation is calculated
  and related to lattice QCD investigations. With a universal energy
  dependence motivated by the two-pomeron (soft + hard) picture that
  respects the unitarity condition in impact parameter space, a
  unified description of $pp$, $\pi p$, $Kp$, $\gamma^* p$, and
  $\gamma\gamma$ reactions is achieved in good agreement with
  experimental data for cross sections, slope parameters, and
  structure functions. Impact parameter profiles for $pp$ and
  $\gamma^*_{L}p$ reactions and the gluon distribution of the proton
  $xG(x,Q^2,|\vec{b}_{\!\perp}|)$ are calculated and found to saturate
  in accordance with $S$-matrix unitarity. The c.m.\ energies and
  Bjorken $x$ at which saturation sets in are determined.

  \vspace{1.cm}


\medskip


\end{abstract}


%
%
\newpage
\pagestyle{empty}
\tableofcontents
\addtocontents{toc}{\protect\enlargethispage{2.cm}}
%
%
\newpage
\pagenumbering{arabic}
\pagestyle{plain}
%
\makeatletter
\@addtoreset{equation}{section}
\makeatother
\renewcommand{\theequation}{\thesection.\arabic{equation}}
%
%
\section{Introduction}
\label{Sec_Introduction}

One of the challenges in quantum chromodynamics (QCD) is the
description and understanding of hadronic high-energy scattering.
Since the momentum transfers can be small, the QCD coupling constant
is too large for a reliable perturbative treatment. Non-perturbative
QCD is needed which is rigorously only available as a computer
simulation on Euclidean lattices. Since lattice QCD is limited to
Euclidean space-time, it cannot be applied in Minkowski space-time to
describe particles moving near the light-cone. Only static properties
of hadrons such as the hadron spectrum or the phenomenon of
confinement can be accessed within lattice QCD until now.

An interesting phenomenon observed in hadronic high-energy scattering
is the rise of the total cross sections with increasing c.m.\ energy.
While the rise is slow in hadronic reactions of {\em large} particles
such as protons, pions, kaons, or real photons~\cite{Groom:2000in}, it
is steep if only one {\em small} particle is involved such as an
incoming virtual photon~\cite{Adloff:1997mf+X,Adloff:2001qk} or an
outgoing charmonium~\cite{Breitweg:1997rg+X}.

This energy behavior is best displayed in the proton structure function
$F_2(x,Q^2)$ that is equivalent to the total $\gamma^* p$ cross
section. With increasing photon virtuality $Q^2$, the increase of
$F_2(x,Q^2)$ towards small Bjorken $x$ becomes significantly stronger.
Together with the steep rise of the gluon distribution in the proton
$xG(x,Q^2)$ with decreasing $x$, the rise of the structure function
$F_2(x,Q^2)$ towards small $x$~\cite{Adloff:2001qk,Adloff:1997mf+X} is
one of the most exciting results of HERA.

Postulating the unitarity of the $S$-matrix, $SS^{\dagger} =
S^{\dagger}S = \Identity$, asymptotic limits on the growth of total
hadronic cross sections have been derived such as the
Froissart-Lukaszuk-Martin bound~\cite{Froissart:1961ux+X}. This
limit allows at most a logarithmic energy dependence at asymptotic
energies. Analogously, the rise of the total $\gamma^* p$ cross
section is expected to slow down due to parton saturation effects
reflecting $S$-matrix unitarity. In fact, it is a key issue to
determine the energy at which unitarity limits become significant.

A phenomenologically very successful and economical description of the
energy dependence in both hadron-hadron reactions and $\gamma^* p$
reactions is given by the two-pomeron model of Donnachie and
Landshoff~\cite{Donnachie:1998gm+X}. In this picture, the energy
dependence of the cross sections at high energies results from the
exchange of a soft and a hard pomeron, the first of which dominates in
hadron-hadron and $\gamma^* p$ reactions at low
$Q^2$~\cite{Donnachie:1992ny} and the second one in $\gamma^* p$
reactions at high $Q^2$~\cite{Donnachie:1998gm+X}. Both pomerons carry
by definition the quantum numbers of the vacuum and may be related to
a glueball trajectory~\cite{Donnachie:1998gm+X} or a gluon
ladder~\cite{BFKL}. The two-pomeron model, however, explicitly
violates the Froissart-Lukaszuk-Martin bound~\cite{Froissart:1961ux+X}
at asymptotic energies and does not contain parton saturation. A model
motivated by the concept of parton saturation is the one of
Golec-Biernat and W\"usthoff~\cite{Golec-Biernat:1999js+X} which
allows very successful fits to $\gamma^* p$ data but cannot be applied
to hadron-hadron reactions.
  
In this work, we develop a model combining perturbative and
non-perturbative QCD to compute high-energy reactions of hadrons and
photons with special emphasis on saturation effects that manifest the
$S$-matrix unitarity. Aiming at a unified description of
hadron-hadron, photon-hadron, and photon-photon reactions involving
real and virtual photons as well, we follow the {\em functional
  integral approach} to high-energy scattering in the eikonal
approximation~\cite{Nachtmann:1991ua,Kramer:1990tr,Dosch:1994ym,Nachtmann:ed.kt},
in which the $S$-matrix element factorizes into the universal
correlation of two light-like Wegner-Wilson loops $S_{DD}$ and
reaction-specific light-cone wave functions. The light-like
Wegner-Wilson loops describe color-dipoles given by the quark and
antiquark in the meson or photon and in a simplified picture by a
quark and diquark in the baryon. Consequently, hadrons and photons are
described as color-dipoles with size and orientation determined by
appropriate light-cone wave
functions~\cite{Dosch:1994ym,Nachtmann:ed.kt}. Thus, the {\em
  loop-loop correlation function} $S_{DD}$ is the basis for our
unified description.
 
We evaluate the loop-loop correlation function $S_{DD}$ in the
approach of Berger and Nachtmann~\cite{Berger:1999gu}. In this
approach, the $S$-matrix unitarity condition is respected as a
consequence of a matrix cumulant expansion and the Gaussian
approximation of the functional integrals. We explicitly review the
Berger-Nachtmann approach as it is crucial for our loop-loop
correlation model and our investigation of saturation effects.

We express the loop-loop correlation function $S_{DD}$ in terms of the
gauge-invariant bilocal gluon field strength correlator integrated
over two connected surfaces. These surfaces enter from an application
of the non-Abelian Stokes' theorem, in which the line integrals are
transformed into integrals over surfaces bounded by the loops. We use
for the first time explicitly {\em minimal surfaces}. This surface
choice is usually used to obtain Wilson's area law in Euclidean
space~\cite{Dosch:1987sk+X,Euclidean_Model_Applications}. The
simplicity of the minimal surfaces is appealing. It allows us to
present the explicit computation of $S_{DD}$ compactly in this work
and to extract an analytic structure of the non-perturbative
contribution to the dipole-dipole cross section
in~\cite{K-Space_Investigations}.

We decompose the gluon field strength correlator into a
non-perturbative and a perturbative component. The {\em stochastic
  vacuum model} (\SVM)~\cite{Dosch:1987sk+X} is used for the
non-perturbative low frequency background field and {\em perturbative
  gluon exchange} for the additional high frequency contributions.
This combination allows us to describe long and short distance
correlations in agreement with Euclidean lattice calculations of the
gluon field strength
correlator~\cite{DiGiacomo:1992df+X,Meggiolaro:1999yn}.  Moreover, if
applied with the minimal surface in Euclidean space-time, this two
component ansatz leads to the static quark-antiquark potential with
color-Coulomb behavior at short distances and confining linear rise at
long distances~\cite{Euclidean_Model_Applications}. In this way, a
connection of high-energy scattering to lattice simulations of QCD and
the QCD string tension is established.
 
We use in the non-perturbative component the {\em exponential
  correlation function} directly from lattice QCD investigations of
the correlator~\cite{Meggiolaro:1999yn}. This correlation function
stays positive for all Euclidean distances and, thus, is compatible
with a spectral representation of the correlation
function~\cite{Dosch:1998th}. This means a conceptual improvement
since the correlation function that has been used in earlier
applications of the \SVM\ becomes negative at large
distances~\cite{Dosch:1994ym,Rueter:1996yb+X,Dosch:1998nw,Rueter:1998up,D'Alesio:1999sf,Berger:1999gu,Dosch:2001jg}.
Besides the conceptual and computational advantages, the new
combination --- exponential correlation function and minimal surfaces
--- allows a successful phenomenological description of the slope
parameter $B(s)$, the differential elastic cross section
$d\sigma^{el}/dt(s,t)$, and the elastic cross section $\sigma^{el}(s)$
as shown in this work.

We introduce the energy dependence into the loop-loop correlation
function $S_{DD}$ in order to describe simultaneously the energy
behavior in hadron-hadron, photon-hadron, and photon-photon reactions
involving real and virtual photons as well. Motivated by the
two-pomeron picture of Donnachie and
Landshoff~\cite{Donnachie:1998gm+X}, we ascribe to the soft and hard
component a weak and strong energy dependence, respectively. Including
{\em multiple gluonic interactions}, we obtain an $S$-matrix element
with a universal energy dependence that respects unitarity in impact
parameter space.

We adjust the model parameters to reproduce a wealth of high-energy
scattering data, i.e.\ total, differential, and elastic cross
sections, structure functions, and slope parameters for many different
reactions over a large range of c.m.\ energies. In this way, we have
confidence in our model predictions for future experiments (LHC,
THERA) and for energies beyond the experimentally accessible range.

To study saturation effects that manifest the $S$-matrix unitarity, we
consider the scattering amplitudes in impact parameter space, where
the $S$-matrix unitarity imposes rigid limits on the impact parameter
profiles such as the {\em black disc limit}. Having confirmed that our
model respects the unitarity condition in impact parameter space, we
compute profile functions for proton-proton and longitudinal
photon-proton scattering that explicitly saturate at the black disc
limit at high energies.  These profiles provide also an intuitive
geometrical picture for the energy dependence of the cross sections.

Using a leading twist, next-to-leading order DGLAP relation, we
estimate the {\em impact parameter dependent gluon distribution} of
the proton $xG(x,Q^2,|\vec{b}_{\perp}|)$ from the profile function for
longitudinal photon-proton scattering. We find low-$x$ saturation of
$xG(x,Q^2,|\vec{b}_{\perp}|)$ as a manifestation of the $S$-matrix
unitarity. The implications on the integrated gluon distribution
$xG(x,Q^2)$ are studied and compared with complementary investigations
of gluon saturation.

With the profile function saturation and the intuitive geometrical 
picture gained in impact parameter space, we turn to experimental
observables to localize saturation effects in the cross sections and
to interpret the energy dependence of the cross sections. We compare
the model results with the experimental data and provide predictions
for future experiments and saturation effects. Total cross sections
$\sigma^{tot}$, the structure function of the proton $F_2$, slope
parameters $B$, differential elastic cross sections $d\sigma^{el}/dt$,
elastic cross sections $\sigma^{el}$, and the ratios
$\sigma^{el}/\sigma^{tot}$ and $\sigma^{tot}/B$ are considered for
proton-proton, pion-proton, kaon-proton, photon-proton, and
photon-photon reactions involving real and virtual photons as well.

The outline of the paper is as follows: In Sec.~\ref{Sec_The_Model},
the model is developed and the model parameters are given. Going to
impact parameter space in Sec.~\ref{Sec_Impact_Parameter}, we confirm
the unitarity condition in our model and study the impact parameter
profiles for proton-proton and photon-proton scattering. The impact
parameter dependent gluon distribution of the proton
$xG(x,Q^2,|\vec{b}_{\perp}|)$ and gluon saturation are discussed in
Sec.~\ref{Sec_Gluon_Saturation}. Finally, in
Sec.~\ref{Sec_Comparison_Data}, we present the phenomenological
performance of the model and the saturation effects in the
experimental observables. The appendices present explicitly the used
hadron and photon light-cone wave functions and the analytic
continuation of the non-perturbative correlation functions from
Euclidean to Minkowski space-time.


%
\vspace*{1cm}
%
%
\section{The Loop-Loop Correlation Model}
\label{Sec_The_Model}

The $T$-matrix is the central quantity in scattering processes. It
enters every observable we intend to look at and is obtained from the
$S$-matrix by subtracting the trivial case in which the final state
equals the initial state,
\be
        S_{fi} = \delta_{fi} + i (2 \pi)^4 \delta^4(P_f - P_i) T_{fi} 
        \ ,
\label{Eq_T_matrix_element}
\ee
where $P_i$ and $P_f$ represent the sum of incoming and outgoing
momenta, respectively. We compute the $T$-matrix in a {\em functional
  integral approach} developed for parton-parton
scattering~\cite{Nachtmann:1991ua} in the {\em eikonal approximation}
and extended to gauge-invariant loop-loop
scattering~\cite{Kramer:1990tr,Dosch:1994ym,Nachtmann:ed.kt}. In this
approach, the $T$-matrix element for the reaction $ab \rightarrow cd$
factorizes as follows
\bea
        \!\!\!\!\!\!
        &&\hspace{-1cm}
        T_{ab \rightarrow cd}(s,t) =
        2is \int \!\!d^2b_{\!\perp} 
        e^{i {\vec q}_{\!\perp} {\vec b}_{\!\perp}}
        \int \!\!dz_1 d^2r_1 \!\int \!\!dz_2 d^2r_2      
        \hphantom{\hspace*{5.cm}}   
        \nonumber \\ 
        & \!\! \times \!\! & 
        \psi_c^*(z_1,{\vec r}_1)\,\psi_d^*(z_2,{\vec r}_2)
        \left[1-S_{DD}({\vec b}_{\!\perp},z_1,{\vec r}_1,z_2,{\vec r}_2)\right]
        \psi_a(z_1,{\vec r}_1)\,\psi_b(z_2,{\vec r}_2) 
        \ , 
\label{Eq_model_T_amplitude}
\eea
where the {\em loop-loop correlation function} 
\be
        S_{DD}({\vec b}_{\!\perp},z_1,{\vec r}_1,z_2,{\vec r}_2)
        = \Big\langle W[C_1] W[C_2] \Big\rangle_G
\label{Eq_loop_loop_correlation_function}
\ee
describes the elastic scattering of two color-dipoles (DD) with
transverse size and orientation ${\vec r}_i$ and longitudinal quark
momentum fraction $z_i$ at impact parameter ${\vec b}_{\!\perp}$,
transverse momentum transfer ${\vec q}_{\!\perp}$ ($t = -{\vec
  q}_{\!\perp}^{\,\,2}$) and c.m.\ energy squared $s$. In this
framework, the color-dipoles are given by the quark and antiquark in
the meson or photon and in a simplified picture by a quark and diquark
in the baryon. Consequently, the hadrons and photons are characterized
by the {\em light-cone wave functions} $\psi_{a,b}$ and $\psi_{c,d}$
that describe the ${\vec r}_i$ and $z_i$ distribution of the
color-dipoles. Concentrating in this work on reactions with $a = c$
and $b = d$, only squared wave functions
$|\psi_1(z_1,\vec{r}_1)|^2:=\psi_c^*(z_1,{\vec r}_1)\,\psi_a(z_1,{\vec
  r}_1)$ and $|\psi_2(z_2,\vec{r}_2)|^2:=\psi_d^*(z_2,{\vec
  r}_2)\,\psi_b(z_2,{\vec r}_2)$ are needed. We use for hadrons the
phenomenological Gaussian wave
function~\cite{Dosch:2001jg,Wirbel:1985ji} and for photons the
perturbatively derived wave function with running quark masses
$m_f(Q^2)$ to account for the non-perturbative region of low photon
virtuality $Q^2$~\cite{Dosch:1998nw}, as discussed explicitly in
Appendix~\ref{Sec_Wave_Functions}.

The path of each color-dipole is represented by a {\em light-like QCD
  Wegner-Wilson loop}~\cite{Wilson:1974sk+X}
\be
        W[C_{1,2}] = 
        \inv{N_c} \Tr\,\Pc
        \exp\!\left[-i g\!\oint_{\scriptsize C_{1,2}}\!\!dz^{\mu}
        \G_{\mu}(z) \right]      
        \ ,
\label{Eq_Wegner-Wilson_loop}
\ee
where $N_c$ is the number of colors, $\Tr$ the trace in color space,
$g$ the strong coupling, and $\G_{\mu}(z) = \G_{\mu}^a(z) t^a$ the
gluon field with the $SU(N_c)$ group generators $t^a$ that demand the
path ordering indicated by $\Pc$. Quark-antiquark dipoles\footnote{or
  equivalently quark-diquark ($q-qq$) or antiquark-diantiquark systems
  ($\qbar-\qbar\qbar$)} are represented by loops in the fundamental
$SU(N_c = 3)$ representation. In the eikonal approximation to
high-energy scattering the $q$ and ${\bar q}$ paths form straight
light-like trajectories.
Figure~\ref{Fig_loop_loop_scattering_surfaces} illustrates the
space-time (a) and transversal (b) arrangement of these loops.  The
world line $C_1$ ($C_2$) is characterized by its light-cone coordinate
$x^- = x^0 - x^3 = 0$ ($x^+ = x^0 + x^3 = 0$), the transverse size and
orientation ${\vec r}_1$ (${\vec r}_2$) and the longitudinal quark
momentum fraction $z_1$($z_2$) of the corresponding dipole. The impact
parameter between the loops is
\be
        {\vec b}_{\!\perp} 
        \,=\, {\vec r}_{1q} + (1-z_1) {\vec r}_{1} 
            - {\vec r}_{2q} - (1-z_2) {\vec r}_{2} 
        \,=\, {\vec r}_{1\,cm} - {\vec r}_{2\,cm} 
        \ ,
\label{Eq_impact_vector}
\ee
as shown in Fig.~\ref{Fig_loop_loop_scattering_surfaces}b, where
${\vec r}_{iq}$ (${\vec r}_{i\qbar}$) is the transverse position of
the quark (antiquark) in loop $i$, ${\vec r}_{i} = {\vec r}_{i\qbar} -
{\vec r}_{iq}$, and ${\vec r}_{i\,cm} = z_i {\vec r}_{iq} +
(1-z_i){\vec r}_{i\qbar}$.
\befig[p!]
  \begin{center}
        \epsfig{file=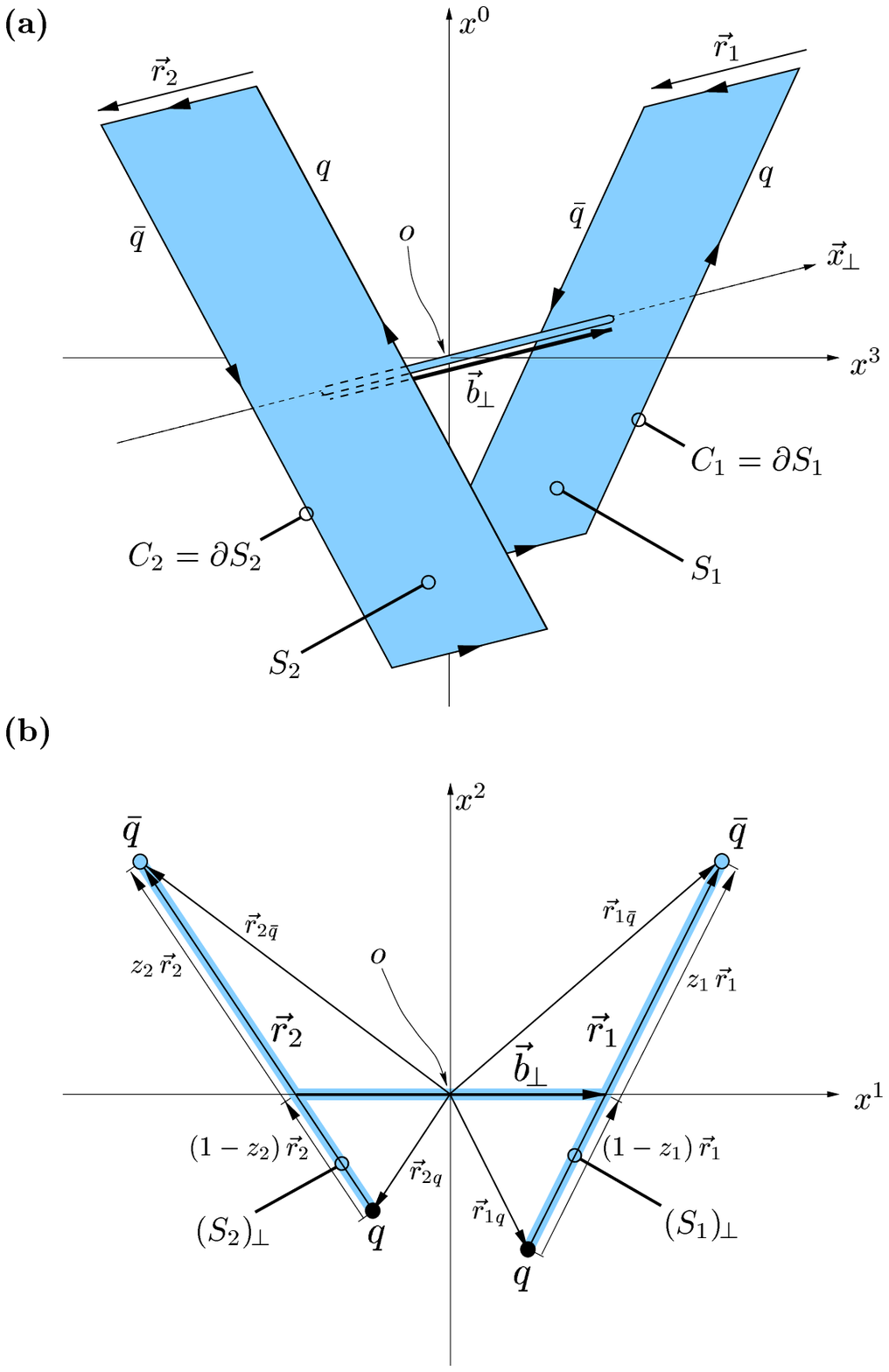,width=10.cm}
  \end{center}
\caption{\small Space-time (a) and transverse (b) arrangement of the Wegner-Wilson loops.}
\label{Fig_loop_loop_scattering_surfaces}
\efig

The QCD vacuum expectation value $\langle \ldots \rangle_G$ in the
loop-loop correlation function
(\ref{Eq_loop_loop_correlation_function}) represents functional
integrals \cite{Nachtmann:ed.kt} in which the functional integration over the fermion fields has already been carried out as indicated by the subscript $G$.  The model we use for the QCD vacuum (see Sec.~\ref{Sec_Non-pert_Pert_Cont}) describes only gluon dynamics and, thus, implies the
quenched approximation that does not allow string breaking through
dynamical quark-antiquark production.\footnote{The quenched
  approximation becomes explicit in the linear rise of the
  dipole-proton and dipole-dipole cross-section with growing dipole
  size obtained in our model.}

\subsection{The Loop-Loop Correlation Function}
\label{Sec_loop_loop_func}

To compute the loop-loop correlation function
(\ref{Eq_loop_loop_correlation_function}), we transform the line
integrals over the loops $C_{1,2}$ into integrals over surfaces
$S_{1,2}$ with $\partial S_{1,2} = C_{1,2}$ by applying the {\em
  non-Abelian Stokes' theorem}~\cite{Bralic:1980ra+X,Nachtmann:ed.kt}
\bea
        \Big\langle W[C_{1}] W[C_{2}] \Big\rangle_G = 
        \Big\langle 
          \inv{N_c} \Tr\,\Ps 
          \exp \left[-i \frac{g}{2} 
                \int_{S_{1}} \! d\sigma^{\mu\nu}(x_{1}) 
                \G^a_{\mu\nu}(o_{1},x_{1};C_{x_{1} o_{1}})\,t^a 
          \right] 
        && \nonumber \\
        \hphantom{\langle W[C_{1}] W[C_{2}] \rangle_G = }
          \times \inv{N_c} \Tr\,\Ps 
          \exp \left[-i \frac{g}{2} 
                \int_{S_{2}} \! d\sigma^{\rho\sigma}(x_{2}) 
                \G^b_{\rho\sigma}(o_{2},x_{2};C_{x_{2} o_{2}})\,t^b 
          \right]
        \Big\rangle_G
        \ ,
        && 
\label{Eq_Non-Abelian_Stokes_<W[C1]W[C2]>}
\eea
where the gluon field strength tensors, $\G_{\mu\nu}(x) =
\G_{\mu\nu}^{a}(x)t^{a}$, are parallel transported to the point
$o$ along the path $C_{xo}$
\be
        \G_{\mu\nu}(o,x;C_{xo}) 
        = \Phi(x,o;C_{xo})^{-1} \G_{\mu\nu}(x) \Phi(x,o;C_{xo})
\label{Eq_gluon_field_strength_tensor}
\ee
with the QCD Schwinger string
\be
        \Phi(x,o;C_{xo}) 
        = \Pc \exp 
        \left[-i g \int_{C_{xo}} dz^{\mu}\G_{\mu}(z)\right] 
        \ .
\label{Eq_parallel_transport}
\ee
In~(\ref{Eq_Non-Abelian_Stokes_<W[C1]W[C2]>}), $\Ps$ indicates surface
ordering and $o_{1}$ and $o_{2}$ are the reference points on the
surfaces $S_{1}$ and $S_{2}$, respectively, that enter through the
non-Abelian Stokes' theorem. In order to ensure gauge invariance in
our model, the gluon field strengths associated with the loops must be
compared at {\em one} reference point $o$. Therefore, we require the
surfaces $S_{1}$ and $S_{2}$ to touch at a common reference point
$o_{1} = o_{2} = o$.

Following the approach of Berger and Nachtmann~\cite{Berger:1999gu},
the product of the two traces ($\Tr$) over $N_c \times N_c$ matrices
in (\ref{Eq_Non-Abelian_Stokes_<W[C1]W[C2]>}) is expressed as one
trace ($\Tr_2$) that acts in the $N_c^2$-dimensional tensor product
space of two fundamental $SU(N_c)$ representations
\bea
        \Big\langle W[C_{1}] W[C_{2}] \Big\rangle_G
        & \!\!\! = \!\!\! &
        \Big\langle 
        \inv{N_c^2} \Tr_2
          \left\{
            \Big[\,\,\Ps \exp 
                \big[-i \frac{g}{2} 
                \int_{S_{1}} \! d\sigma^{\mu\nu}(x_{1}) 
                \G^a_{\mu\nu}(o,x_{1};C_{x_{1} o}) \, t^a \big] 
                \,\,\otimes\,\,\Identity\,\,\Big]
          \right.
        \nonumber \\
        &\!\!\!\! \times \!\!\!\!& \!\!
          \left. 
          \Big[\,\,\Identity\,\,\otimes\,\,
          \Ps \exp \big[-i \frac{g}{2} 
                \int_{S_{2}} \! d\sigma^{\rho\sigma}(x_{2}) 
                \G^b_{\rho\sigma}(o,x_{2};C_{x_{2} o}) \, t^b \big]
          \,\,\Big]
          \right\}
        \Big\rangle_G .
\label{Eq_trace_trick_<W[C1]W[C2]>}
\eea
Using the identities
\bea
        \exp\left(\,t^a\,\right) \,\otimes\, \Identity 
        & = & \exp\left(\,t^a \,\otimes\, \Identity\,\right) , \\
        \Identity \,\otimes\, \exp\left(\,t^a\,\right) 
        & = & \exp\left(\,\Identity \,\otimes\, t^a\,\right)
        ,
\label{Eq_exp(t^a)_times_1_identities}
\eea
the tensor products can be shifted into the exponents. With the
matrix multiplication in the tensor product space
\bea
        \big( t^a \,\otimes\, \Identity \big)
        \big( t^b \,\otimes\, \Identity \big) 
        & = & t^a t^b \,\otimes\, \Identity \ ,
        \nonumber
        \\
        \big( t^a \,\otimes\, \Identity \big)
        \big( \Identity \,\otimes\, t^b \big) 
        & = & t^a \,\otimes\, t^b \ ,
\label{Eq_matrix_multiplication_in_tensor_product_space}
\eea
and the vanishing commutator 
\be
        \left[t^a \otimes \Identity , \Identity \otimes t^b\right] 
        = 0
        \ ,
\label{Eq_[t_x_1,1_x_t]}
\ee
the two exponentials in (\ref{Eq_trace_trick_<W[C1]W[C2]>}) commute
and can be written as one exponential
\be
        \Big\langle W[C_{1}] W[C_{2}] \Big\rangle_G =
        \Big\langle 
        \inv{N_c^2} \Tr_2\,
          \Ps \exp\!
            \left[-i \frac{g}{2} 
                \int_{S} \! d\sigma^{\mu\nu}(x) 
                \GG_{\mu\nu}(o,x;C_{xo}) 
            \right]
        \Big\rangle_G    
\label{Eq_<W[C1]W[C2]>_analogous_to_<W[C]>}
\ee
with the following gluon field strength tensor acting in the
$N_c^2$-dimensional tensor product space
\be
        \GG_{\mu\nu}(o,x;C_{xo})
        := \left\{ \begin{array}{cc}
            \G_{\mu\nu}^a(o,x;C_{xo})
                \big( t^a \,\otimes\, \Identity \big)
            & \mbox{for $\,\,\, x\,\,\, \in\,\,\, S_1$} \\
            \G_{\mu\nu}^a(o,x;C_{xo})
                \big( \Identity \,\otimes\, t^b \big)
            & \mbox{for $\,\,\, x\,\,\, \in \,\,\, S_2$}
        \end{array}\right. \ .
\label{Eq_GG} 
\ee
In~(\ref{Eq_<W[C1]W[C2]>_analogous_to_<W[C]>}), the surface integrals
over $S_1$ and $S_2$ are written as one integral over the combined
surface $S = S_1 + S_2$.  For the evaluation
of~(\ref{Eq_<W[C1]W[C2]>_analogous_to_<W[C]>}), the linearity of the functional integral, $\langle\Tr \cdots \rangle = \Tr\langle \cdots \rangle$, and a {\em matrix cumulant
  expansion} is used as explained in~\cite{Nachtmann:ed.kt} (cf.\ 
also~\cite{VAN_KAMPEN_1974_1976+X})
\bea
        && \Big\langle 
                \Ps \, \exp \left[-i\frac{g}{2} \int_{S} \! d\sigma(x) 
                  \GG(o,x;C_{xo})\right] 
           \Big\rangle_G \nonumber \\
        && 
        = \exp[\,\,\sum_{n=1}^{\infty}\frac{1}{n !}(-i\frac{g}{2})^n
        \int d\sigma(x_1)\cdots d\sigma(x_n)\,K_n(x_1,\cdots ,x_n)]
        \ ,
\label{Eq_matrix_cumulant_expansion}
\eea
where Lorentz indices are suppressed to lighten notation. The
cumulants $K_n$ consist of expectation values of {\em ordered}
products of the non-commuting matrices $\GG(o,z;C_{zo})$. The leading
matrix cumulants are
\bea
        K_1(x)       
        & = & \langle \GG(o,x;C_x) \rangle_G, 
\label{Eq_K_1_matrix_cumulant}\\
        K_2(x_1,x_2) 
        & = & \langle \Ps\,
        \GG(o,x_1;C_{x_1})\GG(o,x_2;C_{x_2})\rangle_G 
        \nonumber\\
        &   & - \frac{1}{2}\left(\langle \GG
        (o,x_1;C_{x_1})\rangle_G \langle \GG(o,x_2;C_{x_2})\rangle_G 
                                 + (1 \leftrightarrow 2)\,\right) \ .
\label{Eq_K_2_matrix_cumulant}
\eea
Since the vacuum does not prefer a specific color direction, $K_1$
vanishes and $K_2$ becomes
\be
        K_2(x_1,x_2) 
        = \langle\Ps\, \GG(o,x_1;C_{x_1})\GG(o,x_2;C_{x_2})\rangle_G
        \ .
\label{Eq_K_2_matrix_cumulant<-no_color_direction_preferred}
\ee
Now, we restrict the functional integral associated with the
expectation values $\langle \ldots \rangle_G$ to be a {\em Gaussian
  functional integral}. Consequently, all higher cumulants, $K_n$ with
$n>2$, vanish\footnote{We are going to use the cumulant expansion in
  the Gaussian approximation also for perturbative gluon exchange.
  Here certainly the higher cumulants are non-zero.} and the loop-loop
correlation function can be expressed in terms of $K_2$
\bea
&& \!\!\!\!\!\!\!\!\!\!\!
        \Big\langle W[C_{1}] W[C_{2}] \Big\rangle_G 
        \nonumber \\
&& \!\!\!\!\!\!\!\!\!\!\!
        = \inv{N_c^2} \Tr_2
        \exp\!\left[-\frac{g^2}{8} \!
          \int_{S} \! d\sigma^{\mu\nu}(x_1) \!
          \int_{S} \! d\sigma^{\rho\sigma}(x_2) 
          \Big\langle \Ps\, 
          \GG_{\mu\nu}(o,x_1;C_{x_1 o}) 
          \GG_{\rho\sigma}(o,x_2;C_{x_2 o})
          \Big\rangle_G 
        \right]
        . \nonumber \\   
\label{Eq_matrix_cumulant_expansion_<W[C1]W[C2]>}
\eea

Using definition~(\ref{Eq_GG}) and the
relations~(\ref{Eq_matrix_multiplication_in_tensor_product_space}), we
now redivide the exponent in
(\ref{Eq_matrix_cumulant_expansion_<W[C1]W[C2]>}) into integrals of
the ordinary parallel transported gluon field strengths over the
separate surfaces $S_{1}$ and $S_{2}$
%
\bea
        && \Big\langle W[C_{1}] W[C_{2}] \Big\rangle_G = 
        \inv{N_c^2} \Tr_2
        \exp \Bigg[             \nonumber \\
        &&\hspace{-0.8cm}
          -\frac{g^2}{8} 
          \int_{S_1} \! d\sigma^{\mu\nu}(x_1) 
          \int_{S_2} \! d\sigma^{\rho\sigma}(x_2)\, 
          \Ps\,\Big\langle 
          \G^a_{\mu\nu}(o,x_1;C_{x_1 o}) 
          \G^b_{\rho\sigma}(o,x_2;C_{x_2 o})\Big\rangle_G
                \big(t^a  \,\otimes\, t^b\big)
          \nonumber \\
        &&\hspace{-0.8cm}
          -\frac{g^2}{8} 
          \int_{S_2} \! d\sigma^{\mu\nu}(x_1) 
          \int_{S_1} \! d\sigma^{\rho\sigma}(x_2)\, 
          \Ps\,\Big\langle 
            \G^a_{\mu\nu}(o,x_1;C_{x_1 o}) 
            \G^b_{\rho\sigma}(o,x_2;C_{x_2 o})\Big\rangle_G
                \big(t^a  \,\otimes\, t^b\big) 
            \nonumber \\
        &&\hspace{-0.8cm}
          -\frac{g^2}{8} 
          \int_{S_1} \! d\sigma^{\mu\nu}(x_1) 
          \int_{S_1} \! d\sigma^{\rho\sigma}(x_2)\, 
          \Ps\,\Big\langle 
            \G^a_{\mu\nu}(o,x_1;C_{x_1 o}) 
            \G^b_{\rho\sigma}(o,x_2;C_{x_2 o})\Big\rangle_G
                \big(t^a t^b \,\otimes\, \Identity\big)
        \nonumber \\
        &&\hspace{-0.8cm}
        \left.
          -\frac{g^2}{8} 
          \int_{S_2} \! d\sigma^{\mu\nu}(x_1) 
          \int_{S_2} \! d\sigma^{\rho\sigma}(x_2)\, 
           \Ps\,\Big\langle 
             \G^a_{\mu\nu}(o,x_1;C_{x_1 o}) 
             \G^b_{\rho\sigma}(o,x_2;C_{x_2 o})\Big\rangle_G
                \big(\Identity \,\otimes\, t^a t^b\big)
              \,\right]\!. \nonumber \\   
\label{Eq_exponent_decomposition_<W[C1]W[C2]>}
\eea
Due to the color-neutrality of the vacuum, the gauge-invariant bilocal
gluon field strength correlator contains a $\delta$-function in
color-space,
\be
        \Big\langle
        \frac{g^2}{4\pi^2}\,
        \G^a_{\mu\nu}(o,x_1;C_{x_1 o})
        \G^b_{\rho\sigma}(o,x_2;C_{x_2 o})
        \Big\rangle_G
        =: \inv{4}\delta^{ab} 
        F_{\mu\nu\rho\sigma}(x_1,x_2,o;C_{x_1 o},C_{x_2 o}) \ ,
\label{Eq_Ansatz}
\ee
which makes the surface ordering $\Ps$
in~(\ref{Eq_exponent_decomposition_<W[C1]W[C2]>}) irrelevant.  The
quantity $F_{\mu\nu\rho\sigma}$ will be specified below.  With ansatz
(\ref{Eq_Ansatz}) and the definition
\be
        \chi_{S_i S_j}
        := - \, i \frac{\pi^2}{4} 
        \int_{S_i} \! d\sigma^{\mu\nu}(x_1) 
        \int_{S_j} \! d\sigma^{\rho\sigma}(x_2)
        F_{\mu\nu\rho\sigma}(x_1,x_2,o;C_{x_1 o},C_{x_2 o}) \ ,
\label{Eq_chi_SS}        
\ee
Eq.~(\ref{Eq_exponent_decomposition_<W[C1]W[C2]>}) reads  
\bea
        \Big\langle W[C_{1}] W[C_{2}] \Big\rangle_G &=& 
        \inv{N_c^2} \Tr_2
        \exp\!\Bigg[-i\, \frac{1}{2} 
                \Big\{\,
                \left(\chi_{S_1 S_2}+\chi_{S_2 S_1}\right)
                \big(t^a \,\otimes\, t^a\big) 
        \nonumber \\
        &&  + \,\chi_{S_1 S_1} 
                \big(t^a t^a \,\otimes\, \Identity\big) 
            + \chi_{S_2 S_2}
                \big(\Identity \,\otimes\, t^a t^a\big) 
                \,\Big\}\,
              \Bigg]
        \ . 
\label{Eq_eikonal_functions_<W[C1]W[C2]>}
\eea
Our ansatz for the tensor structure of $F_{\mu\nu\rho\sigma}$ ---
see~(\ref{Eq_F_decomposition}), (\ref{Eq_MSV_Ansatz_F}),
and~(\ref{Eq_PGE_Ansatz_F}) --- leads to $\chi_{S_1 S_1} = \chi_{S_2
  S_2} = 0$ for light-like loops, as explained in
Sec.~\ref{Sec_Chi_Computation}, and also to $\chi_{S_1 S_2} =
\chi_{S_2 S_1} =: \chi$.  For the evaluation of the trace of the
remaining exponential, we employ the projectors
\bea
        &&
        (P_s)_{(\alpha_1 \alpha_2)( \beta_1 \beta_2)} =
        \frac{1}{2}
        (\delta_{\alpha_1 \beta_1} \delta_{\alpha_2 \beta_2} 
        +\delta_{\alpha_1 \beta_2} \delta_{\alpha_2 \beta_1}),
        \\
        &&
        (P_a)_{(\alpha_1 \alpha_2)( \beta_1 \beta_2)} =
        \frac{1}{2}
        (\delta_{\alpha_1 \beta_1} \delta_{\alpha_2 \beta_2} 
        -\delta_{\alpha_1 \beta_2} \delta_{\alpha_2 \beta_1}),
\label{Eq_projectors}
\eea
that decompose the direct product space of two fundamental $SU(N_c)$
representations, in short $\mbox{N}_c$, into the irreducible representations
\be
        \mbox{N}_c \,\otimes\, \mbox{N}_c 
        = (\mbox{N}_c + 1)\mbox{N}_c/2 \,\oplus\, \overline{\mbox{N}_c(\mbox{N}_c - 1)/2}
        \ .
\label{Eq_tensor_product_decomposition}
\ee
With the identity
\be
        t^a \,\otimes\, t^a 
        = \frac{N_c-1}{2N_c} P_s - \frac{N_c+1}{2N_c} P_a
        \ ,
\label{Eq_projector_ta_x_ta_relation}
\ee
and the projector properties
\be
        P^2_{s,a} = P_{s,a}
        \ , \quad 
        \Tr_2 \,P_s = (N_c + 1)N_c/2
        \ , \quad \mbox{and} \quad
        \Tr_2 \,P_a = (N_c - 1)N_c/2
        \ ,
\label{Eq_projector_properties}
\ee 
we find for the loop-loop correlation function in the fundamental
$SU(N_c)$ representation
\be       
        \Big\langle W[C_{1}] W[C_{2}] \Big\rangle_G
         =  \frac{N_c+1}{2N_c}
        \exp\!\left[- i \frac{N_c-1}{2N_c}\chi \right]
        +
        \frac{N_c-1}{2N_c}
        \exp\!\left[i \frac{N_c+1}{2N_c}\chi \right] 
        \ 
\label{Eq_final_result_<W[C1]W[C2]>}
\ee
and recover, of course, for $N_c = 3$ the result
from~\cite{Berger:1999gu}.

\subsection{Perturbative and Non-Perturbative QCD Components}
\label{Sec_Non-pert_Pert_Cont}

We decompose the gauge-invariant bilocal gluon field strength
correlator~(\ref{Eq_Ansatz}) into a perturbative ($\pert$) and
non-perturbative ($\nprt$) component
\be
        F_{\mu\nu\rho\sigma} 
        = F_{\mu\nu\rho\sigma}^{\nprt} + F_{\mu\nu\rho\sigma}^{\pert} 
        \ .
\label{Eq_F_decomposition}
\ee
Here, $F_{\mu\nu\rho\sigma}^{\nprt}$ gives the low frequency
background field contribution modelled by the non-perturbative {\em
  stochastic vacuum model} (\SVM)~\cite{Dosch:1987sk+X} and
$F_{\mu\nu\rho\sigma}^{\pert}$ the additional high frequency
contributions described by {\em perturbative gluon exchange}.  Such a
decomposition is supported by lattice QCD computations of the
Euclidean field strength
correlator~\cite{DiGiacomo:1992df+X,Meggiolaro:1999yn}.

In the \SVM, one makes the approximation that the correlator
$F_{\mu\nu\rho\sigma}^{\nprt}$ depends only on the difference $z:= x_1
- x_2$ but not on the reference point $o$ and the curves $C_{x_1 o}$
and $C_{x_2 o}$~\cite{Dosch:1987sk+X}. Then, the most general form of
the correlator that respects translational, Lorentz, and parity
invariance reads in four-dimensional Minkowski
space-time~\cite{Kramer:1990tr,Dosch:1994ym}
\bea
        F_{\mu\nu\rho\sigma}^{\nprt}(z) 
        & := & F_{\mu\nu\rho\sigma}^{\nprt_{(c)}}(z) +  
               F_{\mu\nu\rho\sigma}^{\nprt_{(nc)}}(z)
        \nonumber\\
        &  = & \inv{3(N_c^2-1)}\,G_2\, \Bigl\{
        \kappa\, \left(g_{\mu\rho}g_{\nu\sigma}
          -g_{\mu\sigma}g_{\nu\rho}\right) \,
        D(z^2/a^2)                                
        \nonumber\\
        &   &  +\,(1-\kappa)\,\inv{2}\Bigl[
        \frac{\partial}{\partial z_\nu}
        \left(z_\sigma g_{\mu\rho}
          -z_\rho g_{\mu\sigma}\right)
        +\frac{\partial}{\partial z_\mu}
        \left(z_\rho g_{\nu\sigma}
          -z_\sigma g_{\nu\rho}\right)\Bigr]\,
        D_1(z^2/a^2) \Bigr\}
        \nonumber\\
        &  = & \inv{3(N_c^2-1)}\,G_2 
        \int \frac{d^4k}{(2\pi)^4} \,e^{-ikz}\,\Bigl\{
        \kappa \,\left(g_{\mu\rho}g_{\nu\sigma}
          -g_{\mu\sigma}g_{\nu\rho}\right)\, 
        \tilde{D}(k^2)                                
        \nonumber\\
        &   &  -\,(1-\kappa)\,\Bigl[
        k_\nu k_\sigma g_{\mu\rho}   - k_\nu k_\rho   g_{\mu\sigma}
        + k_\mu k_\rho  g_{\nu\sigma} - k_\mu k_\sigma g_{\nu\rho} \Bigr]\,
               \tilde{D}_{1}^{\prime}(k^2) \Bigr\} 
        \ .
\label{Eq_MSV_Ansatz_F}
\eea
Here, $a$ is the {\em correlation length}, $G_2 := \langle
\frac{g^2}{4\pi^2} \G^a_{\mu\nu}(0) \G^a_{\mu\nu}(0) \rangle$ is the
{\em gluon condensate}~\cite{Shifman:1979bx+X}, $\kappa$
determines the non-Abelian character of the correlator, $D$ and $D_1$
are {\em correlation functions} in four dimensional Minkowski
space-time, and
\be
        \tilde{D}_{1}^{\prime}(k^2) 
        := \frac{d}{dk^2} \int d^4z D_{1}(z^2/a^2) \ e^{ikz} 
        \ .
\label{Eq_D1_prime}
\ee

In the case of $\kappa \neq 0$, the Euclidean version of
$F_{\mu\nu\rho\sigma}^{\nprt_{(c)}}(z)$ in (\ref{Eq_MSV_Ansatz_F})
leads to {\em confinement} and does not fulfill the Bianchi identity.
In contrast, the Euclidean version of
$F_{\mu\nu\rho\sigma}^{\nprt_{(nc)}}(z)$ fulfills the Bianchi identity
but does not lead to confinement~\cite{Dosch:1987sk+X}. Therefore, we
call the tensor structure multiplied by $\kappa$ non-Abelian or
confining ($c$) and the one multiplied by $(1-\kappa)$ Abelian or
non-confining ($nc$).

The non-perturbative correlator was originally constructed in
Euclidean space-time~\cite{Dosch:1987sk+X}.  The transition to
Minkowski space-time is performed by the substitution
$\delta_{\mu\rho} \rightarrow - g_{\mu\rho}$ and the analytic
continuation of the Euclidean correlation functions to real time,
$D^{E} \rightarrow D$ and $D_1^{E} \rightarrow
D_1$~\cite{Kramer:1990tr,Dosch:1994ym}. Euclidean correlation
functions are accessible together with the Euclidean correlator in
lattice QCD~\cite{DiGiacomo:1992df+X,Meggiolaro:1999yn}.  We adopt for
our calculations the simple {\em exponential correlation functions}
specified in four dimensional Euclidean space-time
\be
        D^{E}(Z^2/a^2) = D^{E}_1(Z^2/a^2) = \exp(-|Z|/a)
        \ ,
\label{Eq_MSV_correlation_functions}
\ee
that are motivated by lattice QCD measurements of the gluon field
strength correlator~\cite{DiGiacomo:1992df+X,Meggiolaro:1999yn}. These
correlation functions stay positive for all Euclidean distances $Z$.
In earlier applications of the \SVM, a different correlation function
$D^{E}$ has been used that becomes negative at large
distances~\cite{Dosch:1994ym,Rueter:1996yb+X,Dosch:1998nw,Rueter:1998up,D'Alesio:1999sf,Berger:1999gu,Dosch:2001jg}.
Such a negative part is not compatible with a spectral representation
of the correlation function~\cite{Dosch:1998th}. By analytic
continuation of (\ref{Eq_MSV_correlation_functions}) we obtain the
Minkowski correlation functions in~(\ref{Eq_MSV_Ansatz_F}) as shown in
Appendix~\ref{Sec_Correlation_Functions}.

Treating the vacuum fluctuations as a Gaussian random process, the
non-perturbative Euclidean correlator leads to the following explicit
expression for the {\em QCD string tension}~\cite{Dosch:1987sk+X}
\be
        \sigma 
        = \frac{\pi^3 \kappa G_2}{36} 
          \int_0^\infty dZ^2 D^{E}(Z^2/a^2) 
        = \frac{\pi^3 \kappa\,G_2\,a^2}{18}
\label{Eq_string_tension}
\ee
with the exponential correlation
function~(\ref{Eq_MSV_correlation_functions}) used in the final step.
The QCD string tension $\sigma$ characterizes the confining
quark-antiquark potential and can be computed from first principles
within lattice QCD~\cite{Bali:2001gf}. Thus,
relation~(\ref{Eq_string_tension}) puts an important constraint on the
three fundamental parameters of the non-perturbative QCD vacuum ---
$a$, $G_2$, and $\kappa$ --- and eliminates one degree of freedom.

While a non-perturbative model must be used to describe the low
frequency contributions, the perturbative component
$F_{\mu\nu\rho\sigma}^{\pert}$ is computed from the gluon propagator
in Feynman-'t~Hooft gauge
\be
        \Big\langle  \G^a_{\mu}(x_1)\G^b_{\nu}(x_2) \Big\rangle
        = \int
        \frac{d^4k}{(2\pi)^4}
        \,\frac{-i\delta^{ab}g_{\mu\nu}}{k^2-m_G^2}
        \, e^{-ik(x_1-x_2)}
        \ ,
\label{Eq_massive_gluon_propagator}
\ee
with an {\em effective gluon mass} $m_G$ introduced to limit the range
of the perturbative interaction in the infrared (IR) region.

In leading order in the strong coupling $g$, the bilocal gluon field
strength correlator is gauge-invariant already without the parallel
transport to a common reference point so that
$F_{\mu\nu\rho\sigma}^{\pert}$ depends only on the difference $z:= x_1
- x_2$. In this order, $\Order(g^2)$, we obtain
\bea
        F_{\mu\nu\rho\sigma}^{\pert}(z)
        \!\!&=&\!\!\frac{g^2}{\pi^2}\, \inv{2}\Bigl[
                       \frac{\partial}{\partial z_\nu}
                         \left(z_\sigma g_{\mu\rho}
                         -z_\rho g_{\mu\sigma}\right)
                       +\frac{\partial}{\partial z_\mu}
                         \left(z_\rho g_{\nu\sigma}
                         -z_\sigma g_{\nu\rho}\right)\Bigr]\,
              D_{\pert}(z^2)
        \nonumber \\
        \!\!&=&\!\! -\,\frac{g^2}{\pi^2}\!
                \int \!\!\frac{d^4k}{(2\pi)^4} \,e^{-ikz}\,\Bigl[
                k_\nu k_\sigma g_{\mu\rho}  - k_\nu k_\rho   g_{\mu\sigma}
              + k_\mu k_\rho  g_{\nu\sigma} - k_\mu k_\sigma g_{\nu\rho} \Bigr]\,
           \tilde{D}_{\pert}^{\prime}(k^2)
        \nonumber\\
\label{Eq_PGE_Ansatz_F}
\eea
with the {\em perturbative correlation function}
\be
        \tilde{D}_{\pert}^{\prime}(k^2) 
        := \frac{d}{dk^2} \int d^4z \,D_{\pert}(z^2)\, e^{ikz} 
         = \frac{i}{k^2 - m_G^2}
         \ .
\label{Eq_massive_D_pge_prime}
\ee

The tensor structure in~(\ref{Eq_PGE_Ansatz_F}) is identical to the
non-confining tensor structure in the non-perturbative
component~(\ref{Eq_MSV_Ansatz_F}). Together with the perturbative
correlation function in Euclidean space-time, it leads to the
non-confining color-Coulomb potential that is dominant for small
quark-antiquark separations~\cite{Kogut:1979wt}.

In the final step of the computation of $\chi$ in
the next section, the constant coupling $g^2$ 
is replaced by the {\em running coupling}
\be
        g^2(\vec{z}_{\!\perp})
        = 4 \pi \alphaS(\vec{z}_{\!\perp})
        = \frac{12 \pi}
        {(33-2 N_f) 
        \ln\left[
                (|\vec{z}_{\!\perp}|^{-2} + M^2)/\Lambda_{QCD}^2
        \right]}
\label{Eq_g2(z_perp)}
\ee
with the renormalization scale provided by $|\vec{z}_{\!\perp}|$ that
represents the spatial separation of the interacting dipoles in
transverse space.\footnote{Time-like or light-like separations do not
  appear in the final expression for $\chi$. They are integrated out
  as explained in Sec.~\ref{Sec_Chi_Computation}.}
In~(\ref{Eq_g2(z_perp)}), $N_f$ denotes the number of dynamical quark
flavors, which is set to $N_f = 0$ in agreement with the quenched
approximation, $\Lambda_{QCD} = 0.25\;\GeV$, and $M^2$ allows us to
freeze $g^2$ for $|\vec{z}_{\!\perp}| \rightarrow \infty$. Relying on
a low energy theorem~\cite{Rueter:1995cn+X}, we freeze $g^2$ at the
value at which the results for the potential and the total flux tube
energy of a static quark-antiquark pair coincide in our
model~\cite{Euclidean_Model_Applications}.

\subsection[Evaluation of the $\chi$-Function with Minimal Surfaces]{Evaluation of the \boldmath$\chi$-Function with Minimal Surfaces}
\label{Sec_Chi_Computation}

For the computation of the $\chi$-function~(\ref{Eq_chi_SS})
\bea    
        \chi 
        &:=& \chi^{\nprt}_c + \chi^{\nprt}_{nc} + \chi^{\pert} 
        \nonumber \\
        &=& - \,i\,\frac{\pi^2}{4}\!
        \int_{S_1} \! d\sigma^{\mu\nu}(x_1) 
        \int_{S_2} \! d\sigma^{\rho\sigma}(x_2)
        \left( 
          F_{\mu\nu\rho\sigma}^{\nprt_{(c)}} 
        + F_{\mu\nu\rho\sigma}^{\nprt_{(nc)}} 
        + F_{\mu\nu\rho\sigma}^{\pert} 
        \right) \ ,
\label{chi_amplitude}
\eea
one has to specify surfaces $S_{1,2}$ with the restriction $\partial
S_{1,2} = C_{1,2}$ according to the non-Abelian Stokes' theorem. As
illustrated in Fig.~\ref{Fig_loop_loop_scattering_surfaces}, we put
the reference point $o$ at the origin of the coordinate system and
choose for $S_{1,2}$ the {\em minimal surfaces} that are built from
the areas spanned by the corresponding loops $C_{1,2}$ and the
infinitesimally thin tube which connects the two surfaces $S_1$ and
$S_2$.  Since the tube contributions cancel mutually, this choice
makes the calculation explicitly independent of the reference point
$o$ and of the paths $C_{x_1 o}$ and $C_{x_2 o}$.

The minimal surfaces $S_1$ and $S_2$ shown in
Fig.~\ref{Fig_loop_loop_scattering_surfaces} can be parametrized with
the upper (lower) subscripts and signs referring to $S_1$ ($S_2$) as
follows
\be
        S_{\!\!{1 \atop (2)}} =  
        \left\{ 
            \Big(x^{\mu}_{\!\!\!\!{1\atop(2)}}(u,v)\Big) 
            = \Big(
            r^{\mu}_{\!\!\!\!{1q\atop(2q)}}
            + u\,n^{\mu}_{\!\!\!{\oplus\atop(\ominus)}} 
            + v\,r^{\mu}_{\!\!\!{1\atop(2)}}
            \Big),
            \;u \in [-T,T], \;v \in [0,1] \right\}
        \ ,
\label{Eq_S1(S2)_parameterization}
\ee
where 
\be
        \Big(n^{\mu}_{\!\!\!{\oplus\atop(\ominus)}}\Big) 
        := \left( \barray{c} 1 \\ \vec{0} \\ {\scriptscriptstyle{+\atop(-)}}\!1 \earray \right)
        , \quad
        \Big( r^{\mu}_{\!\!\!\!{1q\atop(2q)}} \Big)
        := \left( \barray{c} 0 \\ \vec{r}_{\!\!\!{1q\atop(2q)}} \\ 0 \earray \right)
        , \quad \mbox{and} \quad
        \Big( r^{\mu}_{\!\!\!{1\atop(2)}} \Big)
        := \left( \barray{c} 0 \\ \vec{r}_{\!\!\!{1\atop(2)}}   \\ 0 \earray \right)
        \ .
\label{Eq_C1(C2)_four_vectors}
\ee
The infinitesimally thin tube is neglected since it does not
contribute to the $\chi$-function as already mentioned. The
computation of the $\chi$-function requires only the parametrized
parts of the minimal surfaces~(\ref{Eq_S1(S2)_parameterization}), the
corresponding infinitesimal surface elements
\be
        d\sigma^{\mu\nu} 
        = \left( \frac{\partial x^{\mu}}{\partial u} 
                 \frac{\partial x^{\nu}}{\partial v}
               - \frac{\partial x^{\mu}}{\partial v} 
                 \frac{\partial x^{\nu}}{\partial u} \right)\,du\,dv
        = \left( n^{\mu}_{\!\!\!{\oplus\atop(\ominus)}} 
                 r^{\nu}_{\!\!\!{1\atop(2)}}
               - r^{\mu}_{\!\!\!{1\atop(2)}} 
                 n^{\nu}_{\!\!\!{\oplus\atop(\ominus)}} \right)\,du\,dv
        \ ,
\label{Eq_S1(S2)_surface_element}
\ee
and the limit $T \to \infty$ which is appropriate since the
correlation length $a$ is much smaller (see
Sec.~\ref{Sec_Model_Parameters}) than the longitudinal extension of
the loops.


Starting with the confining component 
\bea
        \!\!\!\!\!\!\!\!\!\!&&\!\!\!\!\! 
        \chi_{c}^{\nprt} 
        := - \, i \,\frac{\pi^2}{4} 
                \int_{S_1} \!\! d\sigma^{\mu\nu}(x_1) 
                \int_{S_2} \!\! d\sigma^{\rho\sigma}(x_2)\,
                F_{\mu\nu\rho\sigma}^{\nprt_{(c)}}(z = x_1 - x_2)
        \nonumber \\ 
        \!\!\!\!\!\!\!\!\!\!&& 
        = -\,\frac{\pi^2 G_2 \kappa}{12(N_c^2-1)}
        \int_{S_1} \!\! d\sigma^{\mu\nu}(x_1) 
        \int_{S_2} \!\! d\sigma^{\rho\sigma}(x_2)
        \left(g_{\mu\rho}g_{\nu\sigma}
                -g_{\mu\sigma}g_{\nu\rho}\right)
        iD(z^2/a^2) 
        \ ,
\label{Eq_chi_MSV_confining_definition}
\eea
one exploits the anti-symmetry of the surface elements,
$d\sigma^{\mu\nu} = - d\sigma^{\nu\mu}$, and applies the surface
parametrization~(\ref{Eq_S1(S2)_parameterization}) with the
corresponding surface elements~(\ref{Eq_S1(S2)_surface_element}) to
obtain
\be
        \chi_{c}^{\nprt} = 
        \frac{\pi^2 G_2 \kappa}{3(N_c^2-1)}\,2
        \left(\vec{r}_1\cdot\vec{r}_2\right)
        \int_0^1 \!\! dv_1 \int_0^1 \!\! dv_2
        \lim_{T\to\infty} \int_{-T}^T\!\!du_1 \int_{-T}^T\!\!du_2\,
        iD(z^2/a^2)
        \ ,
\label{Eq_chi_MSV_confining_intermediate}
\ee
where
\be
        z^{\mu} = x_1^{\mu} - x_2^{\mu} 
        = u_1 n^{\mu}_{\oplus} - u_2 n^{\mu}_{\ominus}
        + r^{\mu}_{1q} - r^{\mu}_{2q}
        + v_1 r^{\mu}_{1} - v_2 r^{\mu}_{2}
        \ ,
\label{Eq_z_S1_S2}
\ee
and the identities $n_{\oplus}\cdot r_{2} = r_{1}\cdot n_{\ominus} =
0$ and $n_{\oplus}\cdot n_{\ominus} = 2$, evident
from~(\ref{Eq_C1(C2)_four_vectors}), have been used. Next, one Fourier
transforms the correlation function and performs the $u_1$ and $u_2$
integrations in the limit $T \to \infty$
\bea
        && \lim_{T\to\infty} 
                \int_{-T}^T \!\! du_1 
                \int_{-T}^T \!\! du_2\,iD(z^2/a^2)
        \nonumber \\
        && = \int \frac{d^4k}{(2\pi)^4}\,i\tilde{D}(k^2) 
                \lim_{T\to\infty} \int_{-T}^T \!\! du_1 
                \int_{-T}^T \!\! du_2 \,e^{-ikz}
        \nonumber \\
        && = \int \frac{d^4k}{(2\pi)^2}\,i\tilde{D}(k^2)\, 
        \exp[-ik_{\mu}
        (r^{\mu}_{1q}-r^{\mu}_{2q}+v_1 r^{\mu}_{1}-v_2 r^{\mu}_{2})]
        \,\delta(k^0-k^3)\,\delta(k^0+k^3)
        \nonumber \\
        && = \inv{2} \, iD^{(2)}
        \left(\vec{r}_{1q}+v_1\vec{r}_1-\vec{r}_{2q}-v_2\vec{r}_2\right)
        \ ,
\label{Eq_light-cone_coordinates_integrated_out}
\eea
where $iD^{(2)}$ is the confining correlation function in the
two-dimensional transverse space
(cf.~Appendix~\ref{Sec_Correlation_Functions})
\be
        D^{(2)}(\vec{z}_{\!\perp}) 
        =  \int \frac{d^2k_{\!\perp}}{(2\pi)^2} 
        e^{i\vec{k}_{\!\perp}\vec{z}_{\!\perp}}
        \tilde{D}^{(2)}(\vec{k}_{\!\perp}) \ .
\label{Eq_transverse_Fourier_transform}
\ee
The contributions along the light-cone coordinates have been
integrated out so that $\chi_{c}^{\nprt}$ is completely determined by
the transverse projection of the minimal surfaces. Inserting
(\ref{Eq_light-cone_coordinates_integrated_out}) into
(\ref{Eq_chi_MSV_confining_intermediate}), one finally obtains
\be
      \chi_{c}^{\nprt} = 
        \frac{\pi^2 G_2}{3(N_c^2-1)}\,\kappa
        \left(\vec{r}_1\cdot\vec{r}_2\right) \,
        \int_0^1 \! dv_1 \int_0^1 \! dv_2 \, 
        iD^{(2)}\left(\vec{r}_{1q} + v_1\vec{r}_1 
        - \vec{r}_{2q} - v_2\vec{r}_2\right)
        \ .
\label{Eq_chi_MSV_confining} 
\ee
With $\tilde{D}^{(2)}(\vec{k}_{\!\perp})$ obtained from the
exponential correlation function~(\ref{Eq_MSV_correlation_functions}),
cf.~Appendix~\ref{Sec_Correlation_Functions}, we find
\be
        iD^{(2)}(\vec{z}_{\!\perp})
        =  2 \pi \, a^2 
        \left[1+(|\vec{z}_{\!\perp}|/a)\right] 
        \exp\!\left( -|\vec{z}_{\!\perp}|/a\right)
\label{Eq_F2[i_D_confining]}
\ee
which is positive for all transverse distances. 

As evident from the $v_1$ and $v_2$ integrations in
(\ref{Eq_chi_MSV_confining}) and
Fig.~\ref{Fig_loop_loop_scattering_surfaces}b, there are
contributions from the transverse projections of the minimal surfaces
$(S_{1,2})_{\!\perp}$ connecting the quark and antiquark in each of
the two dipoles. We interpret these contributions as a manifestation
of the strings that confine the quarks and antiquarks in the dipoles
and understand, therefore, the confining component $\chi_{c}^{\nprt}$
as a {\em string-string interaction}. This component gives the main
contribution to the scattering amplitude in the non-perturbative
region~\cite{K-Space_Investigations}.

Due to the truncation of the cumulant expansion or, equivalently, the
Gaussian approximation, a considerable dependence of $\chi_{c}^{\nprt}$
on the specific surface choice is observed. In fact, a different and
more complicated result for $\chi_{c}^{\nprt}$ was obtained with the
pyramid mantle choice for the surfaces $S_{1,2}$ in earlier
applications of the \SVM\ to high-energy
scattering~\cite{Dosch:1994ym,Rueter:1996yb+X,Dosch:1998nw,Rueter:1998up,D'Alesio:1999sf,Berger:1999gu,Dosch:2001jg}.
However, we use minimal surfaces in line with model applications in
Euclidean space-time: If one considers the potential of a static
quark-antiquark pair, usually the minimal surface is used to obtain
Wilson's area law~\cite{Dosch:1987sk+X,Euclidean_Model_Applications}.
Moreover, the simplicity of the minimal surfaces allows us to give an
analytic expression for the leading term of the non-perturbative
dipole-dipole cross section~\cite{K-Space_Investigations}.
Phenomenologically, in comparison with pyramid mantles, the
description of the slope parameter $B(s)$, the differential elastic
cross section $d\sigma^{el}/dt(s,t)$, and the elastic cross section
$\sigma^{el}(s)$ can be improved with minimal surfaces as shown in
Sec.~\ref{Sec_Comparison_Data}.

Continuing with the computation of the non-confining component
\bea
        \chi_{nc}^{\nprt}\!\!\!\!\!\!  
        &&:= -\, i \,\frac{\pi^2}{4} 
                \int_{S_1} \!\! d\sigma^{\mu\nu}(x_1) 
                \int_{S_2} \!\! d\sigma^{\rho\sigma}(x_2)\,
                F_{\mu\nu\rho\sigma}^{\nprt_{(nc)}}(z = x_1 - x_2)
        \nonumber \\ 
        && = \frac{\pi^2 G_2 (1-\kappa)}{12(N_c^2-1)} \,
                \int_{S_1} \!\! d\sigma^{\mu\nu}(x_1) \,
                \int_{S_2} \!\! d\sigma^{\rho\sigma}(x_2) 
        \\
\label{Eq_chi_MSV_non-confining_definition}
        && \hphantom{=} 
        \times \int \!\! \frac{d^4k}{(2\pi)^4} \,e^{-ikz}\,
        \Bigl[k_\nu k_\sigma g_{\mu\rho} 
        - k_\nu k_\rho g_{\mu\sigma}
        + k_\mu k_\rho  g_{\nu\sigma} 
        - k_\mu k_\sigma g_{\nu\rho} \Bigr]\,
        i\tilde{D}_{1}^{\prime}(k^2) \nonumber \ ,
\eea
we exploit again the anti-symmetry of both surface elements to obtain
\bea
        \chi_{nc}^{\nprt} 
        & = & \frac{\pi^2 G_2 (1-\kappa)}{3(N_c^2-1)} \,
                \int_0^1 \!\! dv_1 \int_0^1 \!\! dv_2 \,
                \int \!\! \frac{d^4k}{(2\pi)^4} \,
                \lim_{T\to\infty}\,
                \int_{-T}^T\!\!du_1\int_{-T}^T\!\!du_2\,e^{-ikz}
        \nonumber \\ 
        && \times
        \Bigl[2\,(r_1\cdot k)\,(r_2\cdot k)\,
        -\,(\vec{r}_1\cdot\vec{r}_2)\,(k^0 - k^3)(k^0 + k^3)\Bigr]\,
        i\tilde{D}_{1}^{\prime}(k^2) 
\label{Eq_chi_MSV_non-confining_intermediate_1}
\eea
with $z$ as given in~(\ref{Eq_z_S1_S2}). Again the identities
$n_{\oplus}\cdot r_{2} = r_{1}\cdot n_{\ominus} = 0$ and
$n_{\oplus}\cdot n_{\ominus} = 2$ have been used. Performing the $u_1$
and $u_2$ integrations in the limit $T \rightarrow \infty$, one
obtains --- as in~(\ref{Eq_light-cone_coordinates_integrated_out}) ---
two $\delta$-functions which allow us to carry out the integrations
over $k^0$ and $k^3$ immediately. This leads to
\bea
        \chi_{nc}^{\nprt} 
        & \!\!\!\! = \!\!\! & \frac{\pi^2 G_2 (1-\kappa)}{3(N_c^2-1)}
                \!\int_0^1 \!\! dv_1 \!\int_0^1 \!\! dv_2
                \! \int \!\! \frac{d^2k_{\!\perp}}{(2\pi)^2}
                i\tilde{D}_{1}^{\prime\,(2)}(\vec{k}_{\!\perp}^2)
        (\vec{r}_1\cdot\vec{k}_{\!\perp})\,(\vec{r}_2\cdot\vec{k}_{\!\perp})
        e^{i\vec{k}_{\!\perp}(\vec{r}_{1q}+v_1\vec{r}_1-\vec{r}_{2q}-v_2\vec{r}_2)}
        \nonumber \\
        & \!\!\!\! = \!\!\! & \frac{\pi^2 G_2 (1-\kappa)}{3(N_c^2-1)} 
                \int_0^1 \!\! dv_1 \frac{\partial}{\partial v_1} 
                \int_0^1 \!\! dv_2 \frac{\partial}{\partial v_2}\, 
                iD_{1}^{\prime\,(2)}
                (\vec{r}_{1q} + v_1\vec{r}_1 - \vec{r}_{2q} - v_2\vec{r}_2) 
        \ , 
\label{Eq_chi_MSV_non-confining_intermediate_2}
\eea
where $iD_{1}^{\prime\,(2)}$ is the non-confining correlation function
in transverse space defined analogously
to~(\ref{Eq_transverse_Fourier_transform}). The $v_1$ and $v_2$
integrations are trivial and lead (cf.\ 
Fig.~\ref{Fig_loop_loop_scattering_surfaces}b) to
\bea
     &&\!\!\!\!\!\!\!\!\! 
        \chi_{nc}^{\nprt} = 
        \frac{\pi^2 G_2}{3(N_c^2-1)}\,(1-\kappa) 
        \left[ 
        iD^{\prime\,(2)}_1
        \left(\vec{r}_{1q}-\vec{r}_{2q}\right) 
        +iD^{\prime\,(2)}_1
        \left(\vec{r}_{1\qbar}-\vec{r}_{2\qbar}\right)
        \right.
        \nonumber \\
     &&\!\!\!\!\!\!\!\!\! 
     \hphantom{\chi_{nc}^{\nprt}=\frac{\pi^2 G_2}{3(N_c^2-1)}\,(1-\kappa)}
        \left.
        -\,iD^{\prime\,(2)}_1
        \left(\vec{r}_{1q}-\vec{r}_{2\qbar}\right)
        -iD^{\prime\,(2)}_1
        \left(\vec{r}_{1\qbar}-\vec{r}_{2q}\right)
        \right] \ .
\label{Eq_chi_MSV_non-confining}
\eea
Using $\tilde{D}_{1}^{\prime\,(2)}(\vec{k}_{\!\perp}^2)$, derived from
the exponential correlation
function~(\ref{Eq_MSV_correlation_functions}) in
Appendix~\ref{Sec_Correlation_Functions}, we obtain
\be
         iD^{\prime\,(2)}_1(\vec{z}_{\!\perp})
         =  
        \pi \, a^4  \left[3 + 3(|\vec{z}_{\!\perp}|/a) + (|\vec{z}_{\!\perp}|/a)^2 \right]
        \exp\!\left( -|\vec{z}_{\!\perp}|/a\right)
        \ .
\label{Eq_F2[i_D_non-confining_prime]}
\ee

The non-perturbative components, $\chi^{\nprt}_c$ and
$\chi^{\nprt}_{nc}$, lead to {\em color transparency} for small dipoles, i.e.\ 
a dipole-dipole cross section with $\sigma_{DD}(\vec{r}_1,\vec{r}_2)
\propto |\vec{r}_1|^2|\vec{r}_2|^2$ for $|\vec{r}_{1,2}| \to 0$, as
known for the perturbative case~\cite{Nikolaev:1991ja}. This can be
seen by squaring~(\ref{Eq_chi_MSV_confining})
and~(\ref{Eq_chi_MSV_non-confining}) to obtain the leading terms in
the $T$-matrix element for small dipoles
(see~(\ref{Eq_model_purely_imaginary_T_amplitude_small_chi_limit})).

The perturbative component $\chi^{\pert}$ is defined as
\bea
        \chi^{\pert} 
        &:=& -\,i\,\frac{\pi^2}{4} 
                \int_{S_1} \!\! d\sigma^{\mu\nu}(x_1) 
                \int_{S_2} \!\! d\sigma^{\rho\sigma}(x_2)\,
                F_{\mu\nu\rho\sigma}^{\pert}(z = x_1 - x_2) 
        \hphantom{i\tilde{D}_{\pert}^{\prime\,(4)}k}
        \nonumber \\ 
        & =& \,\frac{g^2}{4}
                \int_{S_1} \!\! d\sigma^{\mu\nu}(x_1)\,
                \int_{S_2} \!\! d\sigma^{\rho\sigma}(x_2) 
        \\ && 
        \times \int \!\! \frac{d^4k}{(2\pi)^4} \,e^{-ikz}\,
                \Bigl[k_\nu k_\sigma g_{\mu\rho}   
                - k_\nu k_\rho   g_{\mu\sigma}
                + k_\mu k_\rho  g_{\nu\sigma} 
                - k_\mu k_\sigma g_{\nu\rho} \Bigr]\,
                i\tilde{D}_{\pert}^{\prime}(k^2) 
        \ , \nonumber
\label{Eq_chi_PGE_definition}
\eea
and shows a structure identical to the one of $\chi_{nc}^{\nprt}$ given
in~(\ref{Eq_chi_MSV_non-confining_definition}).  Accounting for the
different prefactors and the different correlation function, the
result for $\chi_{nc}^{\nprt}$~(\ref{Eq_chi_MSV_non-confining}) can be
used to obtain
\bea
     &&\!\!\!\!\!\!\!\!\!
        \chi^{\pert} = 
        \left[ 
        g^2\!\left(\vec{r}_{1q}-\vec{r}_{2q}\right)
        iD^{\prime\,(2)}_{\pert}
        \left(\vec{r}_{1q}-\vec{r}_{2q}\right)
        +g^2\!\left(\vec{r}_{1\qbar}-\vec{r}_{2\qbar}\right)
        iD^{\prime\,(2)}_{\pert}
        \left(\vec{r}_{1\qbar}-\vec{r}_{2\qbar}\right)
        \right.
        \nonumber \\
     &&\!\!\!\!\!\!\!\!\!\!\!\!\!\!\!\!\!\!\!\!\!
        \hphantom{\chi^{\pert} = }
        \left.
        -\,g^2\!\left(\vec{r}_{1q}-\vec{r}_{2\qbar}\right)
        iD^{\prime\,(2)}_{\pert}
        \left(\vec{r}_{1q}-\vec{r}_{2\qbar}\right)
        -g^2\!\left(\vec{r}_{1\qbar}-\vec{r}_{2q}\right)
        iD^{\prime\,(2)}_{\pert}
        \left(\vec{r}_{1\qbar}-\vec{r}_{2q}\right)
        \right]
        \ , 
\label{Eq_chi_PGE}
\eea
where the running coupling $g^2(\vec{z}_{\!\perp})$ is understood as
given in~(\ref{Eq_g2(z_perp)}). With~(\ref{Eq_massive_D_pge_prime})
one obtains the perturbative correlation function in transverse space
\bea
        iD^{\prime\,(2)}_{\pert}
        (\vec{z}_{\!\perp})
        = \inv{2\pi} K_0\left(m_G |\vec{z}_{\!\perp}|\right)
        \ ,
\label{Eq_F2[i_massive_D_pge_prime]}
\eea
where $K_0$ denotes the $0^{th}$ modified Bessel function (McDonald
function).

In contrast to the confining component $\chi_{c}^{\nprt}$, the
non-confining components, $\chi_{nc}^{\nprt}$ and $\chi^{\pert}$, depend
only on the transverse position between the quark and antiquark of the
two dipoles and are therefore independent of the surface choice.

Finally, we explain that the vanishing of $\chi_{S_1S_1}$ and
$\chi_{S_2S_2}$ anticipated in Sec.~\ref{Sec_loop_loop_func} results
from the light-like loops and the tensor structures in
$F_{\mu\nu\rho\sigma}$. Concentrating --- without loss of generality
--- on $\chi_{S_1 S_1}$, the appropriate infinitesimal surface
elements~(\ref{Eq_S1(S2)_surface_element}) and the
$F_{\mu\nu\rho\sigma}$--ansatz given in~(\ref{Eq_F_decomposition}),
(\ref{Eq_MSV_Ansatz_F}), and~(\ref{Eq_PGE_Ansatz_F}) are inserted
into~(\ref{Eq_chi_SS}). Having simplified the resulting expression by
exploiting the anti-symmetry of the surface elements, one finds only
terms proportional to $n_{\oplus}^2$, $n_{\oplus}\cdot r_{1}$, and
$n_{\oplus}\cdot z$ with $z^{\mu} = x_1^{\mu} - x_2^{\mu} = (u_1 -
u_2) n^{\mu}_{\oplus} + (v_1 - v_2) r^{\mu}_{1}$. Since $n_{\oplus}^2
= 0$ and $n_{\oplus}\cdot r_{1} = 0$, which is evident
from~(\ref{Eq_C1(C2)_four_vectors}), all terms vanish and $\chi_{S_1
  S_1} = 0$ is derived.

Note that $\chi = \chi_{c}^{\nprt} + \chi_{nc}^{\nprt} + \chi^{\pert}$ is
a real-valued function. Since, in addition, the wave functions
$|\psi_i(z_i,\vec{r}_i)|^2$ used in this work (cf.\ 
Appendix~\ref{Sec_Wave_Functions}) are invariant under the replacement
$(\vec{r}_i \rightarrow -\vec{r}_i, z_i \rightarrow 1-z_i)$, the
$T$-matrix element becomes purely imaginary and reads for $N_c=3$
\bea
        \!\!\!\!\!\!\!\!\!\!\!\!\!
        T(s,t) 
        & = & 2is \int \!\!d^2b_{\!\perp} 
                e^{i {\vec q}_{\!\perp} {\vec b}_{\!\perp}}
                \int \!\!dz_1 d^2r_1 \!
                \int \!\!dz_2 d^2r_2 \,\,
                |\psi_1(z_1,\vec{r}_1)|^2   \,\,
                |\psi_2(z_2,\vec{r}_2)|^2       
        \nonumber \\    
        && \!\!\!\!\!\!\!\!\!\!
        \times 
        \left[1-\frac{2}{3} 
        \cos\!\left(\frac{1}{3}
        \chi({\vec b}_{\!\perp},z_1,\vec{r}_1,z_2,\vec{r}_2)\!\right)
        - \frac{1}{3}
        \cos\!\left(\frac{2}{3}
        \chi({\vec b}_{\!\perp},z_1,\vec{r}_1,z_2,\vec{r}_2)\!\right)
        \right].
\label{Eq_model_purely_imaginary_T_amplitude}
\eea
The real part averages out in the integration over ${\vec r}_i$ and
$z_i$ since the $\chi$-function changes sign
\be
        \chi(\vec{b}_{\!\perp},1-z_1,-\vec{r}_1,z_2,\vec{r}_2)
        = - \chi(\vec{b}_{\!\perp},z_1,\vec{r}_1,z_2,\vec{r}_2)
        \ ,
\label{Eq_odd_eikonal_function}
\ee
which can be seen directly
from~(\ref{Eq_chi_MSV_confining}),(\ref{Eq_chi_MSV_non-confining}) and
(\ref{Eq_chi_PGE}) as $(\vec{r}_1 \rightarrow -\vec{r}_1, z_1
\rightarrow 1-z_1)$ implies $\vec{r}_{1q} \rightarrow
\vec{r}_{1\qbar}$. In physical terms, $(\vec{r}_i \rightarrow
-\vec{r}_i, z_i \rightarrow 1-z_i)$ corresponds to {\em charge
  conjugation}\, i.e.\ the replacement of each parton with its
antiparton and the associated reversal of the loop direction.

Consequently, the
$T$-matrix~(\ref{Eq_model_purely_imaginary_T_amplitude}) describes
only charge conjugation $C = +1$ exchange. Since in our quenched
approximation purely gluonic interactions are modelled,
(\ref{Eq_model_purely_imaginary_T_amplitude}) describes only
pomeron\footnote{Odderon $C = -1$ exchange is excluded in our model.
  It would survive in the following cases: (a) Wave functions are used
  that are not invariant under the transformation $(\vec{r}_i
  \rightarrow -\vec{r}_i, z_i \rightarrow 1-z_i)$. (b) The proton is
  described as a system of three quarks with finite separations
  modelled by three loops with one common light-like line. (c) The
  Gaussian approximation that enforces the truncation of the cumulant
  expansion is relaxed and additional higher cumulants are taken into
  account.}  but not reggeon exchange.

\subsection{Energy Dependence}
\label{Sec_Energy_Dependence}

Until now, the derived $T$-matrix element leads to energy independent
total cross sections in contradiction to the experimental observation. In
this section, we introduce the energy dependence in a phenomenological
way inspired by other successful models.

Most models for high-energy scattering are constructed to describe
either hadron-hadron or photon-hadron reactions.  For example,
Kopeliovich et al.~\cite{Kopeliovich:2001pc} as well as Berger and
Nachtmann~\cite{Berger:1999gu} focus on hadron-hadron scattering. In
contrast, Golec-Biernat and W\"usthoff~\cite{Golec-Biernat:1999js+X}
and Forshaw, Kerley, and Shaw~\cite{Forshaw:1999uf} concentrate on
photon-proton reactions. A model that describes the energy dependence
in both hadron-hadron and photon-hadron reactions up to large photon
virtualities is the two-pomeron model of Donnachie and
Landshoff~\cite{Donnachie:1998gm+X}. Based on Regge theory, they find
a soft pomeron trajectory with intercept $ 1 + \epsilon_{soft} \approx
1.08$ that governs the weak energy dependence of hadron-hadron or
$\gamma^{*} p$ reactions with low $Q^2$ and a hard pomeron trajectory
with intercept $1 + \epsilon_{hard} \approx 1.4$ that governs the
strong energy dependence of $\gamma^*p$ reactions with high $Q^2$.
Similarly, we aim at a simultaneous description of hadron-hadron,
photon-proton, and photon-photon reactions involving real and virtual
photons as well. 

In line with other two-component (soft $+$ hard)
models~\cite{Donnachie:1998gm+X,D'Alesio:1999sf,Forshaw:1999uf,Rueter:1998up,Donnachie:2001wt}
and the different hadronization mechanisms in soft and hard
collisions, our physical ansatz demands that the perturbative and
non-perturbative contributions do not interfere. Therefore, we modify the
cosine-summation in~(\ref{Eq_model_purely_imaginary_T_amplitude})
allowing only even numbers of soft and hard correlations, $\left
  (\chi^{\nprt} \right)^{2n} \left ( \chi^{\pert} \right)^{2m}$ with
$n,m \in I\!\!N$.  Interference terms with odd numbers of soft and
hard correlations are subtracted by the replacement
\be
        \cos\left[ c\chi \right] =  
        \cos\left[c\left( \chi^{\nprt} + \chi^{\pert} \right)\right] 
        \rightarrow 
        \cos\left[c\chi^{\nprt}\right]\cos\left[c\chi^{\pert}\right] 
        \ ,
\label{Eq_interference_term_subtraction}
\ee
where $c = 1/3$ or $2/3$. This prescription leads to the following
factorization of soft and hard physics in the $T$-matrix
element,
\bea
        && \!\!\!\!\!\!\!\!\!\!
        T(s,t) 
        = 2is \int \!\!d^2b_{\!\perp} 
        e^{i {\vec q}_{\!\perp} {\vec b}_{\!\perp}}
        \int \!\!dz_1 d^2r_1 \!\int \!\!dz_2 d^2r_2\,\,
        |\psi_1(z_1,\vec{r}_1)|^2 \,\, 
        |\psi_2(z_2,\vec{r}_2)|^2       
        \nonumber \\    
        &&\!\!\!\!\!\!\!\!\!\!\!\! 
        \times \left[ 1 - \frac{2}{3} 
        \cos\!\left(\!\frac{1}{3}\chi^{\nprt}\!\right)
        \cos\!\left(\!\frac{1}{3}\chi^{\pert}\!\right)         
        - \frac{1}{3}
        \cos\!\left(\!\frac{2}{3}\chi^{\nprt}\!\right)
        \cos\!\left(\!\frac{2}{3}\chi^{\pert}\!\right)
        \right] \ . 
\label{Eq_model_purely_imaginary_T_amplitude_almost_final_result}
\eea
In the limit of small $\chi$-functions, $\chi^{\nprt}
\ll 1$ and $\chi^{\pert} \ll 1$, one gets
\bea
        T(s,t) 
        & = & 2is \!\int \!\!d^2b_{\!\perp} 
        e^{i {\vec q}_{\!\perp} {\vec b}_{\!\perp}}
        \!\int \!\!dz_1 d^2r_1 \!\int \!\!dz_2 d^2r_2\,\,
        |\psi_1(z_1,\vec{r}_1)|^2 \,\,
        |\psi_2(z_2,\vec{r}_2)|^2       
        \nonumber \\  
        && \times \frac{1}{9}\left[
        \left(\chi^{\nprt}\right)^2
        +\left(\chi^{\pert}\right)^2
        \right].
\label{Eq_model_purely_imaginary_T_amplitude_small_chi_limit}
\eea
In this limit, the $T$-matrix element evidently becomes a sum of a
perturbative and a non-perturbative component. Of course, the
perturbative component, $(\chi^{\pert})^2$, coincides with the
well-known perturbative {\em two-gluon
  exchange}~\cite{K-Space_Investigations}.  Correspondingly, the
non-perturbative component, $(\chi^{\nprt})^2$, represents the
non-perturbative gluonic interaction on the ``two-gluon-exchange''
level.

As the two-component structure of
(\ref{Eq_model_purely_imaginary_T_amplitude_small_chi_limit}) reminds
of the two-pomeron model of Donnachie and
Landshoff~\cite{Donnachie:1998gm+X}, we adopt the powerlike energy
increase and ascribe a weak energy dependence to the non-perturbative
component $\chi^{\nprt}$ and a strong one to the perturbative component
$\chi^{\pert}$
\bea
        \left(\chi^{\nprt}\right)^2 \quad & \rightarrow & \quad 
        \left(\chi^{\nprt}(s)\right)^2 := \left(\chi^{\nprt}\right)^2 
        \left(\frac{s}{s_0}
        \frac{\vec{r}_1^{\,2}\,\vec{r}_2^{\,2}}{R_0^4}\right)^{\epsilon^{\nprt}}
        \nonumber \\
        \left(\chi^{\pert}\right)^2 \quad & \rightarrow & \quad 
        \left(\chi^{\pert}(s)\right)^2 := \left(\chi^{\pert}\right)^2
        \left(\frac{s}{s_0} 
        \frac{\vec{r}_1^{\,2}\,\vec{r}_2^{\,2}}{R_0^4}\right)^{\epsilon^{\pert}}
\label{Eq_energy_dependence}
\eea
with the scaling factor $s_0 R_0^4$. The powerlike energy
dependence with the exponents $0\approx \epsilon^{\nprt} <
\epsilon^{\pert} < 1$ guarantees Regge type behavior at moderately high
energies, where the small-$\chi$
limit~(\ref{Eq_model_purely_imaginary_T_amplitude_small_chi_limit}) is
appropriate. In~(\ref{Eq_energy_dependence}), the energy variable
$s$ is scaled by the factor $\vec{r}_1^{\,2}\,\vec{r}_2^{\,2}$ that
allows to rewrite the energy dependence in photon-hadron scattering in
terms of the appropriate Bjorken scaling variable $x$
\be
        s\,\vec{r}_1^{\,2} \propto \frac{s}{Q^2} = \inv{x}
        \ ,
\label{Eq_x_Bj_<->_s}
\ee
where $|\vec{r}_1|$ is the transverse extension of the $q\qbar $
dipole in the photon. A similar factor has been used before in the
dipole model of Forshaw, Kerley, and Shaw~\cite{Forshaw:1999uf} and
also in the model of Donnachie and Dosch~\cite{Donnachie:2001wt} in
order to respect the scaling properties observed in the structure
function of the proton.\footnote{In the model of Donnachie and
  Dosch~\cite{Donnachie:2001wt}, $s\,|\vec{r}_1|\,|\vec{r}_2|$ is used
  as the energy variable if both dipoles are small, which is in
  accordance with the choice of the typical BFKL energy scale but
  leads to discontinuities in the dipole-dipole cross section. In
  order to avoid such discontinuities, we use the energy
  variable~(\ref{Eq_energy_dependence}) also for the scattering of two
  small dipoles.} In the dipole-proton cross section of Golec-Biernat
and W\"usthoff~\cite{Golec-Biernat:1999js+X}, Bjorken $x$ is used
directly as energy variable which is important for the success of the
model. In fact, also in our model, the
$\vec{r}_1^{\,2}\,\vec{r}_2^{\,2}$ factor improves the description of
$\gamma^{*} p$ reactions at large $Q^2$.

The powerlike Regge type energy dependence introduced
in~(\ref{Eq_energy_dependence}) is, of course, not mandatory but
allows successful fits and can also be derived in other theoretical
frameworks: A powerlike energy dependence is found for hadronic
reactions by Kopeliovich et al.~\cite{Kopeliovich:2001pc} and for hard
photon-proton reactions from the BFKL equation~\cite{BFKL}. However,
these approaches need unitarization since their powerlike energy
dependence will ultimately violate $S$-matrix unitarity at asymptotic
energies. In our model, we use the following $T$-matrix element as the
basis for the rest of this work
\bea
        &&\!\!\!\!\!\!\!\!\!\!
        T(s,t)  
        =  2is \int \!\!d^2b_{\!\perp} 
        e^{i {\vec q}_{\!\perp} {\vec b}_{\!\perp}}
        \int \!\!dz_1 d^2r_1 \!\int \!\!dz_2 d^2r_2\,\, 
        |\psi_1(z_1,\vec{r}_1)|^2 \,\, 
        |\psi_2(z_2,\vec{r}_2)|^2       
        \nonumber \\    
        &&\!\!\!\!\!\!\!\!\!\!\!\! 
        \times \left[1 - \frac{2}{3} 
        \cos\!\left(\!\frac{1}{3}\chi^{\nprt}(s)\!\right)
        \cos\!\left(\!\frac{1}{3}\chi^{\pert}(s)\!\right)         
        - \frac{1}{3}
        \cos\!\left(\!\frac{2}{3}\chi^{\nprt}(s)\!\right)
        \cos\!\left(\!\frac{2}{3}\chi^{\pert}(s)\!\right)
        \right] 
        \ , \nonumber \\
\label{Eq_model_purely_imaginary_T_amplitude_final_result}
\eea
where the cosine functions ensure the unitarity condition in impact
parameter space as shown in Sec.~\ref{Sec_Impact_Parameter}. Indeed,
the multiple gluonic interactions associated with the higher order
terms in the expansion of the cosine functions are important for the
saturation effects observed within our model at ultra-high energies.

Having ascribed the energy dependence to the $\chi$-function, the
energy behavior of hadron-hadron, photon-hadron, and photon-photon
scattering results exclusively from the {\em universal } loop-loop
correlation function $S_{DD}$. 

\subsection{Model Parameters}
\label{Sec_Model_Parameters}

Lattice QCD simulations provide important information and constraints
on the model parameters. The fine tuning of the parameters was,
however, directly performed on the high-energy scattering data for
hadron-hadron, photon-hadron, and photon-photon reactions where an
error ($\chi^2$) minimization was not feasible because of the
non-trivial multi-dimensional integrals in the $T$-matrix
element~(\ref{Eq_model_purely_imaginary_T_amplitude_final_result}).

The parameters $a$, $\kappa$, $G_2$, $m_G$, $M^2$, $s_0R^4_0$,
$\epsilon^{\nprt}$ and $\epsilon^{\pert}$ determine the dipole-dipole
scattering and are universal for all reactions described. In addition,
there are reaction-dependent parameters associated with the wave
functions which are provided in Appendix~\ref{Sec_Wave_Functions}.

The non-perturbative component involves the correlation length $a$,
the gluon condensate $G_2$, and the parameter $\kappa$ indicating the
non-Abelian character of the correlator. With the simple exponential
correlation functions specified in Euclidean space-time
(\ref{Eq_MSV_correlation_functions}), we obtain the following values
for the parameters of the non-perturbative
correlator~(\ref{Eq_MSV_Ansatz_F})
\be
        a =  0.302\,\fm, \quad 
        \kappa = 0.74, \quad 
        G_2 = 0.074\,\GeV^4
        \ ,
\label{Eq_MSV_scattering_fit_parameter_results}
\ee
and, correspondingly, the string tension
\be
        \sigma = 
        \frac{\pi^3 \kappa\,G_2\,a^2}{18} 
        = 0.22\,\GeV^2 \equiv 1.12 \,\GeV/\fm
        \ ,
\label{Eq_sting_tension_from_exp_correlation}
\ee 
which is consistent with hadron spectroscopy~\cite{Kwong:1987mj},
Regge theory~\cite{Goddard:1973qh+X}, and lattice QCD
investigations~\cite{Bali:2001gf}.

Lattice QCD computations of the gluon field strength correlator down
to distances of $0.4\,\fm$ have obtained the following values with the
exponential correlation
function~(\ref{Eq_MSV_correlation_functions})~\cite{Meggiolaro:1999yn}:
$a = 0.219\,\fm$, $\kappa = 0.746$, $G_2 = 0.173\,\GeV^4$. This value
for $\kappa$ is in agreement with the one
in~(\ref{Eq_MSV_scattering_fit_parameter_results}), while the fit to
high-energy scattering data clearly requires a larger value for $a$
and a smaller value for $G_2$.

The perturbative component involves the gluon mass $m_G$ as IR
regulator (or inverse ``perturbative correlation length'') and the
parameter $M^2$ that freezes the running
coupling~(\ref{Eq_g2(z_perp)}) for large distance scales at the value
$\alpha_s = 0.4$, where the non-perturbative component of our model
with the above ingredients is at work according to a low energy
theorem~\cite{Rueter:1995cn+X,Euclidean_Model_Applications}.  We adopt
the parameters
\be
        m_G =  m_{\rho} = 0.77\,\GeV 
        \quad \mbox{and }\quad 
        M^2 = 1.04\,\GeV^2
        \ .
\label{Eq_PGE_scattering_fit_parameter_results}
\ee

The energy dependence of the model is associated with the energy
exponents $\epsilon^{\nprt}$ and $\epsilon^{\pert}$, and the  scaling
parameter $s_0R^4_0$ 
\be
        \epsilon^{\nprt} = 0.125, \quad 
        \epsilon^{\pert} = 0.73, \quad \mbox{and} \quad 
        s_0 R_0^4 = (\,47\,\GeV\,\fm^2\,)^2
        \ .
\label{Eq_energy_dependence_scattering_fit_parameter_results}
\ee
In comparison with the energy exponents of Donnachie and
Landshoff~\cite{Donnachie:1992ny,Donnachie:1998gm+X}, $\epsilon_{soft}
\approx 0.08$ and $\epsilon_{hard} \approx 0.4$, our exponents are
larger. However, the cosine functions in our $T$-matrix
element~(\ref{Eq_model_purely_imaginary_T_amplitude_final_result})
reduce the large exponents so that the energy dependence of the cross
sections agrees with the experimental data as illustrated in
Sec~\ref{Sec_Comparison_Data}.


%
%
\section[Impact Parameter Profiles and \boldmath$S$-Matrix Unitarity]{Impact Parameter Profiles and \boldmath$S$-Matrix Unitarity}
\label{Sec_Impact_Parameter}

In this section, the $S$-matrix unitarity is analysed in our model. On
the basis of the impact parameter dependence of the scattering
amplitude, saturation effects can be exposed that manifest the
unitarity of the $S$-matrix. For each impact parameter the energy at
which the unitarity limit becomes important can be determined. This is
used to show the saturation of the gluon distribution and to localize
saturation effects in experimental observables.

The impact parameter dependence of the scattering amplitude is given
by $\impactT(s,|\vec{b}_{\!\perp}|)$,
\be
        T(s,t=-{\vec q}_{\!\perp}^{\,\,2}) \;=\;
        4s\!\int \!\!d^2b_{\!\perp}\,
        e^{i {\vec q}_{\!\perp} {\vec b}_{\!\perp}}\,
        \impactT(s,|\vec{b}_{\!\perp}|)
\label{Eq_Fourier_transformed_T-matrix_element}
\ee
and in particular by the {\em profile function}
\be
        J(s,|\vec{b}_{\!\perp}|) 
        = 2\,\im\impactT(s,|\vec{b}_{\!\perp}|)
        \ ,
\label{Eq_profile_function_def}
\ee 
which describes the {\em blackness} or {\em opacity} of the
interacting particles as a function of the impact parameter $|{\vec
  b}_{\!\perp}|$ and the c.m.\ energy $\sqrt{s}$. In fact, the profile
function~(\ref{Eq_profile_function_def}) determines every observable
if the $T$-matrix is --- as in our model --- purely imaginary.

The $S$-matrix unitarity, $SS^{\dagger} = S^{\dagger}S = \Identity$,
leads directly to the {\em unitarity condition} in impact parameter
space~\cite{Amaldi:1976gr,Castaldi:1985ft}
\be
        \im\impactT(s,|\vec{b}_{\!\perp}|)
        = |\impactT(s,|\vec{b}_{\!\perp}|)|^2 + G_{inel}(s,|\vec{b}_{\!\perp}|)
        \ ,
\label{Eq_unitarity_condition}
\ee 
where $G_{inel}(s,|\vec{b}_{\!\perp}|) \ge 0$ is the inelastic overlap
function~\cite{VanHove:1964rp}.\footnote{Integrating
  (\ref{Eq_unitarity_condition}) over the impact parameter space and
  multiplying by a factor of $4$ one obtains the relation
  $\sigma^{tot}(s) = \sigma^{el}(s) + \sigma^{inel}(s)$.} This
unitarity condition imposes an absolute limit on the profile function
\be
        0 \;\;\leq\;\;
        2\,|\impactT(s,|\vec{b}_{\!\perp}|)|^2
        \;\;\leq\;\; 
        J(s,|\vec{b}_{\!\perp}|) 
        \;\;\leq\;\; 2
\label{Eq_absolute_unitarity_limit}
\ee 
and the inelastic overlap function, $G_{inel}(s,|\vec{b}_{\!\perp}|)
\le 1/4$.  At high energies, however, the elastic amplitude is
expected to be purely imaginary.  Consequently, the solution
of~(\ref{Eq_unitarity_condition}) reads
\be
        J(s,|\vec{b}_{\!\perp}|) = 1 \pm \sqrt{1-4\,G_{inel}(s,|\vec{b}_{\!\perp}|)}
\label{Eq_solution_unitarity_condition}
\ee
and leads with the minus sign corresponding to the physical situation
to the {\em reduced unitarity bound}
\be
        0 \;\;\leq\;\;
        J(s,|\vec{b}_{\!\perp}|) 
        \;\;\leq\;\; 1
        \ .
\label{Eq_reduced_unitarity_bound}
\ee 
Reaching the {\em black disc limit} or {\em maximum opacity} at a
certain impact parameter $|\vec{b}_{\!\perp}|$,
$J(s,|\vec{b}_{\!\perp}|) = 1$, corresponds to maximal inelastic
absorption $G_{inel}(s,|\vec{b}_{\!\perp}|) = 1/4$ and equal elastic
and inelastic contributions to the total cross section at that impact
parameter.

In our model, every reaction is reduced to dipole-dipole scattering
with well defined dipole sizes $|{\vec r}_i|$ and longitudinal quark
momentum fractions $z_i$. The unitarity condition in our model
becomes, therefore, most explicit in the profile function
\be
        J_{DD}(s,|\vec{b}_{\!\perp}|,z_1,|\vec{r}_1|,z_2,|\vec{r}_2|)  
        = \int \frac{d\phi_1}{2\pi}  \int \frac{d\phi_2}{2\pi} 
        \left[1 - S_{DD}(s,\vec{b}_{\!\perp},z_1,{\vec r}_1,z_2,{\vec r}_2)\right]
        \ ,
\label{Eq_DD_profile_function}
\ee
where $\phi_i$ describes the dipole orientation, i.e.\ the angle
between ${\vec r}_i$ and $\vec{b}_{\!\perp}$, and $S_{DD}$ describes
{\em elastic dipole-dipole scattering}
\be
        S_{DD}
        = \frac{2}{3} 
        \cos\!\left(\frac{1}{3}\chi^{\nprt}(s)\right)
        \cos\!\left(\frac{1}{3}\chi^{\pert}(s)\right)         
        + \frac{1}{3}
        \cos\!\left( \frac{2}{3}\chi^{\nprt}(s)\right)
        \cos\!\left( \frac{2}{3}\chi^{\pert}(s)\right)
\label{Eq_S_DD_final_result}
\ee
with the purely real-valued eikonal functions $\chi^{\nprt}(s)$ and
$\chi^{\pert}(s)$ defined in~(\ref{Eq_energy_dependence}). Because of
$|S_{DD}| \leq 1$, a consequence of the cosine functions
in~(\ref{Eq_S_DD_final_result}) describing multiple gluonic
interactions, $J_{DD}$ respects the absolute
limit~(\ref{Eq_absolute_unitarity_limit}). Thus, the elastic
dipole-dipole scattering respects the unitarity
condition~(\ref{Eq_unitarity_condition}).  At high energies, the
arguments of the cosine functions in $S_{DD}$ become so large that
these cosines average to zero in the integration over the dipole
orientations. This leads to the black disc limit $J_{DD}^{max} = 1$
reached at high energies first for small impact parameters.

If one considers the scattering of two dipoles with fixed orientation,
the inelastic overlap function obtained from the unitarity
constraint~(\ref{Eq_unitarity_condition}),
\bea
        &&\!\!\!\!\!\!\!\!
        G_{inel}^{DD}(s,|\vec{b}_{\!\perp}|) 
\\
        &&\!\!\!\!\!\!\!\!
        = \inv{4} \left( 1 - \left[
            \frac{2}{3} \cos\!\left(\frac{1}{3}\chi^{\nprt}(s)\right)
            \cos\!\left(\frac{1}{3}\chi^{\pert}(s)\right) + \frac{1}{3}
            \cos\!\left( \frac{2}{3}\chi^{\nprt}(s)\right) \cos\!\left(
              \frac{2}{3}\chi^{\pert}(s)\right) \right]^2\right) \ , 
\nonumber
\eea
shows nonphysical behavior with increasing energy. This behavior is a
consequence of aritifically fixing the orientations of the dipoles. If
one averages over the dipole orientations as in all high-energy
reactions considered in this work, no unphysical behavior is observed.

\subsection{The Profile Function for Proton-Proton Scattering}
\label{Sec_PP_Profile_Function}

The profile function for proton-proton scattering
\be
        J_{pp}(s,|\vec{b}_{\!\perp}|)  = 
        \int \!\!dz_1 d^2r_1 \!\int \!\!dz_2 d^2r_2      
        |\psi_p(z_1,\vec{r}_1)|^2 |\psi_p(z_2,\vec{r}_2)|^2
        \left[1-S_{DD}(s,\vec{b}_{\!\perp},z_1,{\vec r}_1,z_2,{\vec r}_2)\right]
\label{Eq_model_pp_profile_function}
\ee
is obtained from~(\ref{Eq_DD_profile_function}) by weighting the
dipole sizes $|{\vec r}_i|$ and longitudinal quark momentum fractions
$z_i$ with the proton wave function $|\psi_p(z_i,\vec{r}_i)|^2$ from
Appendix~\ref{Sec_Wave_Functions}.

Using the model parameters from Sec.~\ref{Sec_Model_Parameters}, one
obtains the profile function $J_{pp}(s,|\vec{b}_{\!\perp}|)$ shown in
Fig.~\ref{Fig_J_pp(b,s)} for c.m.\ energies from $\sqrt{s} = 10\,\GeV$
to $\sqrt{s} = 10^8\,\GeV$.
\befig[h!]  \centerline{\epsfig{figure=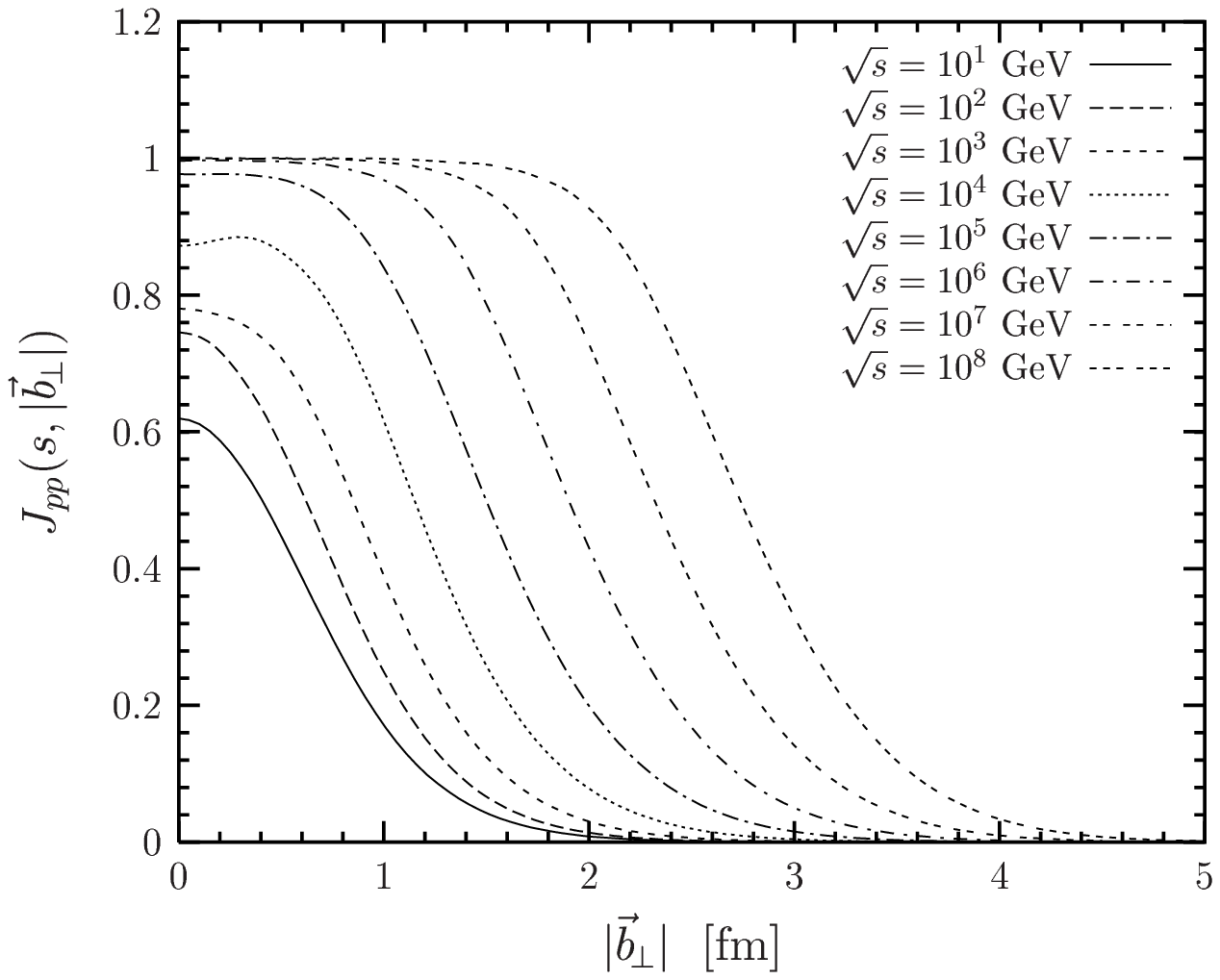}}
\protect\caption{\small The profile function for proton-proton
  scattering $J_{pp}(s,|\vec{b}_{\!\perp}|)$ is shown versus the
  impact parameter $|\vec{b}_{\!\perp}|$ for c.m.\ energies from
  $\sqrt{s} = 10\,\GeV$ to $\sqrt{s} = 10^8\,\GeV$. The unitarity
  limit~(\ref{Eq_absolute_unitarity_limit}) corresponds to
  $J_{pp}(s,|\vec{b}_{\!\perp}|) = 2$ and the black disc
  limit~(\ref{Eq_reduced_unitarity_bound}) to
  $J_{pp}(s,|\vec{b}_{\!\perp}|) = 1$.}
\label{Fig_J_pp(b,s)}
\efig
Up to $\sqrt{s} \approx 100\,\GeV$, the profile has approximately a
Gaussian shape. Above $\sqrt{s}=1\,\TeV$, it significantly develops
into a broader and higher profile until the black disc limit is
reached for $\sqrt{s} \approx 10^6\,\GeV$ and $|\vec{b}_{\!\perp}|=0$.
At this point, the cosine functions in $S_{DD}$ average to zero
\be
        \int \!\!dz_1 d^2r_1 \!\int \!\!dz_2 d^2r_2  
        |\psi_p(z_1,{\vec r}_1)|^2|\psi_p(z_2,{\vec r}_2)|^2
        S_{DD}(\sqrt{s}\gtsim10^6\,\GeV,|\vec{b}_{\!\perp}|=0,\dots) 
        \approx 0
\label{Eq_pp_black_disc_limit_explained}
\ee
so that the proton wave function normalization determines the maximum
opacity
\be
        J_{pp}^{max}
        =\int \!\!dz_1 d^2r_1 \!\int \!\!dz_2 d^2r_2\,        
        |\psi_p(z_1,{\vec r}_1)|^2\,|\psi_p(z_2,{\vec r}_2)|^2
        = 1
        \ .
\label{Eq_pp_black_disc_limit}
\ee
Once the maximum opacity is reached at a certain impact parameter, the
profile function saturates at that $|\vec{b}_{\!\perp}|$ and extends
towards larger impact parameters with increasing energy. Thus, the
multiple gluonic interactions important to respect the $S$-matrix
unitarity constraint~(\ref{Eq_unitarity_condition}) lead to saturation
for $\sqrt{s} \gtsim 10^6\,\GeV$.

The above behavior of the profile function illustrates the evolution
of the proton with increasing c.m.\ energy. The proton is gray and of
small transverse size at small $\sqrt{s}$ but becomes blacker and more
transversally extended with increasing $\sqrt{s}$ until it reaches the
black disc limit in its center at $\sqrt{s} \approx 10^6\,\GeV$.
Beyond this energy, the proton cannot become blacker in its central
region but in its periphery with continuing transverse growth.
Furthermore, the proton boundary seems to stay diffusive as claimed also
in~\cite{Frankfurt:2001nt+X}.

According to our model the black disc limit will not be reached at
LHC. Our prediction of $\sqrt{s} \approx 10^6\,\GeV = 10^3\,\TeV$ for
the onset of the black disc limit in proton-proton collisions is about
two orders of magnitude beyond the LHC energy $\sqrt{s} = 14\,\TeV$.
This is in contrast, for example, with~\cite{Desgrolard:1999pr}, where
the value predicted for the onset of the black disc limit is $\sqrt{s}
= 2\,\TeV$, i.e.\ small enough to be reached at LHC. However, we feel
confidence in our LHC prediction since our profile function
$J_{pp}(s,|\vec{b}_{\!\perp}|)$ yields good agreement with
experimental data for cross sections up to the highest energies as
shown in Sec.~\ref{Sec_Comparison_Data}.

For hadron-hadron reactions in general, the wave function
normalization of the hadrons determines the maximum opacity analogous
to~(\ref{Eq_pp_black_disc_limit}) and the transverse hadron size the
c.m.\ energy at which it is reached. Consequently, the maximum opacity
obtained for $\pi p$ and $K p$ scattering is identical to the one for
$pp$ scattering due to the
normalization~(\ref{Eq_hadron_wave_function_normalization}). Furthermore,
the smaller size of pions and kaons in comparison to protons demands
slightly higher c.m.\ energies to reach this maximum opacity. This
size effect becomes more apparent in longitudinal photon-proton
scattering, where the size of the dipole emerging from the photon can
be controlled by the photon virtuality.

\subsection{The Profile Function for Photon-Proton Scattering}
\label{Sec_GP_Profile_Function}

The profile function for a longitudinal photon $\gamma_L^*$ scattering
off a proton $p$
\bea
        J_{\gamma_L^* p}(s,|\vec{b}_{\!\perp}|,Q^2)  
        & = & 
        \int \!\!dz_1 d^2r_1 \!\int \!\!dz_2 d^2r_2\,
        |\psi_{\gamma_L^*}(z_1,\vec{r}_1,Q^2)|^2 \,
        |\psi_p(z_2,\vec{r}_2)|^2
        \nonumber \\
        && 
        \times
        \left[1-S_{DD}(\vec{b}_{\!\perp},s,z_1,{\vec r}_1,z_2,{\vec r}_2)\right]
\label{Eq_model_gp_profile_function}
\eea
is calculated with the longitudinal photon wave function
$|\psi_{\gamma_L^*}(z_i,\vec{r}_i,Q^2)|^2$ given
in~(\ref{Eq_photon_wave_function_L_squared}). In this way, the profile
function~(\ref{Eq_model_gp_profile_function}) is ideally suited for
the investigation of dipole size effects since the photon virtuality
$Q^2$ determines the transverse size of the dipole into which the
photon fluctuates before it interacts with the proton.

Figure~\ref{Fig_J_gp_(b,s,Q^2)} shows the $|\vec{b}_{\!\perp}|$
dependence of the profile function $J_{\gamma_L^*
  p}(s,|\vec{b}_{\!\perp}|,Q^2)$ divided by $\alphaEM/{\pi}$ for c.m.\ 
energies $\sqrt{s}$ from $10\,\GeV$ to $10^9\,\GeV$ and a photon
virtuality of $Q^2 = 1\,\GeV^2$, where $\alphaEM$ is the
fine-structure constant.
\befig[htb] \centerline{\epsfig{figure=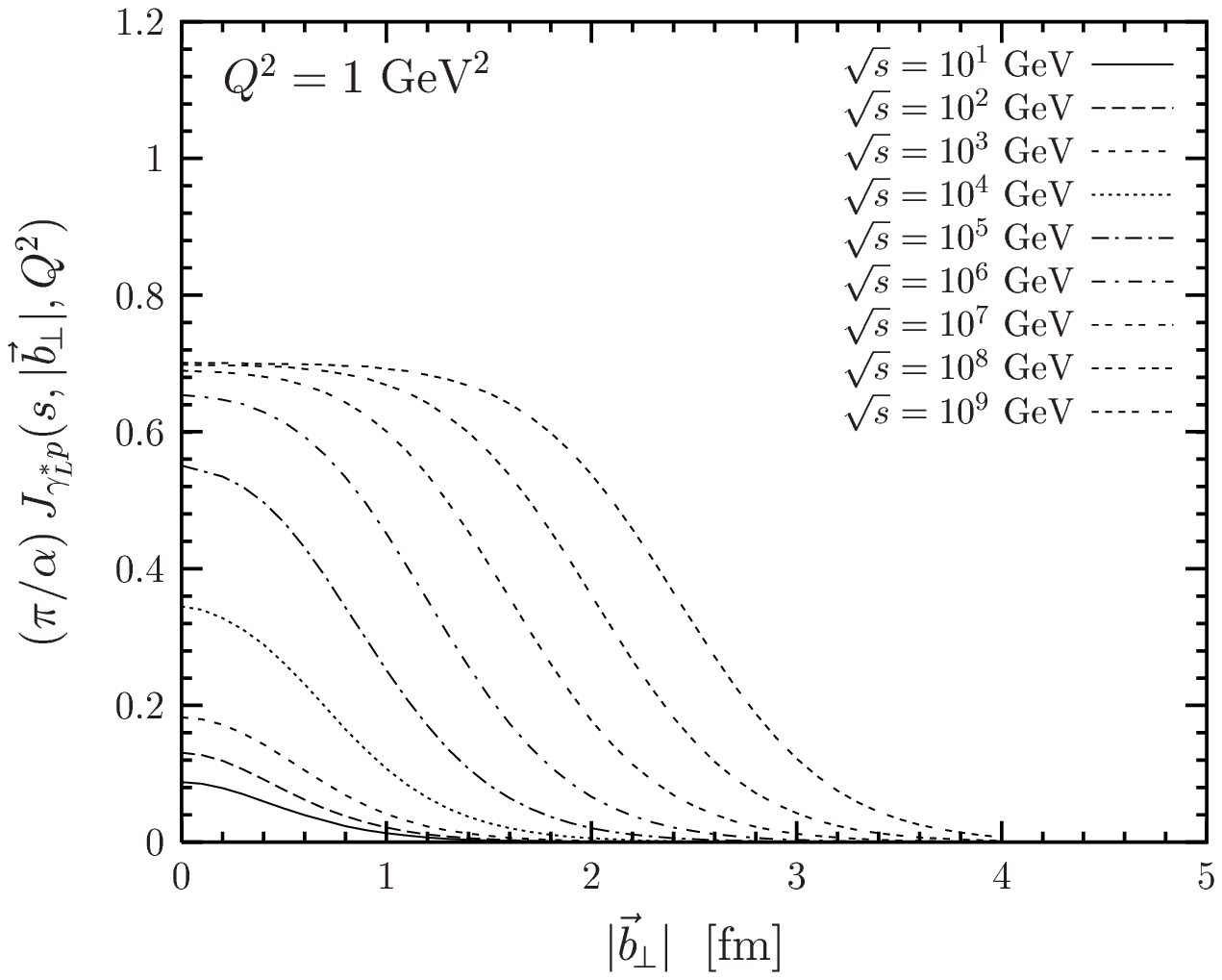}}
\protect\caption{\small The profile function for a longitudinal photon
  scattering off a proton $J_{\gamma_L^*
    p}(s,|\vec{b}_{\!\perp}|,Q^2)$ divided by $\alphaEM/{\pi}$ is
  shown versus the impact parameter $|\vec{b}_{\!\perp}|$ at a photon
  virtuality of $Q^2 = 1\,\GeV^2$ and c.m.\ energies from $\sqrt{s} =
  10\,\GeV$ to $\sqrt{s} = 10^9\,\GeV$. The value of the black disc
  limit is $J_{\gamma_L^*p}^{max}(Q^2=1\,\GeV^2) = 0.00164$\ .}
\label{Fig_J_gp_(b,s,Q^2)}
\end{figure}
One clearly sees that the qualitative behavior of this rescaled
profile function is similar to the one for proton-proton scattering.
However, the black disc limit induced by the underlying dipole-dipole
scattering depends on the photon virtuality $Q^2$ and is given by the
normalization of the longitudinal photon wave function
\bea
        J_{\gamma_L^* p}^{max}(Q^2)  
        = \int \!\!dz d^2r |\psi_{\gamma_L^*}(z,\vec{r},Q^2)|^2
\label{Eq_gp_black_disc_limit}
\eea
since the proton wave function is normalized to one.

The photon virtuality $Q^2$ does not only determine the absolute value
of the black disc limit but also the c.m.\ energy at which it is
reached. This is illustrated in Fig.~\ref{Fig_J_gp_(b=0,s,Q^2)}, where
the $\sqrt{s}$ dependence of $J_{\gamma_L^*
  p}(s,|\vec{b}_{\!\perp}|=0,Q^2)$ divided by $\alphaEM/{\pi}$ is
presented for $Q^2 = 1,\,10,\,\mbox{and}\,100\,\GeV^2$.
\befig[htb]
\centerline{\epsfig{figure=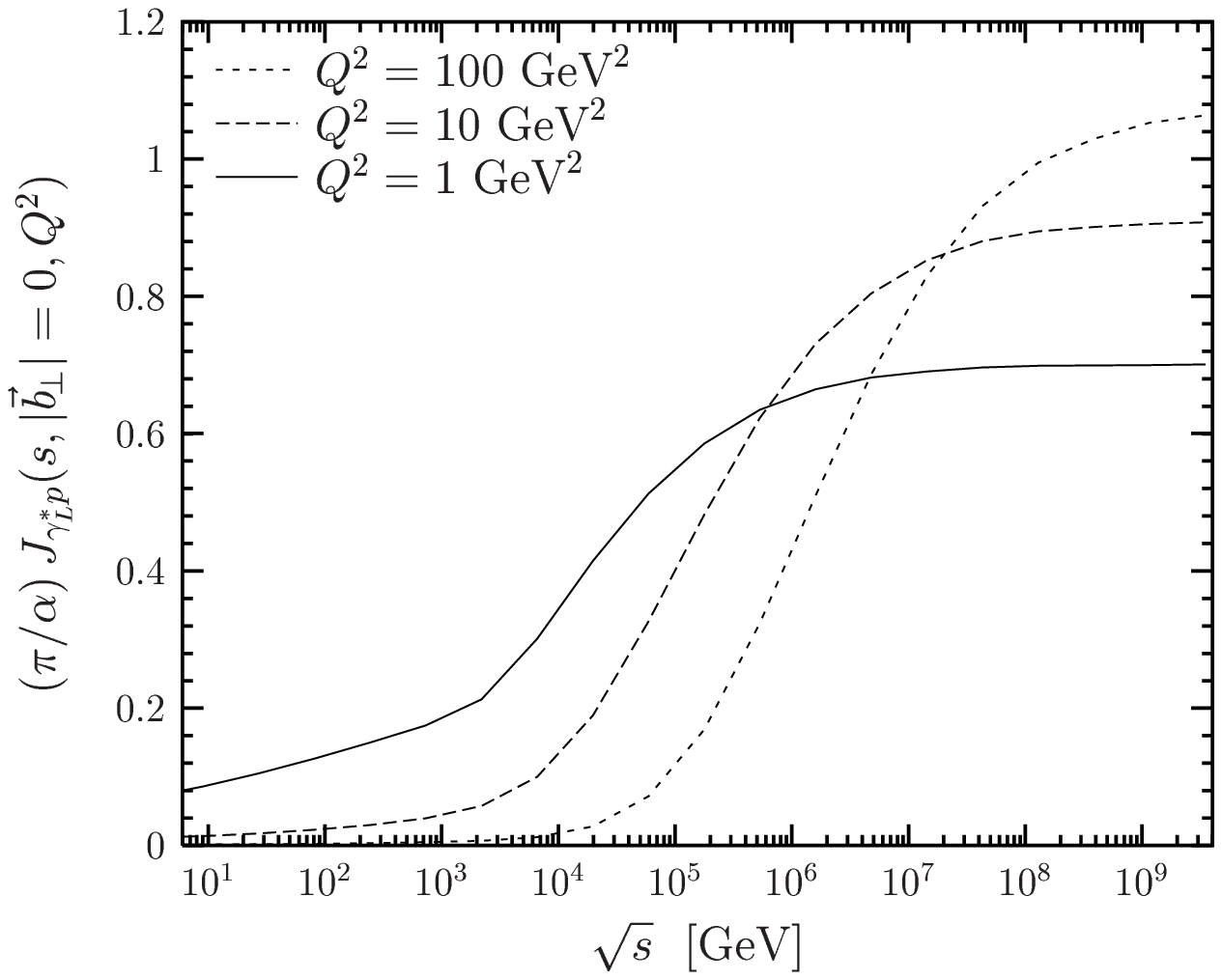}}
\protect\caption{\small The profile function for a longitudinal photon
  scattering off a proton $J_{\gamma_L^*p}(s,|\vec{b}_{\!\perp}|,Q^2)$
  divided by $\alphaEM/{\pi}$ is shown versus the c.m.\ energy
  $\sqrt{s}$ at zero impact parameter ($|\vec{b}_{\!\perp}|=0$) for
  photon virtualities $Q^2 = 1,\,10,\,\mbox{and}\,100\,\GeV^2$.}
\label{Fig_J_gp_(b=0,s,Q^2)}
\end{figure}
With increasing resolution $Q^2$, i.e.\ decreasing dipole sizes,
$|\vec{r}_{\gamma_L^*}|^2 \propto 1/Q^2$, the absolute value of the
black disc limit grows and higher energies are needed to reach this
limit.\footnote{Note that the Bjorken $x$ at which the black disc
  limit is reached decreases with increasing photon virtuality $Q^2$.
  (See also Fig.~\ref{Fig_xg(x,Q^2,b=0)_vs_x})} The growth of the
absolute value of the black disc limit is simply due to the
normalization of the longitudinal photon wave function while the
requirement of higher energies to reach this limit is due to the
decreasing interaction strength with decreasing dipole size. The
latter explains also why the energies needed to reach the black disc
limit in $\pi p$ and $K p$ scattering are higher than in $pp$
scattering. Comparing $\gamma_L^*p$ scattering at $Q^2=1\,\GeV^2$ with
$pp$ scattering quantitatively, the black disc limit
$J_{\gamma_L^*p}^{max}(Q^2=1\,\GeV^2) = 0.00164$ is about three orders
of magnitude smaller because of the photon wave function normalization
($\propto \alphaEM/{\pi}$). At $|\vec{b}_{\!\perp}|=0$ it is reached
at an energy of $\sqrt{s} \approx 10^8\,\GeV$, which is about two orders
of magnitude higher because of the smaller dipoles involved.

The way in which the profile function $J_{\gamma_L^*
  p}(s,|\vec{b}_{\!\perp}|,Q^2)$ approaches the black disc limit at
high energies depends on the shape of the proton and longitudinal
photon wave function at small dipole sizes $|\vec{r}_{1,2}|$. At high
energies, dipoles of typical sizes $0 \leq |\vec{r}_{1,2}| \leq
R_0\,(s_0/s)^{1/4}$ give the main contribution to
$S_{\gamma_L^*p} = 1 - J_{\gamma_L^*p}$ because
of~(\ref{Eq_energy_dependence}) and the fact that the contribution of
the large dipole sizes averages to zero upon integration over the
dipole orientations,
cf.~also~(\ref{Eq_pp_black_disc_limit_explained}). Since
$S_{\gamma_L^*p}$ is a measure of the proton transmittance, this means
that only small dipoles can penetrate the proton at high energies.
Increasing the energy further, even these small dipoles are absorbed
and the black disc limit is reached. However, the dependence of the
profile function on the short distance behavior of normalizable wave functions is
weak which can be understood as follows. Because of color
transparency, the eikonal functions $\chi^{\nprt}(s)$ and
$\chi^{\pert}(s)$ are small for small dipole sizes $0 \leq
|\vec{r}_{1,2}| \leq R_0\,(s_0/s)^{1/4}$ at large
$\sqrt{s}$. Consequently, $S_{DD} \approx 1$ and
\bea
        \!\!\!\!\!\!\!\!
        && 
        \!\!\!\!\!\!\!\!\!\!\!\!\!\!\!\!
        J_{\gamma_L^* p}(s,|\vec{b}_{\!\perp}|,Q^2)  
\nonumber \\
        \!\!\!\!\!\!\!\!
        &&
        \!\!\!\!\!\!\!\!\!\!\!\!\!\!\!\!
         \approx 
        J_{\gamma_L^* p}^{max}(Q^2) - 4\pi^2\!\!
        \int\limits_0^1 \!\!dz_1 \!\!
        \int\limits_0^{r_c(s)}\!\!\!dr_1 r_1 
        |\psi_{\gamma_L^*}(z_1,r_1,Q^2)|^2 
        \int\limits_0^1 \!\!dz_2 \!\!
        \int\limits_0^{r_c(s)}\!\!\!dr_2 r_2 
        |\psi_p(z_2,r_2)|^2
\label{Eq_model_gp_profile_function_wavefunction_independence}
\eea
where $r_c(s) \approx R_0\,(s_0/s)^{1/4}$. Clearly, the
linear behavior from the phase space factors $r_{1,2}$ dominates over
the $r_{1,2}$-dependence of normalizable wave functions.\footnote{For our
  choice of the wave functions
  in~(\ref{Eq_model_gp_profile_function_wavefunction_independence}),
  one sees very explicitly that the specific Gaussian behavior of
  $|\psi_p(z_2,r_2)|^2$ and the logarithmic short distance behavior of
  $|\psi_{\gamma_L^*}(z_1,r_1,Q^2)|^2$ is dominated by the phase space
  factors $r_{1,2}$.} More generally, for any profile function
involving normalizable wave functions, the way in which the black disc
limit is approached depends only weakly on the short distance behavior
of the wave functions.


%
%
\section{A Scenario for Gluon Saturation}
\label{Sec_Gluon_Saturation}

In this section, we estimate the {\em impact parameter dependent gluon
  distribution} of the proton $xG(x,Q^2,|\vec{b}_{\!\perp}|)$.  Using
a leading twist, next-to-leading order QCD relation between
$xG(x,Q^2)$ and the longitudinal structure function $F_L(x,Q^2)$, we
relate $xG(x,Q^2,|\vec{b}_{\!\perp}|)$ to the profile function
$J_{\gamma_L^* p}(s=Q^2/x,|\vec{b}_{\!\perp}|,Q^2)$ and find low-$x$
saturation of $xG(x,Q^2,|\vec{b}_{\!\perp}|)$ as a manifestation of
$S$-matrix unitarity. The resulting $xG(x,Q^2,|\vec{b}_{\!\perp}|)$
is, of course, only an estimate since our profile function contains
also higher twist contributions. Furthermore, in the considered low-$x$
region, the leading twist, next-to-leading order QCD formula may be
inadequate as higher twist contributions~\cite{Martin:1998kk+X} and
higher order QCD corrections~\cite{Gribov:1983tu,Mueller:1986wy} are
expected to become important. Nevertheless, still assuming a close
relation between $F_L(x,Q^2)$ and $xG(x,Q^2)$ at low $x$, we think
that our approach provides some insight into the gluon distribution as
a function of the impact parameter and into its saturation.

The {\em gluon distribution}\ of the proton $~xG(x,Q^2)~$ has the
following meaning: $xG(x,Q^2)dx$ gives the momentum fraction of the
proton which is carried by the gluons in the interval $[x, x+dx]$ as
seen by probes of virtuality $Q^2$. The {\em impact parameter
  dependent gluon distribution} $xG(x,Q^2,|\vec{b}_{\!\perp}|)$ is the
gluon distribution $xG(x,Q^2)$ at a given impact parameter
$|\vec{b}_{\!\perp}|$ so that
\be
        xG(x,Q^2) = \int
        \!\!d^2b_{\!\perp}\,xG(x,Q^2,|\vec{b}_{\!\perp}|) \ .
\label{Eq_def_xg(x,Q^2)}
\ee

In leading twist, next-to-leading order QCD, the gluon distribution
$xG(x,Q^2)$ is related to the structure functions $F_L(x,Q^2)$ and
$F_2(x,Q^2)$ of the proton~\cite{Martin:1988vw}
\be
        F_L(x, Q^2) 
        = \frac{\alphaS}{\pi}\!
        \left[
        \frac{4}{3}\int_x^1 \!
        \frac{dy}{y}\!\left(\frac{x}{y}\right)^{\!\!2} \!F_2(y,Q^2)
        + 2 \sum_f e_f^2\!\int_x^1 \!
        \frac{dy}{y}\!\left(\frac{x}{y}\right)^{\!\!2} \!\!
        \left(\!1-\frac{x}{y}\right) yG(y,Q^2)
        \right]
\label{Eq_FL_QCD_prediction}
\ee
where $\sum_f e_f^2$ is a flavor sum over the quark charges squared.
For four active flavors and $x \ltsim 10^{-3}$,
(\ref{Eq_FL_QCD_prediction}) can be approximated as
follows~\cite{Cooper-Sarkar:1988ds+X}
\be
        xG(x,Q^2) 
        \approx \frac{3}{5}\,5.8\, 
        \left[ 
        \frac{3\pi}{4\alphaS}\, F_L(0.417 x, Q^2)
        - \inv{1.97}\, F_2(0.75 x, Q^2) 
        \right] 
        \ .
\label{Eq_xg(x,Q^2)_approximation}
\ee
For typical $\Lambda_{QCD} = 100-300\,\MeV$ and $Q^2 = 50 -
100\,\GeV^2$, the coefficient of $F_L$ in
(\ref{Eq_xg(x,Q^2)_approximation}), $3\pi/(4\alphaS) = {\cal{O}}(10)$,
is large compared to the one of $F_2$. Taking into account also the
values of $F_2$ and $F_L$, in this $Q^2$ region and for $x \ltsim
10^{-3}$, the gluon distribution is mainly determined by the
longitudinal structure function. The latter can be expressed in terms
of the profile function for longitudinal photon-proton scattering
using the optical theorem (cf.~(\ref{Eq_optical_theorem}))
\be
        F_L(x,Q^2) 
        = \frac{Q^2}{4\,\pi^2\,\alphaEM}\,
        \sigma^{tot}_{\gamma^*_L p}(x,Q^2) 
        = \frac{Q^2}{4\,\pi^2\,\alphaEM}\, 
        2\!\int \!\!d^2b_{\!\perp}\,
        J_{\gamma_L^*p}(x,|\vec{b}_{\!\perp}|,Q^2) 
        \ ,
\label{fl}
\ee
where the $s$-dependence of the profile function is rewritten in terms
of the Bjorken scaling variable, $x = Q^2/s$. Neglecting the $F_2$
term in~(\ref{Eq_xg(x,Q^2)_approximation}), consequently, the gluon
distribution reduces to
\be
        xG(x,Q^2) 
        \approx
        1.305\,\frac{Q^2}{\pi^2 \alphaS}\,\frac{\pi}{\alphaEM}
        \int \!\!d^2b_{\!\perp}\,
        J_{\gamma_L^*p}(0.417 x,|\vec{b}_{\!\perp}|,Q^2)
        \ . 
\label{Eq_xg(x,Q^2)-J_gLp(x,b,Q^2)_connection}
\ee
Comparing (\ref{Eq_def_xg(x,Q^2)}) with
(\ref{Eq_xg(x,Q^2)-J_gLp(x,b,Q^2)_connection}), it seems natural to
relate the integrand of (\ref{Eq_xg(x,Q^2)-J_gLp(x,b,Q^2)_connection})
to the impact parameter dependent gluon distribution
\be
        xG(x,Q^2,|\vec{b}_{\!\perp}|) 
        \approx
        1.305\,\frac{Q^2}{\pi^2 \alphaS}\,\frac{\pi}{\alphaEM}\,
        J_{\gamma_L^*p}(0.417 x,|\vec{b}_{\!\perp}|,Q^2)
        \ .
\label{Eq_xg(x,Q^2,b)-J_gLp(x,b,Q^2)_relation}
\ee

The black disc limit of the profile function for longitudinal
photon-proton scattering~(\ref{Eq_gp_black_disc_limit}) imposes
accordingly an upper bound on $xG(x,Q^2,|\vec{b}_{\!\perp}|)$
\be
        xG(x,Q^2,|\vec{b}_{\!\perp}|)\ \leq \ 
        xG^{max}(Q^2)
        \approx 
        1.305\,\frac{Q^2}{\pi^2 \alphaS}\,\frac{\pi}{\alphaEM}\,
        J_{\gamma^*_L p}^{max}(Q^2)
        \ ,
\label{Eq_low_x_saturation} 
\ee
which is the low-$x$ saturation value of the gluon distribution
$xG(x,Q^2,|\vec{b}_{\!\perp}|)$ in our approach. With $\pi
J_{\gamma^*_L p}^{max}(Q^2)/\alphaEM \approx 1$ as shown in
Fig.~\ref{Fig_J_gp_(b=0,s,Q^2)}, a compact approximation
of~(\ref{Eq_low_x_saturation}) is obtained
\be
        xG(x,Q^2,|\vec{b}_{\!\perp}|)\ \leq \ 
        xG^{max}(Q^2)
        \approx 
        \frac{Q^2}{\pi^2 \alphaS}
        \ ,
\label{Eq_low_x_saturation_approximation} 
\ee
which is consistent with the results
in~\cite{Mueller:1986wy,Mueller:1999wm,Iancu:2001md} and indicates
strong color-field strengths $G^a_{\mu \nu} \sim 1/ \sqrt{\alphaS}$ as
well.

According to our
relations~(\ref{Eq_xg(x,Q^2,b)-J_gLp(x,b,Q^2)_relation})
and~(\ref{Eq_low_x_saturation}), the {\em blackness} described by the
profile function is a measure for the gluon distribution and the {\em
  black disc limit} corresponds to the maximum gluon distribution
reached at the impact parameter under consideration. In accordance
with the behavior of the profile function $J_{\gamma_L^*p}$, see
Fig.~\ref{Fig_J_gp_(b,s,Q^2)}, the gluon distribution
$xG(x,Q^2,|\vec{b}_{\!\perp}|)$ decreases with increasing impact
parameter for given values of $x$ and $Q^2$. The gluon density,
consequently, has its maximum in the geometrical center of the proton,
i.e.\ at zero impact parameter, and decreases towards the periphery.
With decreasing $x$ at given $Q^2$, the gluon distribution
$xG(x,Q^2,|\vec{b}_{\!\perp}|)$ increases and extends towards larger
impact parameters just as the profile function $J_{\gamma_L^*p}$ for
increasing $s$.  The saturation of the gluon distribution
$xG(x,Q^2,|\vec{b}_{\!\perp}|)$ sets in first in the center of the
proton ($|\vec{b}_{\!\perp}|=0$) at very small Bjorken $x$.

In Fig.~\ref{Fig_xg(x,Q^2,b=0)_vs_x}, the gluon distribution
$xG(x,Q^2,|\vec{b}_{\!\perp}|=0)$ is shown as a function of $x$ for
$Q^2 = 1,\,10,\,\mbox{and}\,100\,\GeV^2$, where the
relation~(\ref{Eq_xg(x,Q^2,b)-J_gLp(x,b,Q^2)_relation}) has been used
also for low photon virtualities.
\begin{figure}[htb]
  \centerline{\epsfig{figure=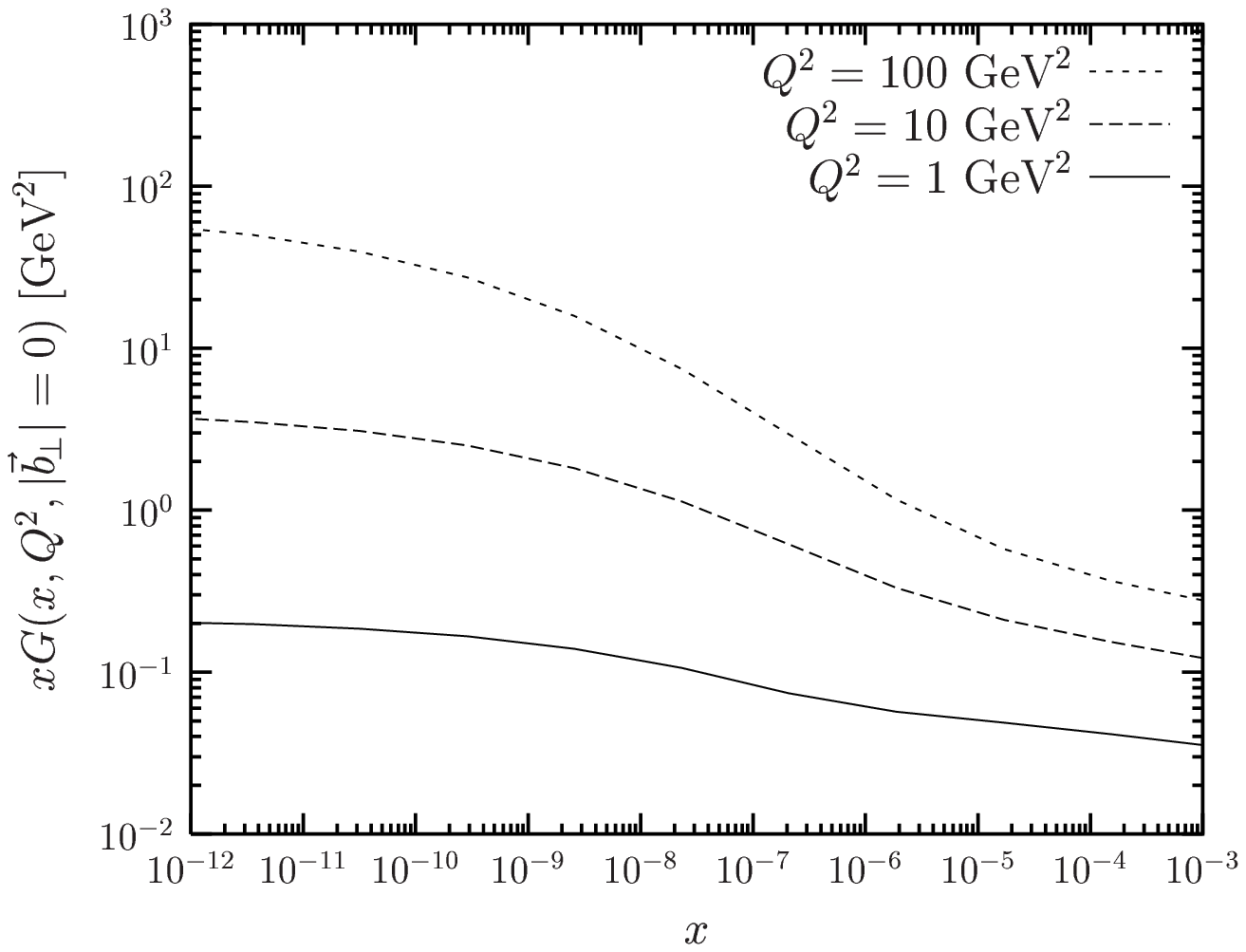}}
  \protect\caption{\small The gluon distribution of the proton at zero
    impact parameter $xG(x,Q^2,|\vec{b}_{\!\perp}|=0)$ is shown as a
    function of $x$ for $Q^2 = 1,\,10,\,\mbox{and}\,100\,\GeV^2$. The
    results are obtained within the
    approximation~(\ref{Eq_xg(x,Q^2,b)-J_gLp(x,b,Q^2)_relation}).}
\label{Fig_xg(x,Q^2,b=0)_vs_x}
\end{figure}
Evidently, the gluon distribution $xG(x,Q^2,|\vec{b}_{\!\perp}|=0)$
saturates at very low values of $x \ltsim 10^{-10}$ for $Q^2 \gtsim
1\,\GeV^2$. The photon virtuality $Q^2$ determines the saturation
value~(\ref{Eq_low_x_saturation}) and the Bjorken-$x$ at which it is
reached (cf. also Fig.~\ref{Fig_J_gp_(b,s,Q^2)}).  For larger $Q^2$,
the low-$x$ saturation value is larger and is reached at smaller
values of $x$, as claimed also in \cite{Gotsman:2001ku}.  Moreover, the
growth of $xG(x,Q^2,|\vec{b}_{\!\perp}|=0)$ with decreasing $x$ becomes
stronger with increasing $Q^2$. This results from the stronger energy
increase of the perturbative component, $\epsilon^{\pert} = 0.73$, that
becomes more important with decreasing dipole size.

According to our approach, the onset of the
$xG(x,Q^2,|\vec{b}_{\!\perp}|)$-saturation appears for $Q^2 \gtsim
1\,\GeV^2$ at $x \ltsim 10^{-10}$, which is far below the $x$-region
accessible at HERA ($x \gtsim 10^{-6}$). Even for THERA ($x\gtsim
10^{-7}$), gluon saturation is not predicted for $Q^2 \gtsim 1
\,\GeV^2$. However, since the HERA data can be described by models
with and without saturation embedded~\cite{Gotsman:2001ku}, the
present situation is not conclusive.\footnote{So far, the most
  striking hint for saturation in the present HERA data at $x\approx
  10^{-4}$ and $Q^2 < 2\,\GeV^2$ has been the turnover of
  $dF_2(x,Q^2)/d\ln(Q^2)$ towards small $x$ in the Caldwell
  plot~\cite{Abramowicz:1999ii}, which is still a controversial issue
  due to the correlation of $Q^2$ and $x$ values.}

Note that the $S$-matrix unitarity
condition~(\ref{Eq_unitarity_condition}) together
with~(\ref{Eq_xg(x,Q^2,b)-J_gLp(x,b,Q^2)_relation}) requires the
saturation of the impact parameter dependent gluon distribution
$xG(x,Q^2,|\vec{b}_{\!\perp}|)$ but not the saturation of the
integrated gluon distribution $xG(x,Q^2)$. Due to multiple gluonic
interactions in our model, this requirement is fulfilled, as can be
seen from Fig.~\ref{Fig_J_gp_(b,s,Q^2)} and
relation~(\ref{Eq_xg(x,Q^2,b)-J_gLp(x,b,Q^2)_relation}). Indeed,
approximating the gluon distribution $xG(x,Q^2,|\vec{b}_{\!\perp}|)$
in the saturation regime of very low $x$ by a step-function
\be
        xG(x,Q^2,|\vec{b}_{\!\perp}|) 
        \approx xG^{max}(Q^2)\,
        \Theta(\,R(x,Q^2)-|\vec{b}_{\!\perp}|\,)
        \ ,
\label{Eq_J_gp_(x,b,Q^2)_Theta-approximation}
\ee
where $R(x,Q^2)$ denotes the full width at half maximum of the profile
function, one obtains with~(\ref{Eq_def_xg(x,Q^2)}),
(\ref{Eq_low_x_saturation}) and
(\ref{Eq_low_x_saturation_approximation}) the integrated gluon
distribution
\be
        xG(x,Q^2) 
        \;\approx\;
        1.305\,\frac{Q^2\,R^2(x,Q^2)}{\pi \alphaS}\,
        \frac{\pi}{\alphaEM}\,
        J_{\gamma^*_L p}^{max}(Q^2)
        \;\approx\;
        \frac{Q^2\,R^2(x,Q^2)}{\pi\alphaS}
        \ ,
\label{Eq_xg(x,Q^2)_saturation_regime}
\ee
which does not saturate because of the increase of the effective
proton radius $R(x,Q^2)$ with decreasing $x$. Nevertheless, although
$xG(x,Q^2)$ does not saturate, the saturation of
$xG(x,Q^2,|\vec{b}_{\!\perp}|)$ leads to a slow-down in its growth
towards small $x$.\footnote{This is analogous to the rise of the total
  $pp$ cross section with growing c.m.\ energy that slows down as the
  corresponding profile function $J_{pp}(s,|\vec{b}_{\!\perp}|)$
  reaches its black disc limit as shown in
  Sec.~\ref{Sec_Total_Cross_Sections}.} Interestingly, our
result~(\ref{Eq_xg(x,Q^2)_saturation_regime}) coincides with the
result of Mueller and Qiu~\cite{Mueller:1986wy}.

Finally, it must be emphasized that the low-$x$ saturation of
$xG(x,Q^2,|\vec{b}_{\!\perp}|)$, required in our approach by the
$S$-matrix unitarity, is realized by {\em multiple gluonic
  interactions}. In other approaches that describe the evolution of
the gluon distribution with varying $x$ and $Q^2$, {\em gluon
  recombination} leads to gluon
saturation~\cite{Gribov:1983tu,Mueller:1986wy,McLerran:1994ni+X,Jalilian-Marian:1999dw+X,Iancu:2001hn+X},
which is reached when the probability of a gluon splitting up into two
is equal to the probability of two gluons fusing into one. A more
phenomenological understanding of saturation is attempted
in~\cite{Golec-Biernat:1999js+X,Capella:2001hq+X}.


%
%
\section{Comparison with Experimental Data}
\label{Sec_Comparison_Data}

In this section, we discuss the phenomenological performance of our
model. We compute total, differential, and elastic cross sections,
structure functions, and diffractive slopes for hadron-hadron,
photon-proton, and photon-photon scattering, compare the results with
experimental data including cosmic ray data, and provide predictions
for future experiments. Having studied the saturation of the impact
parameter profiles, we show here how this manifestation of unitarity
translates into the quantities mentioned above and how it could become
observable.

Using the
$T$-matrix~(\ref{Eq_model_purely_imaginary_T_amplitude_final_result})
with the parameters and wave functions from
Sec.~\ref{Sec_Model_Parameters} and Appendix~\ref{Sec_Wave_Functions},
we compute the {\em pomeron} contribution to $pp$, $p\pbar$,
$\pi^{\pm}p$, $K^{\pm}p$, $\gamma^{*} p$, and $\gamma \gamma$
reactions in terms of the universal dipole-dipole scattering amplitude
$S_{DD}$. This allows one to compare reactions induced by hadrons and
photons in a systematic way. In fact, it is our aim to provide a
unified description of all these reactions and to show in this way
that the pomeron contribution to the above reactions is universal and
can be traced back to the dipole-dipole scattering amplitude $S_{DD}$.

Our model describes pomeron ($C=+1$ gluon exchange) but neither
odderon ($C=-1$ gluon exchange) nor reggeon exchange (quark-antiquark
exchange) as discussed in Sec.~\ref{Sec_Chi_Computation}. Only in the
computation of the hadronic total cross sections the reggeon
contribution is added~\cite{Donnachie:1992ny,Donnachie:2000kp}. This
improves the agreement with the data for $\sqrt{s} \ltsim 100\,\GeV$
and describes exactly the differences between $ab$ and $\bar{a}b$
reactions.

The fine tuning of the model and wave function parameters was
performed on the data shown below. The resulting parameter set given
in Sec.~\ref{Sec_Model_Parameters} and
Appendix~\ref{Sec_Wave_Functions} is used throughout this paper.

\subsection{Total Cross Sections}
\label{Sec_Total_Cross_Sections}

The total cross section for the high-energy reaction $ab \to X$ is
related via the {\em optical theorem} to the imaginary part of the
forward elastic scattering amplitude and can also be expressed in
terms of the profile function~(\ref{Eq_profile_function_def})
\be
        \sigma^{tot}_{ab}(s) 
        \;=\; \inv{s}\,\im\,T(s, t=0) 
        \;=\; 2 \int \!d^2b_{\!\perp}\,J_{ab}(s,|\vec{b}_{\!\perp}|)
        \ ,  
\label{Eq_optical_theorem}
\ee
where $a$ and $b$ label the initial particles whose masses were
neglected as they are small in comparison to the c.m.\ energy
$\sqrt{s}$. 

We compute the pomeron contribution to the total cross section,
$\sigma^{tot, \Pomeron}_{ab}(s)$, from the
$T$-matrix~(\ref{Eq_model_purely_imaginary_T_amplitude_final_result}),
as explained above, and add only here a reggeon contribution of the
form~\cite{Donnachie:1992ny,Donnachie:2000kp}
\be
        \sigma^{tot, \Reggeon}_{ab}(s)
        = X_{ab}\, \left(  \frac{s}{1\,\GeV^2} \right)^{-0.4525} 
        \ ,
\label{Eq_DL_reggeon_contribution}
\ee
where $X_{ab}$ depends on the reaction considered: $X_{pp} =
56.08\,\mb$, $X_{p\pbar} = 98.39\,\mb$, $X_{\pi^+p} = 27.56\,\mb$,
$X_{\pi^-p} = 36.02\,\mb$, $X_{K^+p} = 8.15\,\mb$, $X_{K^-p} =
26.36\,\mb$, $X_{\gamma p} = 0.129\,\mb$, and $X_{\gamma \gamma} =
605\,\nb$. Accordingly, we obtain the total cross section
\be
        \sigma^{tot}_{ab}(s)
        = \sigma^{tot, \Pomeron}_{ab}(s) 
        + \sigma^{tot, \Reggeon}_{ab}(s)
\label{Eq_total_cross_section_final_result}
\ee
for $pp$, $p\pbar$, $\pi^{\pm}p$, $K^{\pm}p$, $\gamma p$ and $\gamma
\gamma$ scattering.

\begin{figure}[p]
\setlength{\unitlength}{1.cm}
\begin{center}
\epsfig{file=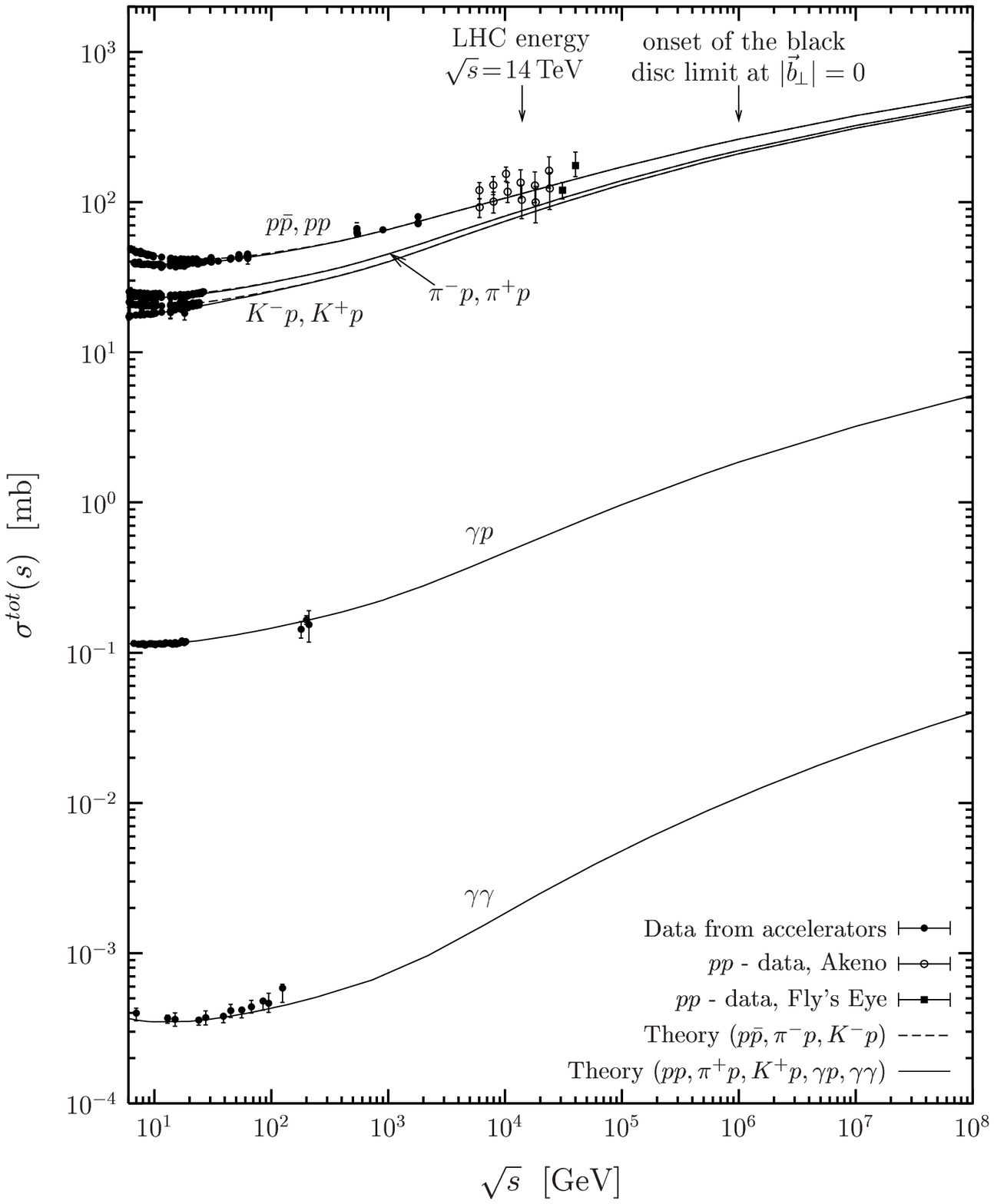, width = 14cm}
\end{center}
\caption{ \small The total cross section $\sigma^{tot}$ is shown as a
  function of the c.m.\ energy $\sqrt{s}$ for $pp$, $p\pbar$,
  $\pi^{\pm}p$, $K^{\pm}p$, $\gamma p$ and $\gamma \gamma$ scattering.
  The solid lines represent the model results for $pp$, $\pi^+p$, $K^+p$,
  $\gamma p$ and $\gamma \gamma$ scattering and the dashed lines the
  ones for $p\pbar$, $\pi^-p$, and $K^-p$ scattering. The $pp$,
  $p\pbar$, $\pi^{\pm}p$, $K^{\pm}p$, $\gamma p$~\cite{Groom:2000in}
  and $\gamma \gamma$ data~\cite{Abbiendi:2000sz+X} taken at
  accelerators are indicated by the closed circles while the closed
  squares (Fly's eye data)~\cite{Baltrusaitis:1984ka+X} and the open
  circles (Akeno data)~\cite{Honda:1993kv+X} indicate cosmic ray data.
  The arrows at the top point to the LHC energy, $\sqrt{s} =
  14\,\TeV$, and to the onset of the black disc limit in $pp$ ($p\pbar$)
  reactions, $\sqrt{s} \approx 10^6\,\GeV$.}
\label{Fig_sigma_tot}
\end{figure}

The good agreement of the computed total cross sections with the
experimental data is shown in Fig.~\ref{Fig_sigma_tot}.
Here, the solid lines represent the theoretical results for $pp$,
$\pi^+p$, $K^+p$, $\gamma p$, and $\gamma \gamma$ scattering and the
dashed lines the ones for $p\pbar$, $\pi^-p$, and $K^-p$ scattering.
The $pp$, $p\pbar$, $\pi^{\pm}p$, $K^{\pm}p$, $\gamma
p$~\cite{Groom:2000in} and $\gamma \gamma$
data~\cite{Abbiendi:2000sz+X} taken at accelerators are indicated by
the closed circles while the closed squares (Fly's eye
data)~\cite{Baltrusaitis:1984ka+X} and the open circles (Akeno
data)~\cite{Honda:1993kv+X} indicate cosmic ray data. Concerning the
photon-induced reactions, only real photons are considered which are,
of course, only transverse polarized.

The prediction for the total $pp$ cross section at LHC ($\sqrt{s} =
14\,\TeV$) is $\sigma^{tot}_{pp} = 114.2\,\mb$ in good agreement with
the cosmic ray data. Compared with other works, our LHC prediction is
close to the one of Block et al.~\cite{Block:1999hu},
$\sigma^{tot}_{pp} = 108 \pm 3.4\,\mb$, but considerably larger than
the one of Donnachie and Landshoff~\cite{Donnachie:1992ny},
$\sigma^{tot}_{pp} = 101.5\,\mb$.

The differences between $ab$ and $\bar{a}b$ reactions for $\sqrt{s}
\ltsim 100\,\GeV$ result solely from the different reggeon
contributions which die out rapidly as the energy increases. The
pomeron contribution to $ab$ and $\bar{a}b$ reactions is, in
contrast, identical and increases as the energy increases. It thus
governs the total cross sections for $\sqrt{s} \gtsim 100\,\GeV$ where
the results for $ab$ and $\bar{a}b$ reactions coincide.

The differences between $pp$ ($p\pbar$), $\pi^{\pm}p$, and $K^{\pm}p$
scattering result from the different transverse extension parameters,
$S_p = 0.86\,\fm > S_{\pi} = 0.607\,\fm > S_{K} = 0.55\,\fm$, cf.\ 
Appendix~\ref{Sec_Wave_Functions}.  Since a smaller transverse
extension parameter favors smaller dipoles, the total cross section
becomes smaller, and the short distance physics described by the
perturbative component becomes more important and leads to a stronger
energy growth due to $\epsilon^{\pert} = 0.73 > \epsilon^{\nprt} =
0.125$. In fact, the ratios $\sigma^{tot}_{pp}/\sigma^{tot}_{\pi p}$
and $\sigma^{tot}_{pp}/\sigma^{tot}_{Kp}$ converge slowly towards
unity with increasing energy as can already be seen in
Fig.~\ref{Fig_sigma_tot}.

For real photons, the transverse size is governed by the constituent
quark masses $m_f(Q^2=0)$, cf.\ Appendix~\ref{Sec_Wave_Functions},
where the light quarks have the strongest effect, i.e.\ 
$\sigma^{tot}_{\gamma p} \propto 1/m_{u,d}^2$ and
$\sigma^{tot}_{\gamma \gamma} \propto 1/m_{u,d}^4$. Furthermore, in
comparison with hadron-hadron scattering, there is the additional
suppression factor of $\alphaEM$ for $\gamma p$ and $\alphaEM^2$ for
$\gamma \gamma$ scattering coming from the photon-dipole transition.
In the $\gamma \gamma$ reaction, also the box diagram
contributes~\cite{Budnev:1975zs,Donnachie:2000kp} but is neglected
since its contribution to the total cross section is less than
1\%~\cite{Donnachie:2001wt}.

It is worthwhile mentioning that total cross sections for $pp$
($p\pbar$), $\pi^{\pm}p$, and $K^{\pm}p$ scattering do not depend on
the longitudinal quark momentum distribution in the hadrons since the
underlying dipole-dipole cross section is independent of the
longitudinal quark momentum fraction $z_i$ for $t = 0$. We show this
analytically on the two-gluon-exchange level
in~\cite{K-Space_Investigations}.

Saturation effects as a manifestation of the $S$-matrix unitarity can
be seen in Fig.~\ref{Fig_sigma_tot}. Having investigated the profile
function for $pp$ ($p\pbar$) scattering, we know that this profile
function becomes higher and broader with increasing energy until it
saturates the black disc limit first for zero impact parameter
($|\vec{b}_{\!\perp}|=0$) at $\sqrt{s} \approx 10^6\,\GeV$.
Beyond this energy, the profile function cannot become higher but
expands towards larger values of $|\vec{b}_{\!\perp}|$. Consequently,
the total cross section~(\ref{Eq_optical_theorem}) increases no longer
due to the growing blackness at the center but only due to the
transverse expansion of the hadrons. This tames the growth of the
total cross section as can be seen for the total $pp$ cross section
beyond $\sqrt{s} \approx 10^6\,\GeV$ in Fig.~\ref{Fig_sigma_tot}.

At energies beyond the onset of the black disc limit at zero impact
parameter, the profile function can be approximated by
\be 
      J_{ab}^{approx}(s,|\vec{b}_{\!\perp}|) =
      N_a\,N_b\,\Theta\left(R(s)-|\vec{b}_{\!\perp}|\right) \,
\label{Eq_J_ab_asymptotic_energies}
\ee
where $N_{a,b}$ denotes the normalization of the wave functions of the
scattered particles and $R(s)$ the full width at half maximum of the
exact profile function $J_{ab}(s,|\vec{b}_{\!\perp}|)$ that reflects
the effective radii of the interacting particles. Thus, the energy
dependence of the total cross section~(\ref{Eq_optical_theorem}) is
driven exclusively by the increase of the transverse extension of the
particles $R(s)$
\be
        \sigma^{tot}_{ab}(s) = 2 \pi N_a N_b R(s)^2
        \ ,
\label{Eq_sigma_tot_ab_asymptotic_energies}
\ee
which is known as {\em geometrical
  scaling}~\cite{Amaldi:1980kd,Castaldi:1983ft}. The growth of $R(s)$
can at most be logarithmic for $\sqrt{s} \to \infty$ because of
the Froissart-Lukaszuk-Martin bound~\cite{Froissart:1961ux+X}.  In
fact, a transition from a power-like to an $\ln^2$-increase of the
total $pp$ cross section seems to set in at about $\sqrt{s} \approx
10^6\,\GeV$ as visible in Fig.~\ref{Fig_sigma_tot}. Moreover, since the
hadronic cross sections join for $\sqrt{s} \to \infty$, $R(s)$ becomes
independent of the hadron-hadron reaction considered at asymptotic
energies as long as $N_{a,b}=1$. Also for photons of different
virtuality $Q_1^2$ and $Q_2^2$ one can check that the ratio of the
total cross sections $\sigma^{tot}_{\gamma^*
  p}(Q_1^2)/\sigma^{tot}_{\gamma^* p}(Q_2^2)$ converges to unity at
asymptotic energies in agreement with the conclusion
in~\cite{Schildknecht:2001qe}.

\subsection{The Proton Structure Function}
\label{Sec_Structure_Functions}

The total cross section for the scattering of a transverse ($T$) and
longitudinally ($L$) polarized photon off the proton,
$\sigma_{\gamma^*_{T\!,L}p}^{tot}(x,Q^2)$, at photon virtuality $Q^2$ and
c.m.\ energy\footnote{Here, $\sqrt{s}$ refers to the c.m.\ energy in
  the $\gamma^* p$ system.} squared, $s=Q^2/x$, is equivalent to the
{\em structure functions} of the proton 
\be
        F_{T,L}(x,Q^2) 
        = \frac{Q^2}{4\pi^2\alphaEM} 
        \sigma_{\gamma^*_{T\!,L}p}^{tot}(x,Q^2)
\label{Eq_FTL}
\ee
and
\be
        F_2(x,Q^2) = F_{T}(x,Q^2) + F_{L}(x,Q^2)
        \ .
\label{Eq_F2}
\ee

Reactions induced by virtual photons are particularly interesting
because the transverse separation of the quark-antiquark pair that
emerges from the virtual photon decreases as the photon virtuality
increases (cf.\ Appendix~\ref{Sec_Wave_Functions})
\be
        |\vec{r}_\gamma| \approx \frac{2}{\sqrt{Q^2+4m_{u,d}^2}} 
        \ ,
\label{pts}
\ee
where $m_{u,d}$ is a mass of the order of the constituent $u$-quark
mass. With increasing virtuality, one probes therefore smaller
transverse distance scales of the proton.

\begin{figure}[p]
  \centerline{\psfig{figure=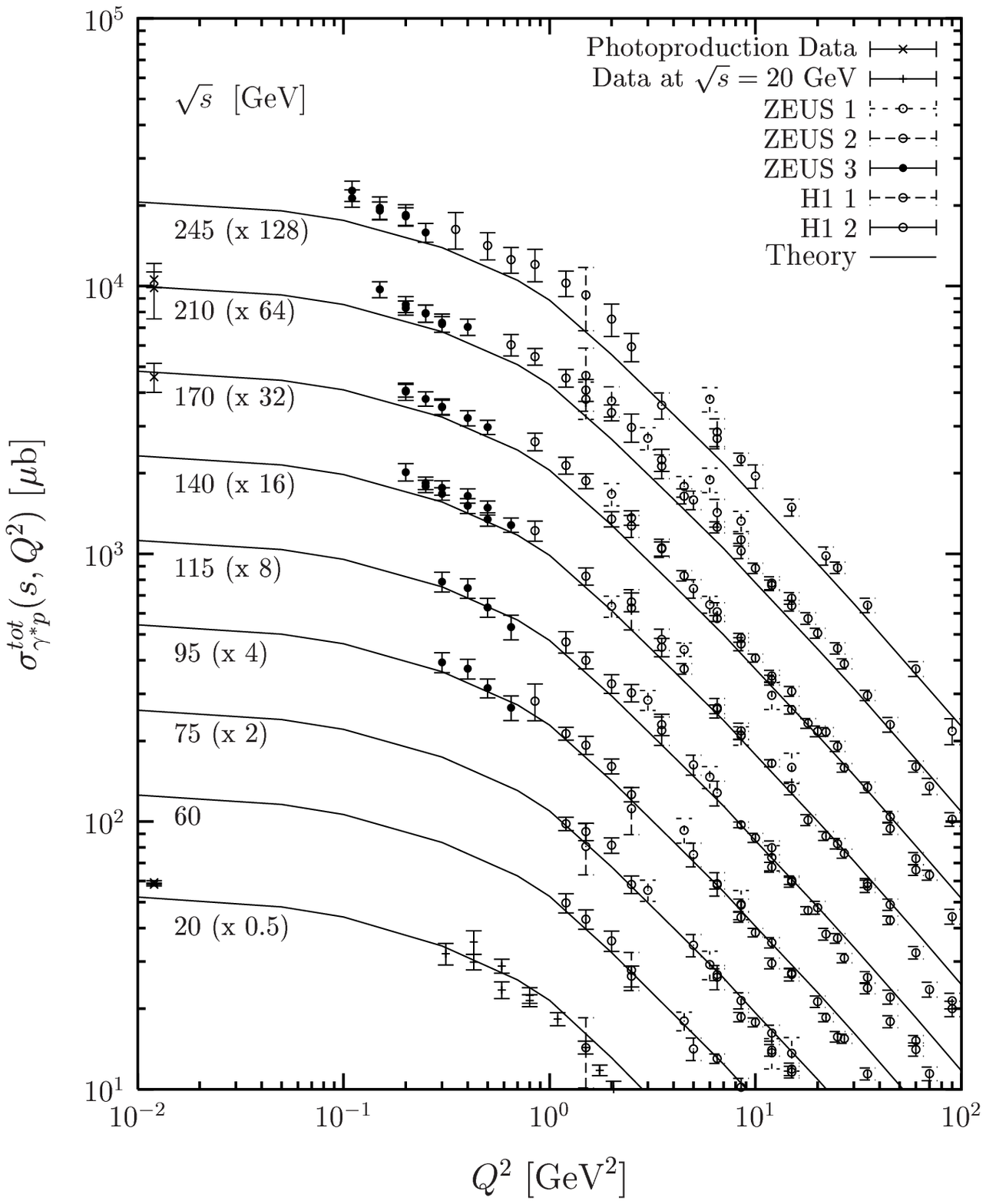,
      width=14cm}} \protect\caption{\small The total $\gamma^* p$
    cross section, $\sigma^{tot}_{\gamma^* p}(s,Q^2)$, is shown as a
    function of the photon virtuality $Q^2$ for c.m.\ energies from
    $\sqrt{s} = 20\,\GeV$ to $\sqrt{s} = 245\,\GeV$, where the model
    results (solid lines) and the experimental data at different
    $\sqrt{s}$ values are scaled with the indicated factors. The low
    energy data at $\sqrt{s} = 20\,\GeV$ are
    from~\cite{Benvenuti:1989rh+X}, the data at higher energies from
    the H1~\cite{Aid:1996au+X} and ZEUS
    collaboration~\cite{Derrick:1996ef+X}. The photoproduction
    ($Q^2=0$) data from~\cite{Caldwell:1978yb+X} are displayed at $Q^2
    = 0.012\,\GeV^2$.}
\label{Fig_sigma_tot_gp_vs_Q^2}
\end{figure}

In Fig.~\ref{Fig_sigma_tot_gp_vs_Q^2}, the $Q^2$-dependence of the
total $\gamma^* p$ cross section
\be
        \sigma^{tot}_{\gamma^*p}(s,Q^2)
        = \sigma^{tot}_{\gamma_T^*p}(s,Q^2)
        + \sigma^{tot}_{\gamma_L^*p}(s,Q^2)
\label{Eq_sigma_tot_gp_=_sigma_T_+_sigma_L}
\ee
is presented, where the model results (solid lines) are compared with
the experimental data for c.m.\ energies from $\sqrt{s} = 20\,\GeV$ up
to $\sqrt{s} = 245\,\GeV$. Note the indicated scaling factors at
different $\sqrt{s}$ values. The low energy data at $\sqrt{s} =
20\,\GeV$ are from~\cite{Benvenuti:1989rh+X} while the data at higher
energies have been measured at HERA by the H1~\cite{Aid:1996au+X} and
ZEUS collaboration~\cite{Derrick:1996ef+X}. At $Q^2 = 0.012\,\GeV^2$,
also the photoproduction ($Q^2=0$) data from~\cite{Caldwell:1978yb+X}
are displayed.

In the window shown in Fig.~\ref{Fig_sigma_tot_gp_vs_Q^2}, the model
results are in reasonable agreement with the experimental data. The
total $\gamma^* p$ cross section levels off towards small values of
$Q^2$ as soon as the photon size $|\vec{r}_\gamma|$, i.e\ the
resolution scale, becomes comparable to the proton size. Our model
reproduces this behavior by using the perturbative photon wave
functions with $Q^2$-dependent quark masses, $m_f(Q^2)$, that
interpolate between the current (large $Q^2$) and the constituent
(small $Q^2$) quark masses as explained in detail in
Appendix~\ref{Sec_Wave_Functions}. The decrease of
$\sigma^{tot}_{\gamma^* p}$ with increasing $Q^2$ results from the
decreasing dipole sizes since small dipoles do not interact as
strongly as large dipoles.

\begin{figure}[htb]
\centerline{\psfig{figure=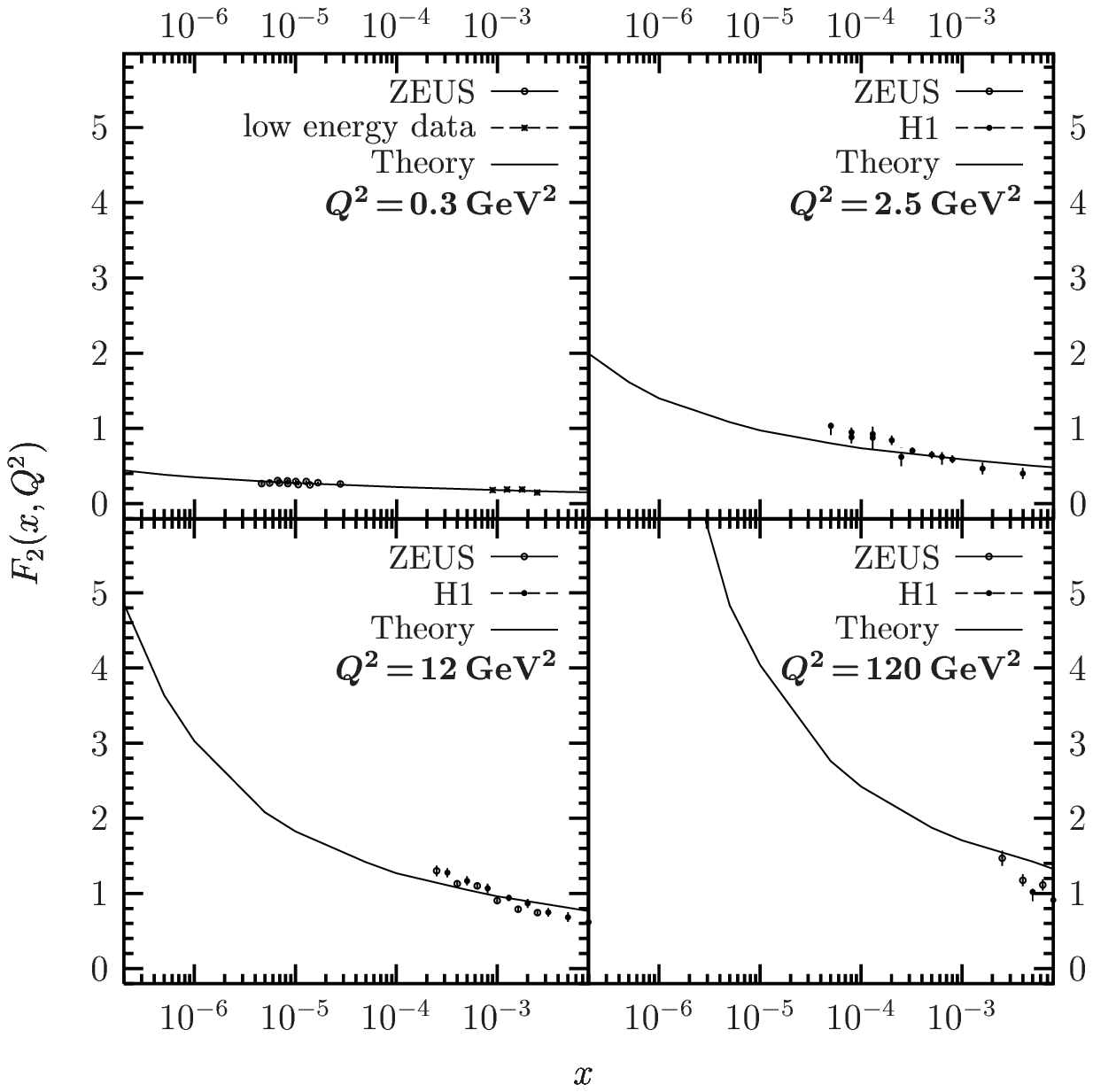, width=11.cm}}
\caption{\small The $x$-dependence of the computed proton structure
  function $F_2(x,Q^2)$ (solid line) is shown for $Q^2 =
  0.3,\,2.5,\,12$ and $120\,\GeV^2$ in comparison to the data measured
  by the H1~\cite{Abt:1993cb+X} and ZEUS~\cite{Derrick:1993ft+X}
  detector, and the low energy data at $\sqrt{s} = 20\,\GeV$
  from~\cite{Benvenuti:1989rh+X}.}
\label{F2_p}
\end{figure}

The $x$-dependence of the computed proton structure function
$F_2(x,Q^2)$ is shown in Fig.~\ref{F2_p} for $Q^2 = 0.3,\,2.5,\,12$
and $120\,\GeV^2$ in comparison to the data measured by the
H1~\cite{Abt:1993cb+X} and ZEUS~\cite{Derrick:1993ft+X} detector.
Within our model, the increase of $F_2(x,Q^2)$ towards small Bjorken
$x$ becomes stronger with increasing $Q^2$ in agreement with the
trend in the HERA data. This behavior results from the fast energy
growth of the perturbative component that becomes more important with
increasing $Q^2$ due to the smaller dipole sizes involved.

As can be seen in Fig.~\ref{F2_p}, the data show a stronger increase
with decreasing $x$ than the model outside the low-$Q^2$ region. This
results from the weak energy boost of the non-perturbative component
that dominates $F_2(x,Q^2)$ in our model. In fact, even for large
$Q^2$ the non-perturbative contribution overwhelms the perturbative
one, which explains also the overestimation of the data for $x \gtsim
10^{-3}$.
 
This problem is typical for the \SVM\ model applied to the scattering
of a small size dipole off a proton. In an earlier application by
R\"uter~\cite{Rueter:1998up}, an additional cut-off was introduced to
switch from the non-perturbative to the perturbative contribution as
soon as one of the dipoles is smaller than $r_{cut} = 0.16\,\fm$. This
yields a better agreement with the data also for large $Q^2$ but leads
to a discontinuous dipole-proton cross section. In the model of
Donnachie and Dosch~\cite{Donnachie:2001wt}, a similar \SVM-based
component is used also for dipoles smaller than $R_c = 0.22\,\fm$ with
a strong energy boost instead of a perturbative component.
Furthermore, their \SVM-based component is tamed for large $Q^2$ by an
additional $\alphaS(Q^2)$ factor.

We did not follow these lines in order to keep a continuous,
$Q^2$-independent dipole-proton cross section and, therefore, cannot
improve the agreement with the $F_2(x,Q^2)$ data without losing
quality in the description of hadronic observables. Since our
non-perturbative component relies on lattice QCD, we are more
confident in describing non-perturbative physics and, thus, put more
emphasis on the hadronic observables. Admittedly, our model misses
details of the proton structure that become visible with increasing
$Q^2$. In comparison, most other existing models provide neither the
profile functions nor a simultaneous description of hadronic and
$\gamma^*$-induced processes.

\subsection[The Slope $B$ of Elastic Forward Scattering]{The Slope \boldmath$B$ of Elastic Forward Scattering}
\label{Sec_Slope_B}

The {\em local slope} of elastic scattering $B(s,t)$ is defined as
\be
        B(s,t) := 
        \frac{d}{dt} \left( \ln \left[ \frac{d\sigma^{el}}{dt}(s,t) \right] \right)
\label{Eq_elastic_local_slope}
\ee
and, thus, characterizes the diffractive peak of the differential
elastic cross section $d\sigma^{el}/dt(s,t)$ discussed below. Here, we
concentrate on the slope for elastic forward ($t=0$) scattering also
called {\em slope parameter}
\be
        B(s) 
        := B(s,t=0) 
        = \inv{2} 
        \frac{\int\!d^2b_{\!\perp}\,|\vec{b}_{\!\perp}|^2\,J(s,|\vec{b}_{\!\perp}|)}
        {\int\!d^2b_{\!\perp}\,J(s,|\vec{b}_{\!\perp}|)}
        = \inv{2} \langle b^2 \rangle \ ,
\label{Eq_elastic_forward_slope}
\ee
which measures the rms interaction radius $\langle b^2 \rangle$ of the
scattered particles, and does not depend on the opacity. 

We compute the slope parameter with the profile function from the
$T$-matrix~(\ref{Eq_model_purely_imaginary_T_amplitude_final_result})
and neglect the reggeon contributions, which are relevant only at
small c.m.\ energies, so that the same result is obtained for $ab$ and
$\bar{a}b$ scattering.

\begin{figure}[htb] 
\centerline{\epsfig{figure=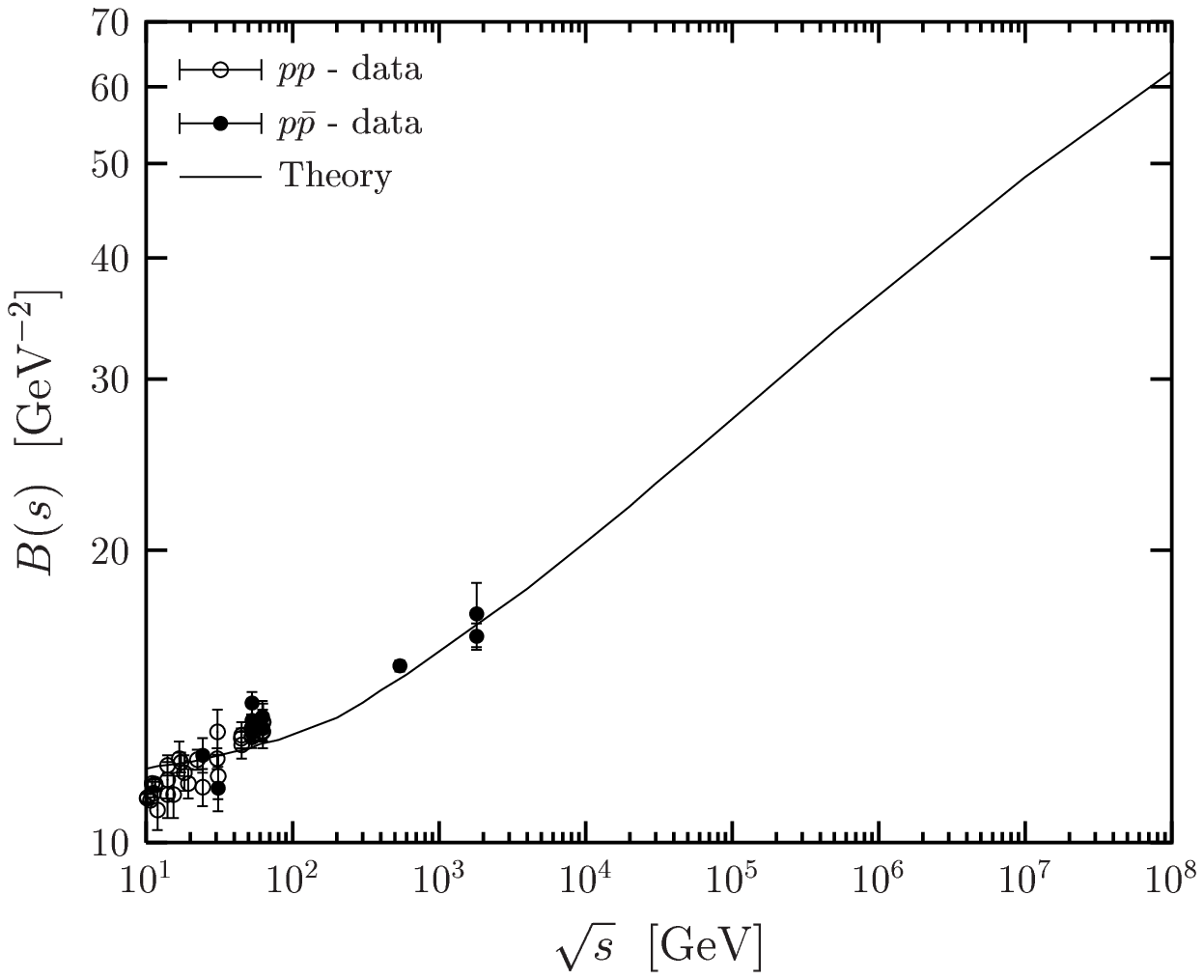, width=11.cm}}
\caption{The elastic slope parameter $B(s)$ is shown as a function of 
  the c.m.\ energy $\sqrt{s}$ for $pp$ and $p\pbar$ forward ($t=0$)
  scattering. The solid line represents the model result that is
  compared with the data for $pp$ (open circles) and $p\pbar$ (closed
  circles) reactions
  from~\cite{Amaldi:1971kt+X,Bozzo:1984ri,Amos:1989at}.}
\label{Fig_B_pp}
\end{figure}

In Fig.~\ref{Fig_B_pp}, the resulting slope parameter $B(s)$ is shown
as a function of $\sqrt{s}$ for $pp$ and $p\pbar$ scattering (solid
line) and compared with the $pp$ (open circles) and $p\pbar$ (closed
circles) data from~\cite{Amaldi:1971kt+X,Bozzo:1984ri,Amos:1989at}.
As expected from the opacity independence of the slope parameter
(\ref{Eq_elastic_forward_slope}), saturation effects as seen in the
total cross sections do not occur. Indeed, one observes an approximate
$B(s) \propto R^2(s) \propto \ln^2(\sqrt{s}/\sqrt{s_0})$ growth for
$\sqrt{s} \gtsim 10^4\,\GeV$.  This behavior agrees, of course, with
the transverse expansion of $J_{pp}(s,|\vec{b}_{\!\perp}|)$ for
increasing $\sqrt{s}$ shown in Fig.~\ref{Fig_J_pp(b,s)}. Analogous
results are obtained also for $\pi p$ and $Kp$ scattering.

For the good agreement of our model with the data, the finite width of
the longitudinal quark momentum distribution in the hadrons, i.e.\ 
$\Delta z_p,\,\Delta z_{\pi},\,\mbox{and}\,\Delta z_{K}\neq 0$
in~(\ref{Eq_hadron_wave_function}), is important as the numerator in
(\ref{Eq_elastic_forward_slope}) depends on this width. In fact,
$B(s)$ comes out more than 10\% smaller with $\Delta z_p,\,\Delta
z_{\pi},\,\mbox{and}\,\Delta z_{K}= 0$. Furthermore, a strong growth
of the perturbative component, $\epsilon^{\pert} = 0.73$, is important
to achieve the increase of $B(s)$ for $\sqrt{s} \gtsim 500\,\GeV$
indicated by the data.

It must be emphasized that only the simultaneous fit of the total
cross section and the slope parameter provides the correct shape of
the profile function $J(s,|\vec{b}_{\!\perp}|)$. This shape leads then
automatically to a good description of the differential elastic cross
section $d\sigma^{el}/dt(s,t)$ as demonstrated below. Astonishingly,
only few phenomenological models provide a satisfactory description of
both quantities~\cite{Block:1999hu,Kopeliovich:2001pc}. In the
approach of~\cite{Berger:1999gu}, for example, the total cross section
is described correctly while the slope parameter exceeds the data by
more than 20\% already at $\sqrt{s} = 23.5\,\GeV$ and, thus, indicates
deficiencies in the form of $J(s,|\vec{b}_{\!\perp}|)$.

\subsection{The Differential Elastic Cross Section}
\label{Sec_Diff_El_Cross_Section}

The {\em differential elastic cross section} obtained from the squared
absolute value of the $T$-matrix element
\be
        \frac{d\sigma^{el}}{dt}(s,t) 
        = \inv{16 \pi s^2}|T(s,t)|^2
\label{Eq_dsigma_el_dt}
\ee
can be expressed for our purely imaginary
$T$-matrix~(\ref{Eq_model_purely_imaginary_T_amplitude_final_result})
in terms of the profile function
\be
        \frac{d\sigma^{el}}{dt}(s,t) 
        = \inv{4\pi} \left[ 
        \int \!\!d^2b_{\!\perp}\,
        e^{i {\vec q}_{\!\perp} {\vec b}_{\!\perp}}\,
        J(s,|\vec{b}_{\!\perp}|)
        \right ]^2
        \ .
\label{Eq_dsigma_el_dt_model}
\ee
and is, thus, very sensitive to the transverse extension {\em and}
opacity of the scattered particles.
Equation~(\ref{Eq_dsigma_el_dt_model}) reminds of optical diffraction,
where $J(s,|\vec{b}_{\!\perp}|)$ describes the distribution of an
absorber that causes the diffraction pattern observed for incident
plane waves.

\begin{figure}[p]
  \centerline{\psfig{figure=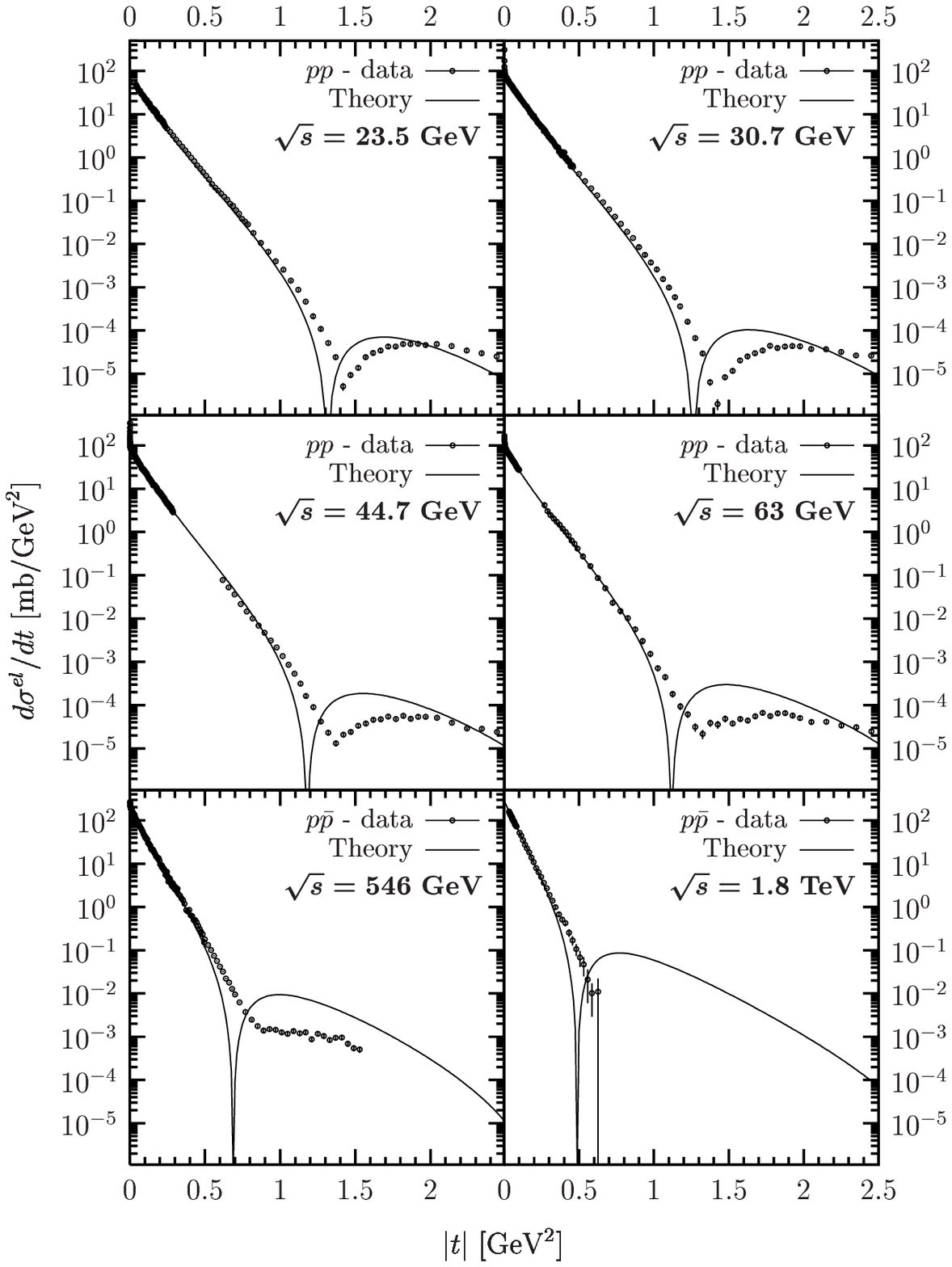,
      width=13.cm}} \protect\caption{ \small The differential elastic
    cross section for $pp$ and $p\pbar$ scattering is shown as a
    function of $|t|$ up to $2.5\,\GeV^2$. The result of our model,
    indicated by the solid line, is compared for $\sqrt{s} =
    23.5,\,30.7,\,44.7,\,\mbox{and}\,63\,\GeV$ with the CERN ISR $pp$
    data~\cite{Amaldi:1980kd}, for $\sqrt{s} = 546\,\GeV$ with the
    CERN $Sp{\pbar}S$ $p\pbar$ data~\cite{Bozzo:1984ri}, and for
    $\sqrt{s} = 1.8\,\TeV$ with the Fermilab Tevatron $p\pbar$
    data~\cite{Amos:1989at,Amos:1990jh}, all indicated by the open
    circles with error bars.}
\label{Fig_dsigma_el_dt_pp}
\end{figure}

\begin{figure}[tb]
  \centerline{\psfig{figure=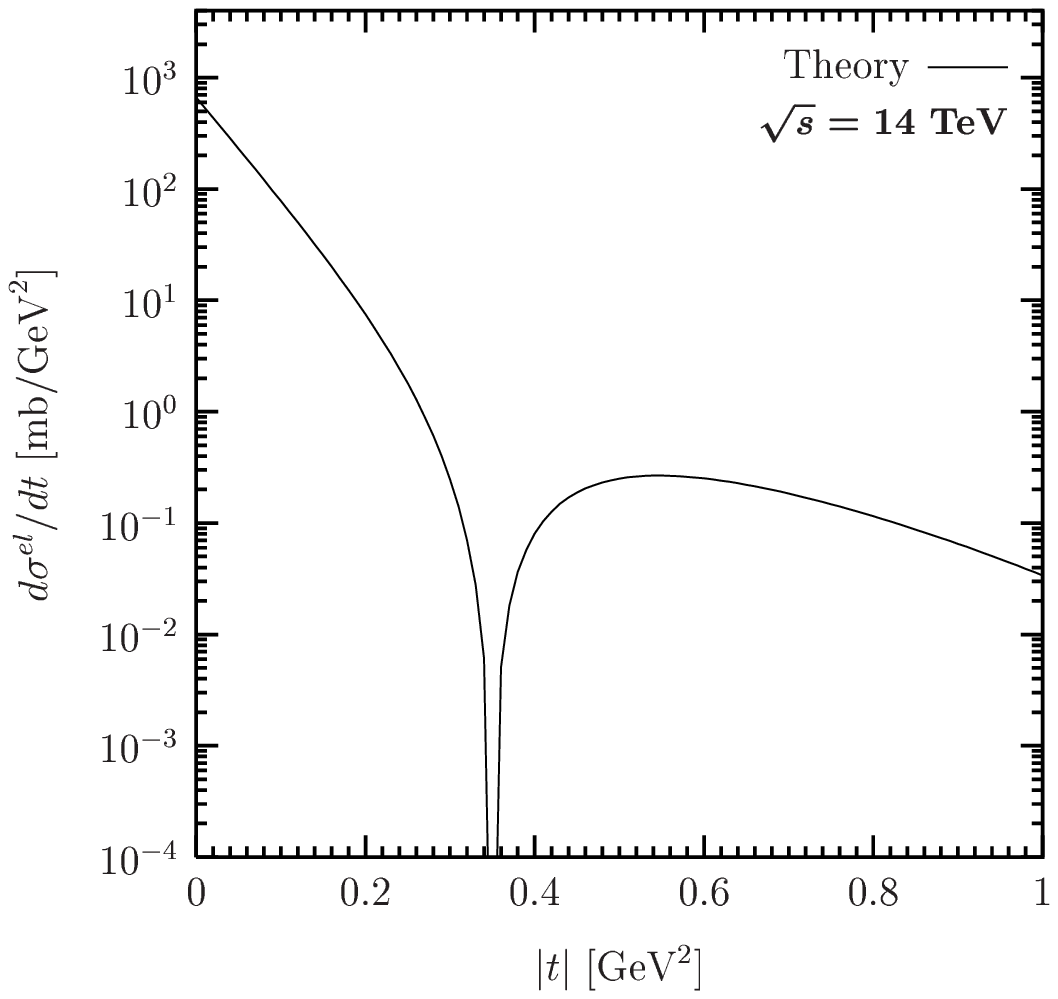,
      width=10.cm}} \protect\caption{ \small The prediction of our
    model for the $pp$ differential elastic cross section at LHC
    ($\sqrt{s} = 14\,\TeV$) is shown as a function of momentum
    transfer $|t|$ up to $1\,\GeV^2$.}
\label{Fig_dsigma_el_dt_pp_LHC}
\end{figure}

In Fig.~\ref{Fig_dsigma_el_dt_pp}, the differential elastic cross
section computed for $pp$ and $p\pbar$ scattering (solid line) is
shown as a function of $|t|=\vec{q}_{\!\perp}^2$ at $\sqrt{s} =
23.5,\,30.7,\,44.7,\,63,\,546$, and $1800\,\GeV$ and compared with
experimental data (open circles), where the $pp$ data at $\sqrt{s} =
23.5,\,30.7,\,44.7,\,\mbox{and}\,63\,\GeV$ were measured at the CERN
ISR~\cite{Amaldi:1980kd}, the $p\pbar$ data at $\sqrt{s} = 546\,\GeV$
at the CERN $Sp{\pbar}S$~\cite{Bozzo:1984ri}, and the $p\pbar$ data at
$\sqrt{s} = 1.8\,\TeV$ at the Fermilab
Tevatron~\cite{Amos:1989at,Amos:1990jh}. The prediction of our model
for the $pp$ differential elastic cross section at the CERN LHC,
$\sqrt{s} = 14\,\TeV$, is given in Fig.~\ref{Fig_dsigma_el_dt_pp_LHC}.

For all energies, the model reproduces the experimentally observed
diffraction pattern, i.e\ the characteristic {\em diffraction peak} at
small $|t|$ and the {\em dip} structure at medium $|t|$. As the energy
increases, also the {\em shrinking of the diffraction peak} is
described which reflects the rise of the slope parameter $B(s,t=0)$
already discussed above. The shrinking of the diffraction peak comes
along with a dip structure that moves towards smaller values of
$|t|$ as the energy increases. This motion of the dip is also
described approximately.

The dip in the theoretical curves reflects a change of sign in the
$T$-matrix
element~(\ref{Eq_model_purely_imaginary_T_amplitude_final_result}). As
the latter is purely imaginary, it is not surprising that there are
deviations from the data in the dip region. Here, the real part is
expected to be important~\cite{Amos:1990jh} which is in the small
$|t|$ region negligible in comparison to the imaginary part.

The difference between the $pp$ and $p\pbar$ data, a deep dip for $pp$
but only a bump or shoulder for $p\pbar$ reactions, requires a $C = -
1$ contribution. Besides the reggeon contribution at small
energies,\footnote{Zooming in on the result for $\sqrt{s} =
  23.5\,\GeV$, one finds further an underestimation of the data for
  all $|t|$ before the dip, which is correct as it leaves room for the
  reggeon contribution being non-negligible at small energies.} one
expects here an additional perturbative $C=-1$ contribution such as
three-gluon exchange~\cite{Fukugita:1979fe,Donnachie:1984hf+X} or an
odderon~\cite{Lukaszuk:1973nt+X,Rueter:1999gj,Dosch:2002ai}. In fact, allowing a
finite size diquark in the (anti-)proton an odderon appears that
supports the dip in $pp$ but leads to the shoulder in $p\pbar$
reactions~\cite{Dosch:2002ai}.

For the differential elastic cross section at the LHC energy,
$\sqrt{s} = 14\,\TeV$, the above findings suggest an accurate
prediction in the small-$|t|$ region but a dip at a position smaller
than the predicted value at $|t| \approx 0.35\,\GeV^2$. Our confidence
in the validity of the model at small $|t|$ is supported additionally
by the total cross section that fixes $d\sigma^{el}/dt(s,t=0)$ and is
in agreement with the cosmic ray data shown in
Fig.~\ref{Fig_sigma_tot}. Concerning our prediction for the dip
position, it is close to the value $|t| \approx 0.41\,\GeV^2$
of~\cite{Block:1999hu} but significantly below the value $|t| \approx
0.55\,\GeV^2$ of~\cite{Berger:1999gu}. Beyond the dip position, the
height of the computed shoulder is always above the data and, thus,
very likely to exceed also the LHC data. In comparison with other
works, the height of our shoulder is similar to the one
of~\cite{Block:1999hu} but almost one order of magnitude above the one
of~\cite{Berger:1999gu}.

Considering Figs.~\ref{Fig_dsigma_el_dt_pp}
and~\ref{Fig_dsigma_el_dt_pp_LHC} more quantitatively in the
small-$|t|$ region, one can use the well known parametrization of the
differential elastic cross section in terms of the slope parameter
$B(s)$ and the {\em curvature} $C(s)$
\be
        d\sigma^{el}/dt(s,t) 
        = d\sigma^{el}/dt(s,t=0)\,\exp\left[B(s)t+C(s)t^2\right]  
        \ .
\label{Eq_dsigma_el_dt_exp_parameterization}
\ee
Using $B(s)$ from the preceding section and assuming for the moment
$C(s) = 0$, one achieves a good description at small momentum
transfers and energies, which is consistent with the approximate
Gaussian shape of $J_{pp}(s,|\vec{b}_{\!\perp}|)$ at small energies
shown in Fig.~\ref{Fig_J_pp(b,s)}. The dip, of course, is generated by
the deviation from the Gaussian shape at small impact parameters.
According to (\ref{Eq_dsigma_el_dt_exp_parameterization}), the
shrinking of the diffraction peak with increasing energy reflects
simply the increasing interaction radius described by $B(s)$.

For small energies $\sqrt{s}$, our model reproduces the experimentally
observed change in the slope at $|t| \approx
0.25\,\GeV^2$~\cite{Barbiellini:1972ua+X} that is characterized by a
positive curvature. For LHC, we find clearly a negative value for the
curvature in agreement with~\cite{Block:1999hu} but in contrast
to~\cite{Berger:1999gu}. The change of sign in the curvature reflects
the transition of $J(s,|\vec{b}_{\!\perp}|)$ from the approximate
Gaussian shape at low energies to the approximate step-function
shape~(\ref{Eq_J_ab_asymptotic_energies}) at high energies.

Important for the good agreement with the data is the longitudinal
quark momentum distribution in the proton. Besides the slope
parameter, which characterizes the diffraction peak, also the dip
position is very sensitive to the distribution width $\Delta z_p$,
i.e.\ with $\Delta z_p= 0$ the dip position appears at more than 10\%
lower values of $|t|$. In the earlier \SVM\ 
approach~\cite{Berger:1999gu}, the reproduction of the correct dip
position was possible without the $z$-dependence of the hadronic wave
functions but a deviation from the data in the low-$|t|$ region had to
be accepted. In this low-$|t|$ region, we achieved a definite
improvement with the new correlation
functions~(\ref{Eq_MSV_correlation_functions}) and the minimal
surfaces used in our model.
\begin{figure}[htb]
  \centerline{\psfig{figure=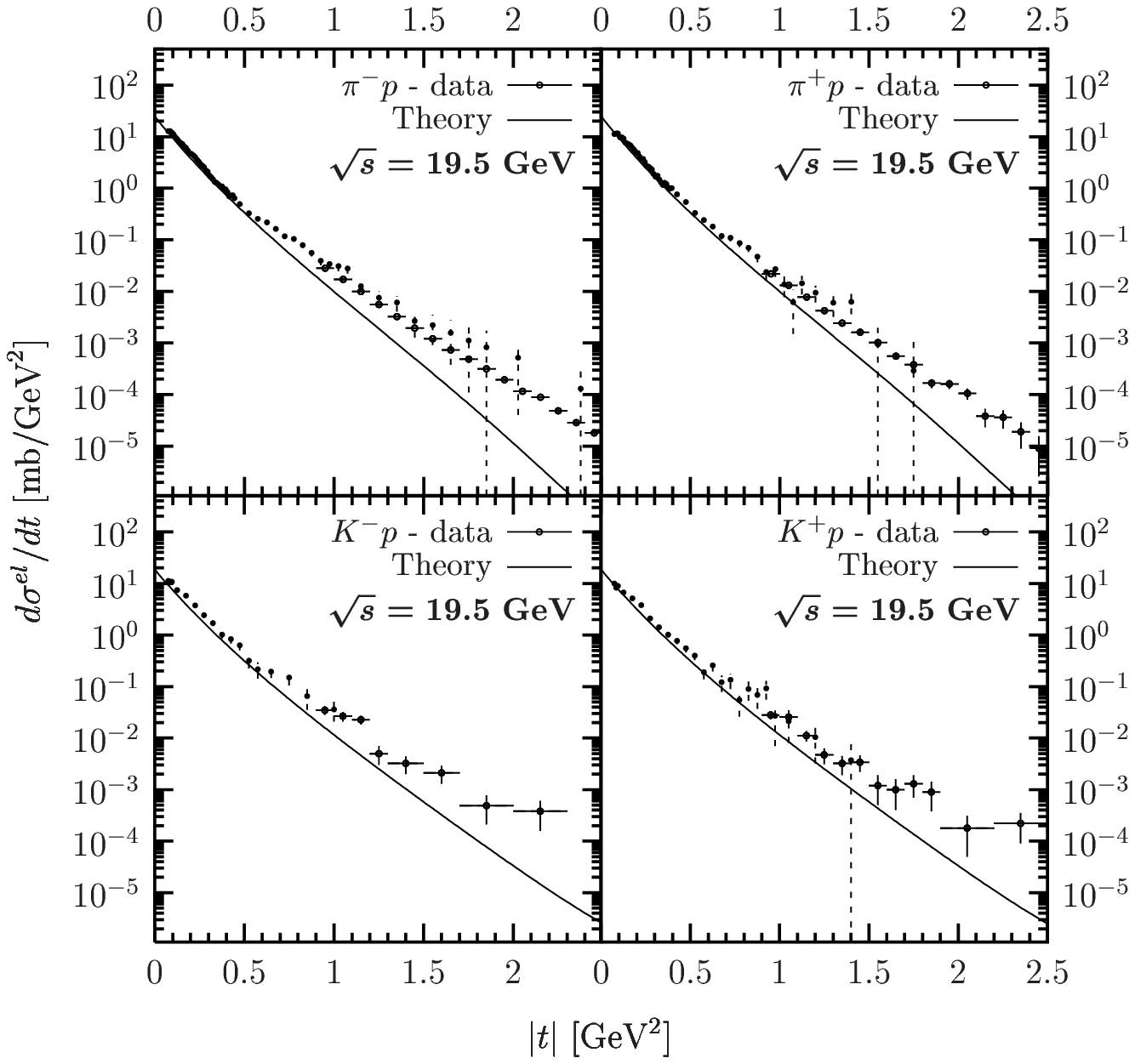,
      width=12.cm}} \protect\caption{ \small The differential elastic
    cross section $d\sigma^{el}/dt(s,t)$ is shown versus the momentum
    transfer $|t|$ for $\pi^{\pm}p$ and $K^{\pm}p$ reactions at the
    c.m.\ energy $\sqrt{s} = 19.5$ GeV. The model results (solid line)
    are compared with the data (closed circles with error bars)
    from~\cite{Akerlof:1976gk+X}.}
\label{Fig_dsigdt_kp_pip}
\end{figure} 

The differential elastic cross section computed for $\pi^{\pm}p$ and
$K^{\pm}p$ reactions has the same behavior as the one for $pp$
($p\pbar$) reactions. The only difference comes from the different
$z$-distribution widths, $\Delta z_{\pi}$ and $\Delta z_{K}$, and the
smaller extension parameters, $S_{\pi}$ and $S_{K}$, which shift the
dip position to higher values of $|t|$. This is illustrated in
Fig.~\ref{Fig_dsigdt_kp_pip}, where the model results (solid line) for
the $\pi^{\pm}p$ and $K^{\pm}p$ differential elastic cross section as
a function of $|t|$ are shown at $\sqrt{s} = 19.5\,\GeV$ in comparison
with experimental data (closed circles) from~\cite{Akerlof:1976gk+X}.
The deviations from the data towards large $|t|$ leave room for
odderon and reggeon contributions. Indeed, with a finite diquark size
in the proton, an odderon occurs that improves the description of the
data at large values of $|t|$~\cite{Berger:PhDthesis:1999}.

\subsection[The Elastic Cross Section $\sigma^{el}$, $\sigma^{el}/ \sigma^{tot}$, and $\sigma^{tot}/B$]{The Elastic Cross Section \boldmath$\sigma^{el}$, \boldmath$\sigma^{el}/ \sigma^{tot}$, and \boldmath$\sigma^{tot}/B$}

The {\em elastic cross section} obtained by integrating the
differential elastic cross section
\be
        \sigma^{el}(s) 
        = \int_0^{-\infty}\!dt\,\frac{d\sigma^{el}}{dt}(s,t) 
        = \int_0^{-\infty}\!dt\,\inv{16 \pi s^2}|T(s,t)|^2
\label{Eq_total_elastic_cross_section}
\ee
reduces for our purely imaginary
$T$-matrix~(\ref{Eq_model_purely_imaginary_T_amplitude_final_result})
to
\be
        \sigma^{el}(s) 
        = \int \!\!d^2b_{\!\perp}\,|J(s,|\vec{b}_{\!\perp}|)|^2 
        \ .
\label{Eq_total_elastic_cross_section_J}
\ee
Due to the squaring, it exhibits the saturation of
$J(s,|\vec{b}_{\!\perp}|)$ at the black disc limit more clearly than
$\sigma^{tot}(s)$. Even more transparent is the saturation in the
following ratios given here for a purely imaginary $T$-matrix
\bea
        \frac{\sigma^{el}}{\sigma^{tot}}(s) 
        & = & 
        \frac
        {\int\!d^2b_{\!\perp}\,|J(s,|\vec{b}_{\!\perp}|)|^2}
        {2\int\!d^2b_{\!\perp}\,J(s,|\vec{b}_{\!\perp}|)}
        \ ,
\label{Eq_sigma_el/sigma_tot} \\
        \frac{\sigma^{tot}}{B}(s) 
        & = & 
        \frac
        {\left(2\int\!d^2b_{\!\perp}\,J(s,|\vec{b}_{\!\perp}|)\right)^2}
        {\int\!d^2b_{\!\perp}\,|\vec{b}_{\!\perp}|^2\,J(s,|\vec{b}_{\!\perp}|)} 
\label{Eq_sigma_tot/B}
        \ ,
\eea
which are directly sensitive to the opacity of the particles. This
sensitivity can be illustrated within the approximation 
\be
        T(s,t) = i\, s\, \sigma^{tot}(s)\, \exp[B(s) t/2]
\label{Eq_T_matrix_exp_parameterization}
\ee
that leads to the differential cross
section~(\ref{Eq_dsigma_el_dt_exp_parameterization}) with $C(s) = 0$,
i.e.\ an exponential decrease over $|t|$ with a slope $B(s)$. As the
purely imaginary $T$-matrix
element~(\ref{Eq_T_matrix_exp_parameterization}) is equivalent to
\be
        J(s,|\vec{b}_{\!\perp}|)
        =(\sigma^{tot}/4\pi B)\,\exp[-|\vec{b}_{\!\perp}|^2/2B] 
        =(4\sigma^{el}/\sigma^{tot})\,\exp[-|\vec{b}_{\!\perp}|^2/2B]
        \ ,
\label{Eq_J(b,s)_exp_parameterization}
\ee
one finds that the above ratios are a direct measure for the opacity
at zero impact parameter
\be
        J(s,|\vec{b}_{\!\perp}|=0) 
        = (\sigma^{tot}/4\pi B)
        = (4\sigma^{el}/\sigma^{tot}) 
        \ .
\label{Eq_J(b=0,s)_exp_parameterization}
\ee
For a general purely imaginary $T$-matrix, $T(s,t) =
i\,s\,\sigma^{tot}\,g(|t|)$ with an arbitrary real-valued function
$g(|t|)$, $J(s,|\vec{b}_{\!\perp}|=0)$ is given by
$(\sigma^{el}/\sigma^{tot})$ times a pure number which depends on the
shape of $g(|t|)$.

\begin{figure}[p]
\centerline{\psfig{figure=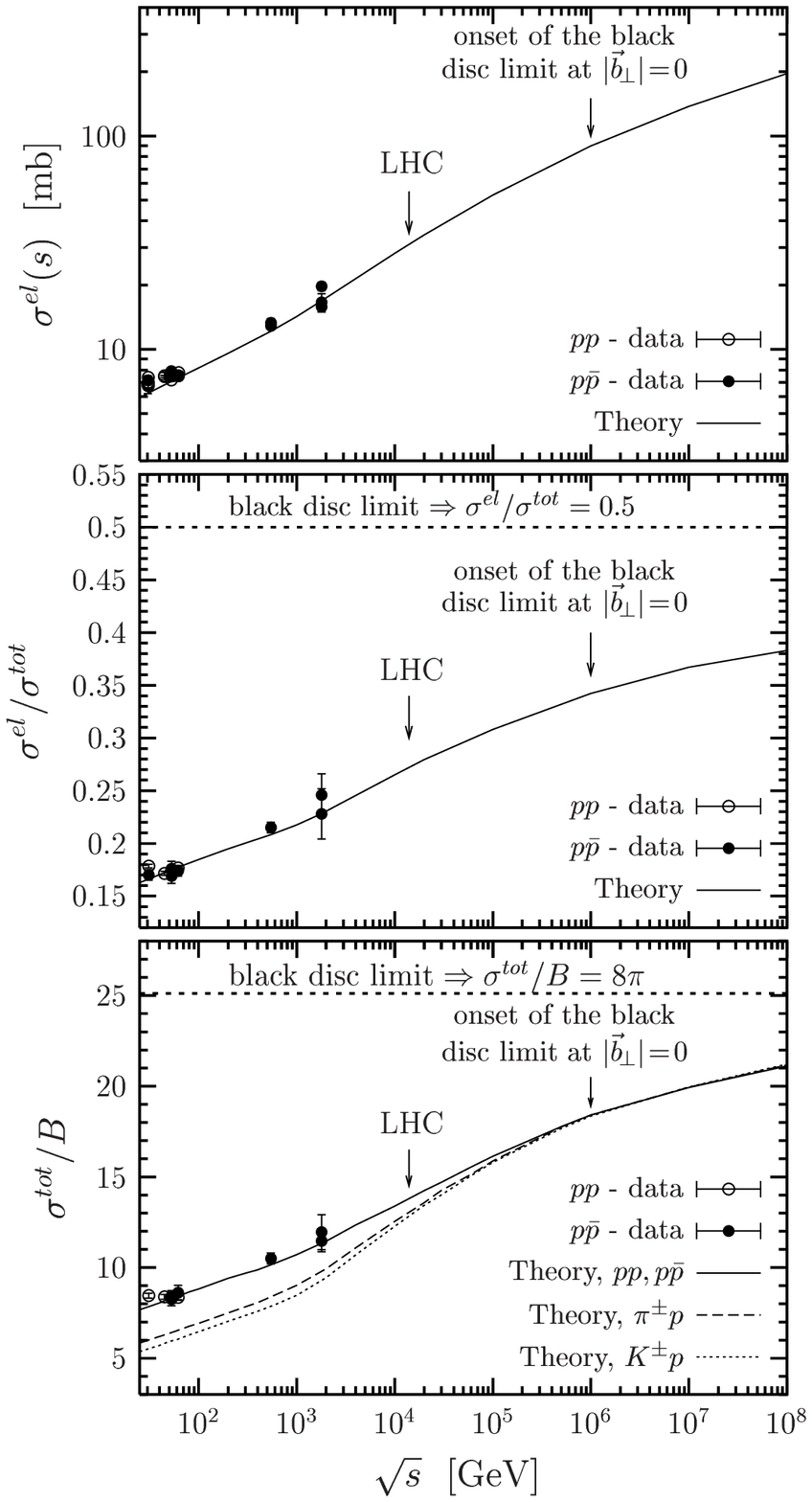, width=10.cm}}
\protect\caption{ \small The elastic cross section $\sigma^{el}$ and
  the ratios $\sigma^{el}/ \sigma^{tot}$ and $\sigma^{tot}/B$ are
  shown as a function of the c.m.\ energy $\sqrt{s}$. The model
  results for $pp$ ($p\pbar$), $\pi p$, and $K p$ scattering are
  represented by the solid, dashed and dotted lines, respectively. The
  experimental data for the $pp$ and $p\pbar$ reactions are indicated
  by the open and closed circles, respectively. The data for the
  elastic cross section are taken from~\cite{Groom:2000in} and the
  data for $\sigma^{tot}$ and $B$ from the references given in
  previous sections.}
\label{Fig_sigtot_el_and_ratios}
\end{figure}

We compute the elastic cross section $\sigma^{el}$ and the ratios
$\sigma^{el}/ \sigma^{tot}$ and $\sigma^{tot}/B$ in our model without
taking into account reggeons. In Fig.~\ref{Fig_sigtot_el_and_ratios},
the results for $pp$ and $p\pbar$ reactions (solid lines) are compared
with the experimental data (open and closed circles). The data for the
elastic cross section are taken from~\cite{Groom:2000in} and the data
for $\sigma^{tot}$ and $B$ from the references given in previous
sections.
For $pp$ ($p\pbar$) scattering, we indicate explicitly the prediction
for LHC at $\sqrt{s}=14\,\TeV$ and the onset of the black disc limit
at $\sqrt{s} = 10^6\,\GeV$. The model results for $\pi p$ and $K p$
reactions are presented as dashed and dotted line, respectively. For
the ratios, the asymptotic limits are indicated: Since the maximum
opacity or black disc limit governs the $\sqrt{s} \to \infty$
behavior, $\sigma^{el}/\sigma^{tot}$ ($\sigma^{tot}/B$) cannot exceed
$0.5$ ($8\pi$) in hadron-hadron scattering.

In the investigation of $pp$ ($p\pbar$) scattering, our theoretical
curves confront successfully the experimental data for the elastic
cross section and both ratios. At low energies, the data are
underestimated as reggeon contributions are not taken into account.
Again, the agreement is comparable to the one achieved
in~\cite{Block:1999hu} and better than in the approach
of~\cite{Berger:1999gu}, where $\sigma^{el}$ comes out too small due
to an underestimation of $d\sigma^{el}/dt$ in the low-$|t|$ region.

Concerning the energy dependence, $\sigma^{el}$ shows a similar
behavior as $\sigma^{tot}$ but with a more pronounced flattening
around $\sqrt{s} \gtsim 10^6\,\GeV$. This flattening is even stronger
for the ratios, drawn on a linear scale, and reflects very clearly the
onset of the black disc limit. As expected from the simple
approximation~(\ref{Eq_J(b=0,s)_exp_parameterization}),
$\sigma^{el}/\sigma^{tot}$ and $\sigma^{tot}/B$ show a similar
functional dependence on $\sqrt{s}$. At the highest energy shown,
$\sqrt{s} = 10^8\,\GeV$, both ratios are still below the indicated
asymptotic limits, which reflects that $J(s,|\vec{b}_{\!\perp}|)$
still deviates from the step-function
shape~(\ref{Eq_J_ab_asymptotic_energies}). The ratios
$\sigma^{el}/\sigma^{tot}$ and $\sigma^{tot}/B$ reach their upper
limits $0.5$ and $8\pi$, respectively, at asymptotic energies,
$\sqrt{s} \to \infty$, where the hadrons become infinitely large,
completely black discs.

Comparing the $pp$ ($p\pbar$) results with the ones for $\pi p$ and
$Kp$, one finds that the results for $\sigma^{tot}/B$ converge at high
energies as shown in Fig.~\ref{Fig_sigtot_el_and_ratios}. This follows
from the identical normalizations of the hadron wave functions that
lead to an identical black disc limit for hadron-hadron reactions.


%
%
\section{Conclusion}
\label{Sec_Conclusion}

We have developed a loop-loop correlation model combining perturbative and non-perturbative
QCD to compute high-energy reactions of hadrons and photons. We have
aimed at a unified description of hadron-hadron, photon-hadron, and
photon-photon reactions involving real and virtual photons as well.
Being particularly interested in saturation effects that manifest the
$S$-matrix unitarity, we have investigated the scattering amplitudes
in impact parameter space since the black disc limit of the profile
function is the most explicit signature of unitarity. Using a leading
twist, next-to-leading order DGLAP relation, we have also estimated
the impact parameter dependent gluon distribution of the proton
$xG(x,Q^2,|\vec{b}_{\perp}|)$ to study gluon saturation as a
manifestation of the $S$-matrix unitarity at small Bjorken $x$. In
addition, the calculated profile functions provide an intuitive
geometrical picture for the energy dependence of the cross sections
and allow us to localize saturation effects in the experimental
observables.

Following the functional integral approach to high-energy scattering
in the eikonal
approximation~\cite{Nachtmann:1991ua,Kramer:1990tr,Dosch:1994ym,Nachtmann:ed.kt},
the scattering hadrons and photons are described by light-like
Wegner-Wilson loops (color-dipoles) with size and orientation weighted
with appropriate light-cone wave functions~\cite{Dosch:1994ym}. The
resulting $S$-matrix element factorizes into the universal correlation
of two light-like Wegner-Wilson loops $S_{DD}$ (loop-loop correlation
function) and reaction-specific light-cone wave functions.  This
factorization has allowed us to provide a unified description of
hadron-hadron, photon-hadron, and photon-photon scattering. We have
used for hadrons the phenomenological Gaussian wave
function~\cite{Dosch:2001jg,Wirbel:1985ji} and for photons the
perturbatively derived wave function with running quark masses
$m_f(Q^2)$ to account for the non-perturbative region of low photon
virtuality $Q^2$~\cite{Dosch:1998nw}.

The loop-loop correlation function $S_{DD}$ has been computed in the
approach of Berger and Nachtmann~\cite{Berger:1999gu}. The loop-loop
correlation function $S_{DD}$ has been expressed in terms of surface
($S_{1,2}$) integrals over the gauge-invariant bilocal gluon field
strength correlator. We have divided this correlator into a
non-perturbative and a perturbative component. The stochastic vacuum
model (\SVM)~\cite{Dosch:1987sk+X} has been used for the
non-perturbative low frequency background field (long-distance
correlations) and perturbative gluon exchange for the additional high
frequency contributions (short-distance correlations) since this
combination is supported by lattice investigations of the gluon field
strength correlator~\cite{DiGiacomo:1992df+X,Meggiolaro:1999yn}. The
{\em exponential correlation function} used in the non-perturbative
component has been adopted directly from a lattice investigation of
the correlator~\cite{Meggiolaro:1999yn}. Since this correlation
function stays positive for all Euclidean distances, it is compatible
with a spectral representation of the correlation
function~\cite{Dosch:1998th}, which means a conceptual improvement
since the correlation function that has been used in earlier
applications of the \SVM\ becomes negative at large
distances~\cite{Dosch:1994ym,Rueter:1996yb+X,Dosch:1998nw,Rueter:1998up,D'Alesio:1999sf,Berger:1999gu,Dosch:2001jg}.
We have presented for the first time an explicit computation of the
surface integrals using minimal surfaces ($S_{1,2}$) bounded by the
Wegner-Wilson loops. This surface choice is usually used to obtain
Wilson's area law in Euclidean
space~\cite{Dosch:1987sk+X,Euclidean_Model_Applications}. Moreover,
the simplicity of the minimal surfaces has allowed us to identify very
clearly a string-string interaction in the non-perturbative component.

The strongest assumption in the presented work is the form of the
energy dependence introduced phenomenologically into the loop-loop
correlation function $S_{DD}$.  Motivated by the two-pomeron model of
Donnachie and Landshoff~\cite{Donnachie:1998gm+X}, we have ascribed to
the non-perturbative and to the perturbative component a weak and a
strong energy dependence, respectively. The constructed $T$-matrix
element shows Regge behavior at moderately high energies and contains
multiple gluonic interactions important to respect unitarity in impact
parameter space at ultra-high energies.

The model parameters have been adjusted to reproduce a wealth of
experimental data (including cosmic ray data) over a large range of
c.m.\ energies: total, differential, and elastic cross sections,
structure functions, and slope parameters --- including cosmic ray
data.  The model parameters that allowed a good fit to high-energy
scattering data are in good agreement with complementary
investigations: The parameters of the non-perturbative component ---
the correlation length $a$, the non-Abelian strength $\kappa$, and the
gluon condensate $G_2$ --- are constrained by lattice QCD
investigations, by the string tension $\sigma$ of a static
quark-antiquark pair, and by the SVZ gluon condensate $G_2$ essential
in QCD sum rule investigations. The parameters of the perturbative
component have not been adjusted.  We have used the $\rho$-meson mass
for the effective gluon mass $m_G$ representing the IR regulator and
have determined $M^2$ so that the strong coupling freezes for large
distance scales at $\alphaS=0.4$, where our non-perturbative component
is at work according to a low energy
theorem~\cite{Rueter:1995cn+X,Euclidean_Model_Applications}.  For the
energy dependence, the exponents of the Donnachie-Landshoff
two-pomeron fit, $\epsilon_{soft}$ and $\epsilon_{hard}$, have been
used as an orientation for our energy exponents $\epsilon^{\nprt}$ and
$\epsilon^{\pert}$. Besides these parameters describing the universal
loop-loop correlation function $S_{DD}$, the reaction-dependent
parameters in the light-cone wave functions are also consistent with
other approaches: In the hadron wave functions, the transverse
extension parameters $S_h$ have been found in good agreement with the
corresponding electromagnetic radii~\cite{Dosch:2001jg} while the
width of the longitudinal quark momentum distributions $\Delta z_h$
has been computed from~\cite{Wirbel:1985ji}. In the photon wave
function, the running quark masses that coincide with the current
quark masses for large $Q^2$ and the constituent quark masses for
small $Q^2$~\cite{Dosch:1998nw} have been chosen in agreement with sum
rule derivations.

Having adjusted the model parameters, we have studied $S$-matrix
unitarity limits of the scattering amplitudes in impact parameter
space. On the basis of dipole-dipole scattering, we have found
explicitly that our model preserves the unitarity condition and
attains the black disc limit at ultra-high energies. The profile
functions have been calculated for proton-proton and photon-proton
scattering and have shown very clearly that the interacting particles
become blacker and larger with increasing energy. At ultra-high
energies, the opacity saturates at the black disc limit while the
transverse expansion of the scattered particles continues. Moreover, in
longitudinal photon-proton scattering, we have observed that with
increasing photon virtuality $Q^2$ not only the maximum opacity
increases but also the energy at which it is reached for zero impact
parameter.

Using a leading twist, next-to-leading order QCD relation between the
gluon distribution of the proton $xG(x,Q^2)$ and the longitudinal
proton structure function
$F_L(x,Q^2)$~\cite{Martin:1988vw,Cooper-Sarkar:1988ds+X}, we have
related the impact parameter dependent gluon distribution
$xG(x,Q^2,|\vec{b}_{\perp}|)$ to the profile function for longitudinal
photon-proton scattering and found low-$x$ saturation of
$xG(x,Q^2,|\vec{b}_{\perp}|)$ as a manifestation of $S$-matrix
unitarity. In accordance with the profile function,
$xG(x,Q^2,|\vec{b}_{\perp}|)$ decreases from the center towards the
periphery of the proton.  With increasing photon virtuality $Q^2$, the
increase of $xG(x,Q^2,|\vec{b}_{\perp}|=0)$ becomes stronger towards
small $x$ and the saturation value of $xG(x,Q^2,|\vec{b}_{\perp}|=0)$
increases but is reached at decreasing values of $x$. In contrast, at
fixed $Q^2$, the integrated gluon distribution $xG(x,Q^2)$ does not
saturate because of the growth of the proton radius with decreasing
$x$ observed in our approach.  Similar results are obtained in
complementary
approaches~\cite{Mueller:1986wy,Mueller:1999wm,Iancu:2001md}.

More model dependent are the specific energies where these saturation
effects set in. The profile function saturates the black disc limit at
zero impact parameter for $\sqrt{s} \gtsim 10^6\,\GeV$ in
proton-proton scattering and for $\sqrt{s} \gtsim 10^7\,\GeV$ in
longitudinal photon-proton scattering with $Q^2 \gtsim 1\,\GeV^2$. In
both reactions, the wave function normalization determines the maximum
opacity. The saturation of $xG(x,Q^2,|\vec{b}_{\perp}|)$ occurs in our
approach for $Q^2\gtsim 1\,\GeV^2$ at values of $x \ltsim 10^{-10}$,
far below the HERA and THERA range.

For proton-proton scattering, we have observed that the rise of the
total and elastic cross section becomes weaker for $\sqrt{s} \gtsim
10^6\,\GeV$ due to the onset of the black disc limit at
$|\vec{b}_{\perp}|=0$ in the profile function. This saturation of the
profile function has become even more apparent in the ratios
$\sigma^{el}/\sigma^{tot}$ and $\sigma^{tot}/B$ which are a measure of
the proton opacity. In contrast, no saturation effect has been
observed in the slope parameter $B(s)$ which is a measure of the
transverse expansion of the proton. Considering the differential
elastic cross section $d\sigma^{el}/dt$, the model has described the
diffraction pattern and also the shrinkage of the diffraction peak
with increasing energy in good agreement with experimental data at
small momentum transfers $|t|$. Around the dip region, where a real
part is expected to be important, deviations from the data have
reflected that our $T$-matrix is purely imaginary. Our predictions for
proton-proton scattering at LHC ($\sqrt{s} = 14\,\TeV$) have been a
total cross section of $\sigma^{tot}_{pp} = 114.2\,\mb$ in good
agreement with the cosmic ray data and a differential elastic cross
section $d\sigma^{el}/dt$ with a slope parameter of $B =
21.26\,\GeV^{-2}$, a negative curvature, $C<0$, and a dip at $|t|
\approx 0.35\,\GeV^2$.

For pion-proton and kaon-proton scattering, results analogous to
proton-proton scattering have been obtained but with a slightly
stronger rise observed in the total cross section. This has been
traced back to the smaller size of pions and kaons in comparison to
protons, $S_p = 0.86\,\fm > S_{\pi} = 0.607\,\fm > S_{K} = 0.55\,\fm$,
and the perturbative component becoming increasingly important with
decreasing dipole sizes involved. Furthermore, a weak convergence of
the ratios $\sigma^{tot}_{pp}/\sigma^{tot}_{\pi p}$ and
$\sigma^{tot}_{pp}/\sigma^{tot}_{Kp}$ towards unity has been observed
as the energy increases. The smaller size of the pion and kaon has
also been reflected in the differential elastic cross sections
$d\sigma^{el}/dt$, where the dip is shifted towards larger values of
$|t|$.

For photon-proton and photon-photon reactions, an even stronger rise
of the total cross section has been observed with increasing energy.
As illustrated in the proton structure function $F_2(x,Q^2)$, this
rise becomes steeper with increasing photon virtuality $Q^2$. Again,
we have traced back the strong energy boost to the growing importance
of the perturbative component with decreasing dipole size.  Besides
some deviations from the experimental data with increasing $Q^2$, our
model has described $\sigma^{tot}_{\gamma^{*} p}(s,Q^2)$ successfully
in the low-$Q^2$ region where the running quark masses become
constituent quark masses.

We plan to present complementary investigations within our model in
future work. Going to momentum space, we calculate on the two-gluon
exchange level the unintegrated gluon distribution of the proton.
Insight into the non-perturbative structure of this distribution
can be gained from the non-perturbative component of our
model~\cite{K-Space_Investigations}. In Euclidean space-time, we
compute the static quark-antiquark potential, the associated flux
tube, and the van der Waals force between two static
color-dipoles~\cite{Euclidean_Model_Applications}. It is a long-range
project to implement the energy dependence more fundamentally. In a
recent attempt, the energy dependence of high-energy scattering has
been related successfully to critical properties of an effective near
light-cone Hamiltonian in a non-perturbative lattice
approach~\cite{Pirner:2001pv}. Furthermore, the correlation of inclined
Euclidean Wegner-Wilson loops generates energy dependence after
analytic continuation to Minkowski
space-time~\cite{Meggiolaro:1996hf+X}. Here, encouraging new results
obtained with instantons~\cite{Shuryak:2000df} and within the AdS/CFT
correspondence~\cite{Janik:1999zk+X} have to be compared to
calculations in the stochastic vacuum model.

\section*{Acknowledgements}

We thank H.~G.~Dosch very much for his continuing interest in this
work which parallelled some of his own investigations. The suggestions
of C.~Ewerz, J.~H\"ufner and O.~Nachtmann have helped us in finding
the right perspective in many cases. A.~Shoshi and F.~Steffen would
also like to thank A.~Polleri, A.~H.~Mueller and J.~Raufeisen for
interesting and insightful discussions. A.~Shoshi wants to express his
gratitude to Yu.~Ivanov for his support in computational issues. The
authors thank S.~Bartsch for the careful reading of the manuscript.


%
%
\begin{appendix}
%
\section{Hadron and Photon Wave Functions}
\label{Sec_Wave_Functions}

The light-cone wave functions $\psi_i(z_i,\vec{r}_i)$ provide the
distribution of transverse size and orientation ${\vec r}_{i}$ and
longitudinal quark momentum fraction $z_i$ to the light-like
Wegner-Wilson loops $W[C_i]$ that represent the scattering
color-dipoles. In this way, they specify the projectiles as mesons,
baryons described as quark-diquark systems, or photons that fluctuate
into a quark-antiquark pair before the interaction.

\subsection*{The Hadron Wave Function}

In this work, mesons and baryons are assumed to have a quark-antiquark
and quark-diquark valence structure, respectively. As quark-diquark
systems are equivalent to quark-antiquark systems~\cite{Dosch:1989hu},
this allows us to model not only mesons but also baryons as
color-dipoles represented by Wegner-Wilson loops. To characterize
mesons and baryons, we use the phenomenological Gaussian
Wirbel-Stech-Bauer ansatz~\cite{Wirbel:1985ji}
\be
        \psi_h(z_i,\vec{r}_i) 
        = \sqrt{\frac{z_i(1-z_i)}{2 \pi S_h^2 N_h}}\, 
        e^{-(z_i-\inv{2})^2 / (4 \Delta z_h^2)}\,  
        e^{-|\vec{r}_i|^2 / (4 S_h^2)} 
        \ ,
\label{Eq_hadron_wave_function}
\ee
where the hadron wave function normalization to unity
\be
        \int \!\!dz_i d^2r_i \ |\psi_i(z_i,\vec{r}_i)|^2 = 1  
        \ ,
\label{Eq_hadron_wave_function_normalization}
\ee
requires the normalization constant
\be
        N_h = \int_0^1 dz_i \ z_i(1-z_i) \ e^{-(z_i-\inv{2})^2 / (2
        \Delta z_h^2)} 
        \ .
\label{Eq_N_h}
\ee
The different hadrons considered --- protons, pions, and kaons --- are
specified by $\Delta z_h$ and $S_h$ providing the width for the
distributions of the longitudinal momentum fraction carried by the
quark $z_i$ and transverse spatial extension $|\vec{r}_i|$,
respectively. In this work, the extension parameter $S_h$ is a fit
parameter that should resemble approximately the electromagnetic
radius of the corresponding hadron~\cite{Dosch:2001jg}, while $\Delta z_h =
w/(\sqrt{2}\,m_h)$~\cite{Wirbel:1985ji} is fixed by the hadron mass
$m_h$ and the value $w = 0.35 - 0.5\,\GeV$ extracted from experimental
data. We find for (anti-)protons $\Delta z_p = 0.3$ and $S_p =
0.86\,\fm$, for pions $\Delta z_{\pi} = 2$ and $S_{\pi} = 0.607\,\fm$,
and for kaons $\Delta z_{K} = 0.57$ and $S_{K} = 0.55\,\fm$ which are
the values used in the main text. For convenience they are summarized
in Table~\ref{Tab_Hadron_Parameters}.
\begin{table}
\caption{\small Hadron Parameters} 
\vspace{0.3cm}
\centering      
\begin{tabular}{|l|l|l|}\hline
Hadron        & $\Delta z_h$    & $S_h\;[\fm]$  \\ [1ex] \hline\hline       
$p, \bar{p}$    & $0.3$           & $0.86$  \\ \hline
$\pi^{\pm}$     & $2$             & $0.607$ \\ \hline
$K^{\pm}$      & $0.57$          & $0.55$ \\ \hline
\end{tabular}
\label{Tab_Hadron_Parameters}
\end{table}

Concerning the quark-diquark structure of the baryons, the more
conventional three-quark structure of a baryon would complicate the
model significantly but would lead to similar predictions once the
model parameters are readjusted~\cite{Dosch:1994ym}. In fact, there
are also physical arguments that favor the quark-diquark structure of
the baryon such as the $\delta I = 1/2$ enhancement in
semi-leptonic decays of baryons~\cite{Dosch:1989hu} and the strong
attraction in the scalar diquark channel in the instanton
vacuum~\cite{Schafer:1994ra}.

\subsection*{The Photon Wave Function}   

The photon wave function $\psi_{\gamma}(z_i,\vec{r}_i,Q^2)$ describes
the fluctuation of a photon with virtuality $Q^2$ into a
quark-antiquark pair with longitudinal quark momentum fraction $z_i$
and spatial transverse size and orientation $\vec{r}_i$. The
computation of the corresponding transition amplitude $\langle
q\qbar(z_i,\vec{r}_i)|\gamma^*(Q^2)\rangle$ can be performed
conveniently in light-cone perturbation theory~\cite{Bjorken:1971ah+X}
and leads to the following squared wave functions for transverse $(T)$
and longitudinally $(L)$ polarized photons~\cite{Nikolaev:1991ja}
\bea
\!\!\!\!\!\!\!\!\!\!\!\!\!|\psi_{\gamma_T^*}(z_i,\vec{r}_i,Q^2)|^2\! 
        &\!\!=\!\!&\!\frac{3\,\alphaEM}{2\,\pi^2} \sum_f e_f^2
                \left\{ 
                  \left[ z_i^2 + (1-z_i)^2\right]\,\epsilon_f^2\,K_1^2(\epsilon_f\,|\vec{r}_i|) 
                  + m_f^2\,K_0^2(\epsilon_f\,|\vec{r}_i|) 
                \right\}
        \label{Eq_photon_wave_function_T_squared} \\
\!\!\!\!\!\!\!\!\!\!\!\!\!|\psi_{\gamma_L^*}(z_i,\vec{r}_i,Q^2)|^2\! 
        &\!\!=\!\!&\!\frac{3\,\alphaEM}{2\,\pi^2} \sum_f e_f^2
                \left\{ 4\,Q^2\,z_i^2(1-z_i)^2\,K_0^2(\epsilon_f\,|\vec{r}_i|) \right\},
        \label{Eq_photon_wave_function_L_squared}
\eea
where $\alphaEM$ is the fine-structure constant, $e_f$ is the electric
charge of the quark with flavor $f$, and $K_0$ and $K_1$ are the modified
Bessel functions (McDonald functions). In the above expressions,
\be
        \epsilon_f^2 = z_i(1-z_i)\,Q^2 + m_f^2
\label{Eq_photon_extension_parameter}
\ee
controlls the transverse size(-distribution) of the emerging dipole,
$|\vec{r}_i| \propto 1/ \epsilon_f$, that depends on the quark flavor
through the current quark mass $m_f$.

For small $Q^2$, the perturbatively derived wave functions,
(\ref{Eq_photon_wave_function_T_squared}) and
(\ref{Eq_photon_wave_function_L_squared}), are not appropriate since
the resulting large color-dipoles, i.e.\ $|\vec{r}_i| \propto 1/m_f
\gg 1\,\fm$, should encounter non-perturbative effects such as
confinement and chiral symmetry breaking. To take these effects into
account the vector meson dominance (VMD) model~\cite{Bauer:1978iq} is
usually used. However, the transition from the ``partonic'' behavior
at large $Q^2$ to the ``hadronic'' one at small $Q^2$ can be modelled
as well by introducing $Q^2$-dependent quark masses, $m_f = m_f(Q^2)$,
that interpolate between the current quarks at large $Q^2$ and the
constituent quarks at small $Q^2$~\cite{Dosch:1998nw}. Following this
approach, we use~(\ref{Eq_photon_wave_function_T_squared})
and~(\ref{Eq_photon_wave_function_L_squared}) also in the low-$Q^2$
region but with the running quark masses
\bea
        m_{u,d}(Q^2) 
        &=& 0.178\,\GeV\,(1-\frac{Q^2}{Q^2_{u,d}})\,\Theta(Q^2_{u,d}-Q^2) 
        \ , 
        \label{Eq_m_ud_(Q^2)}\\
        m_s(Q^2) 
        &=& 0.121\,\GeV + 0.129\,\GeV\,(1-\frac{Q^2}{Q^2_s})\,\Theta(Q^2_s-Q^2) 
        \label{Eq_m_s_(Q^2)}
        \ ,
\eea
and the fixed charm quark mass
\be
        m_c = 1.25\,\GeV
        \ ,
\ee
where the parameters $Q^2_{u,d} = 1.05\,\GeV^2$ and $Q^2_s =
1.6\,\GeV^2$ are taken directly from~\cite{Dosch:1998nw} while we
reduced the values for the constituent quark masses $m_f(Q^2 = 0)$
of~\cite{Dosch:1998nw} by about 20\%. The smaller constituent quark
masses are necessary in order to reproduce the total cross sections
for $\gamma^* p$ and $\gamma^* \gamma^*$ reactions at low $Q^2$.
Similar running quark masses are obtained in a QCD-motivated model of
the spontaneous chiral symmetry breaking in the instanton
vacuum~\cite{Petrov:1998kf} that improve the description of $\gamma^*
p$ scattering at low $Q^2$~\cite{Martin:1999bh+X}.


%
%
\section{Correlation Functions}
\label{Sec_Correlation_Functions}

In this Appendix, we describe explicitly the way from the simple
exponential correlation functions in Euclidean
space-time~(\ref{Eq_MSV_correlation_functions}) to their transverse
Fourier transforms in Minkowski space-time,
(\ref{Eq_F2[i_D_confining]})
and~(\ref{Eq_F2[i_D_non-confining_prime]}). The first step in this
procedure is the Fourier transformation of the exponential correlation
functions~(\ref{Eq_MSV_correlation_functions}) in four-dimensional
Euclidean space-time
\bea
        \tilde{D}^{E}(K^2) & \! = \! & \tilde{D}^{E}_{1}(K^2)
        = \int \! d^4Z \,D^{E}(Z^2/a^2)\,e^{iKZ} 
        \nonumber \\
        & \! = \! &   \int_0^\infty \!\!\! d|Z|\,|Z|^3 
                \int_0^\pi    \!\!\! d\phi_3\,\sin^2\!\phi_3 
                \int_0^\pi    \!\!\! d\phi_2\,\sin\!\phi_2 
                \int_0^{2\pi} \!\!\! d\phi_1 
                \,D^{E}(Z^2/a^2)\,e^{-i|K||Z|\cos\!\phi_3}
        \nonumber\\
        & \! = \! &   \frac{4\pi^2}{|K|} 
                \int_0^\infty \!\!\! d|Z|\,|Z|^2
                \,D^{E}(Z^2/a^2)\, J_1(|K||Z|) \, = \,\frac{12\pi^2}{a\,(K^2+a^{-2})^\frac{5}{2}}
        \ ,
\label{Eq_D^E(K^2)_for_exp_correlation}
\eea
where $J_1$ is the $1st$ order Bessel function of the first kind.
Here, the Euclidean metric $-\delta_{\mu\nu}$ and four-dimensional
polar coordinates and the corresponding four-volume element $d^4Z =
d|Z|\,|Z|^3\,d\phi_3\,\sin^2\!\phi_3\,d\phi_2\,\sin\!\phi_2\,d\phi_1$
have been used. With~(\ref{Eq_D^E(K^2)_for_exp_correlation}), one
obtains 
\be
        \tilde{D}^{\prime\,E}_1(K^2) 
                := \frac{d}{dK^2}\,\tilde{D}^{E}_{1}(K^2) 
                 = -\,\frac{30\pi^2}{a\,(K^2+a^{-2})^\frac{7}{2}}
        \ .
\label{Eq_D_prime^E(K^2)_for_exp_correlation}
\ee
Now,~(\ref{Eq_D^E(K^2)_for_exp_correlation})
and~(\ref{Eq_D_prime^E(K^2)_for_exp_correlation}) are analytically
continued to Minkowski space-time, $K_4 \to i k^0$ or equivalently
$-\delta_{\mu\nu} \to g_{\mu\nu} = \mbox{diag}(1,-1,-1,-1)$,
\bea
        \tilde{D}(k^2) 
        =  -\,\frac{12\,\pi^2\,i}{a\,(-k^2+a^{-2})^\frac{5}{2}} \ , \quad\quad\quad
        \tilde{D}^{\prime}_1(k^2) 
        =  -\,\frac{30\,\pi^2\,i}{a(-k^2+a^{-2})^\frac{7}{2}} \ .
\label{Eq_D_(prime)(k^4)_for_exp_correlation}
\eea
Setting $k^0 = k^3 = 0$, which is enforced in the computation of
$\chi$ by $\delta$-functions, one finds $k^2=-{\vec k}^2_{\!\perp}$ and consequently
\be
        \hphantom{-}
        \tilde{D}^{(2)}({\vec k}^2_{\!\perp}) 
        =  -\,\frac{12\,\pi^2\,i}{a\,({\vec k}^2_{\!\perp} + a^{-2})^\frac{5}{2}} \ , \quad\quad\quad
        \hphantom{-}
        \tilde{D}^{\prime \,(2)}_1({\vec k}^2_{\!\perp}) 
        =  -\,\frac{30\,\pi^2\,i}{a({\vec k}^2_{\!\perp} + a^{-2})^\frac{7}{2}} \ .
\label{Eq_D_(prime)(k^2)_for_exp_correlation}
\ee
The transverse Fourier transformation~(\ref{Eq_transverse_Fourier_transform}) of these two expressions is the remaining step that leads directly to~(\ref{Eq_F2[i_D_confining]}) and~(\ref{Eq_F2[i_D_non-confining_prime]}).


%
\end{appendix}
%
%
  
%
%
\end{document}